\documentclass{aa}  
\usepackage{graphicx}
\usepackage{txfonts}
\usepackage{natbib}
\usepackage{lipsum}

\begin{document} 

\title{VLT/SPHERE robust astrometry of the HR8799 planets \\ at milliarcsecond-level accuracy\thanks{Based on observations collected at the European Organisation for Astronomical Research in the Southern Hemisphere under ESO programme 60.A-9352.}}
\subtitle{Orbital architecture analysis with \texttt{PyAstrOFit}}

\author{O.~Wertz\inst{1}\fnmsep\thanks{Current address: Argelander-Institut f\"ur Astronomie, Auf dem H\"ugel 71, D-53121 Bonn. \email{owertz@astro.uni-bonn.de}}
            \and O.~Absil\inst{1}\fnmsep\thanks{F.R.S.-FNRS Research Associate}
            \and C.~A.~G\'omez Gonz\'alez\inst{1}
            \and J.~Milli\inst{2}
            \and J.~H.~Girard\inst{2}                        
            \and D.~Mawet\inst{3,4}
            \and L.~Pueyo\inst{5}
            }

\institute{Space sciences, Technologies and Astrophysics Research (STAR) Institute, Universit\'e de Li\`ege, 19c All\'ee du Six Ao\^ut, B-4000 Li\`ege, Belgium.
              \and European Southern Observatory, Alonso de Cordova 3107, Vitacura, Casilla 19001, Santiago de Chile, Chile
              \and Department of Astronomy, California Institute of Technology, 1200 E. California Blvd, MC 249-17, Pasadena, CA 91125 USA
              \and Jet Propulsion Laboratory, California Institute of Technology, 4800 Oak Grove Drive, Pasadena, CA 91109, USA
              \and Space Telescope Science Institute, 3700 San Martin Drive, Baltimore, MD 21218 USA
              }

\date{Received April 15, 2016; accepted October 7, 2016}

  \abstract
   {HR8799 is orbited by at least four giant planets, making it a prime target for the recently commissioned Spectro-Polarimetric High-contrast Exoplanet REsearch (VLT/SPHERE). As such, it was observed on five consecutive nights during the SPHERE science verification in December 2014.}
   {We aim to take full advantage of the SPHERE capabilities to derive accurate astrometric measurements based on H-band images acquired with the Infra-Red Dual-band Imaging and Spectroscopy (IRDIS) subsystem, and to explore the ultimate astrometric performance of SPHERE in this observing mode. We also aim to present a detailed analysis of the orbital parameters for the four planets.}
   {We performed a thorough post-processing of the IRDIS images with the Vortex Imaging Processing (\texttt{VIP}) package to derive a robust astrometric measurement for the four planets. This includes the identification and careful evaluation of the different contributions to the error budget, including systematic errors. Combining our astrometric measurements with the ones previously published in the literature, we constrain the orbital parameters of the four planets using \texttt{PyAstrOFit}, our new open-source python package dedicated to orbital fitting using Bayesian inference with Monte-Carlo Markov Chain sampling.}
   {We report the astrometric positions for epoch 2014.93 with an accuracy down to 2.0~mas, mainly limited by the astrometric calibration of IRDIS. For each planet, we derive the posterior probability density functions for the six Keplerian elements and identify sets of highly probable orbits. For planet d, there is clear evidence for nonzero eccentricity $(e \sim 0.35)$, without completely excluding solutions with smaller eccentricities. 
   The three other planets are consistent with circular orbits, although their probability distributions spread beyond $e=0.2$, and show a peak at $e\simeq 0.1$ for planet e. 
   The four planets have consistent inclinations of about $30\degr$ with respect to the sky plane, but the confidence intervals for the longitude of ascending node are disjoint for planets b and c, and we find tentative evidence for non-coplanarity between planets b and c at the $2\sigma$ level.}
   {}


\maketitle

%
%

\titlerunning
\authorrunning

\section{Introduction}
\label{section:introduction}

Since its discovery by \citet{Marois_2008_HR8799}, the HR8799 planetary system has been and still remains one of the most intriguing among the thousands of known planetary systems. Composed of at least four giant planets in a range of angular separations of about $0\farcs 4$ to $1\farcs7$ \citep{Marois_2010b_HR8799}, and of two dusty debris belts \citep{Su_2009_HR8799dd, Hughes_2011_HR8799dd, Matthews_2014_HR8799dd, Booth16}, it has been the focus of many different studies, including dynamical stability analyses to constrain the global orbital motion and estimate the masses of the four planets \citep[see e.g.][]{Gozdziewski_2009_HR8799, Reidemeister_2009_HR8799, Fabrycky_2010_HR8799, Soummer_2011_HR8799, Currie_2012_HR8799, Currie_2014_HR8799, Gozdziewski_2014_HR8799, Maire_2015_HR8799}. This dynamical approach allows the orbits of the four planet to be simultaneously constrained, but requires strong assumptions, such as coplanar (but eccentric) or circular (but not necessarily coplanar) orbits. The individual analysis of each planet offers an alternative method to constraint the orbital architecture. To this aim, nonlinear least-squares fits of Keplerian elements (semi-major axis $a$, eccentricity $e$, inclination $i$, longitude of ascending  node $\Omega$, argument of the periastron $\omega$, and time of periastron passage $t_p$) have been performed \citep[see e.g.][]{Lafreniere_2009_HR8799, Bergfors_2011_HR8799, Esposito_2013_HR8799, Zurlo_2016_HR8799}.

Recently, \citet{Pueyo_2015_HR8799} proposed an in-depth analysis of the HR8799bcde orbital motion. The authors carried out a Bayesian analysis based on Monte-Carlo Markov Chain (MCMC) techniques adopting both a Metropolis Hastings algorithm \citep{MacKay_2003_InformationTheory, Ford_2005_MCMC, Ford_2006_MCMC} and an affine-invariant ensemble sampler \citep{emcee}. This approach echoes the works published in \citet{Chauvin_2012_betaPic} for $\beta$ Pictoris b, in \citet{Kalas_2013_Fomalhaut} for Fomalhault b and more recently in \citet{Beust_2016_MCMC} for Fomalhault b and PZ Telescopii B. Among other things, \citet{Pueyo_2015_HR8799} discussed the coplanarity of the system, the orbital eccentricities of the planets, the possibility for mean motion resonances, and the role of HR8799d in possible dynamical interactions in the youth of this system. They also estimated the dynamical masses of HR8799bcde by computing the fraction of allowable orbits that pass the so-called close-encounter test. As pointed out in \citet{Pueyo_2015_HR8799}, unaccounted biases and/or systematically underestimated error bars on the planets astrometry affect the MCMC results \citep[see e.g.][]{Givens_2012_stat} and may lead to a biased estimation of the confidence intervals for the orbital parameters. Studying the astrometric history of HR8799 reveals indeed that the errors affecting some positions are most probably underestimated, as one can readily identify pairs or sets of positions that are not consistent with each other within their error bars, or cannot be modeled with a unique orbit. This was one of the incentives of the study presented by \citet{Konopacky16}, who very recently re-reduced all the Keck/NIRC2 observations of HR8799 to come up with a self-consistent data set free from variable instrument-related biases. This consistent data set was then used to derive updated probability distributions for the elements of the planetary orbits based on Monte Carlo simulations.

With the advent of second-generation high-contrast planet imagers like the Spectro-Polarimetric High-contrast Exoplanet REsearch \citep[SPHERE,][]{Beuzit_2008_SPHERE} at the Very Large Telescope (VLT), obtaining astrometric measurements of directly imaged planets is now becoming routine. It is therefore more important than ever that the methods used to derive such astrometric measurements include a careful estimation of all error sources, including systematic biases that are expected to affect even the most advanced planet imaging instruments. Here, we propose to derive the astrometry of the four HR8799 planets based on a data set obtained with SPHERE during its science verification phase in December 2014. While this data set was already analyzed and presented in \citet{Zurlo_2016_HR8799} and \citet{Apai_2016}, our aim here is to propose a detailed description of all individual contributions to the astrometric error budget, including systematic biases, and to derive general recommendations for future studies aiming at an accurate estimation of astrometric error bars. In Sect.~\ref{section:obs and redu}, we start by describing the observations, data reduction and image processing steps that allow the four planets to be revealed with a high signal-to-noise ratio (S/N). Then, Sect.~\ref{section:RA} discusses our method to derive the astrometry of the four planets, gives a thorough description of all major sources of astrometric errors, and evaluates their respective contribution. We present in Sect.~\ref{section:PyAstrOFit} the new open-source \texttt{PyAstrOFit} package, fully dedicated to orbital fitting based on Bayesian inference using the MCMC approach, which we use to perform an updated analysis of the individual orbits of the four planets. Some aspects of our results differ from previous analyses published in the literature. A short discussion of their implication on the orbital dynamics of the system is given before concluding in Sect.~\ref{section:Conclusions}.

\section{Observations and data reduction}
\label{section:obs and redu}

\subsection{Observations}
\label{subsection:observations}

SPHERE performs high-contrast imaging by combining an extreme adaptive optics system \citep{Fusco_2006_SPHERE}, several coronographic masks, and three science sub-systems including the Infra-Red Dual-band Imager and Spectrograph \citep[IRDIS,][]{Dohlen_2008_IRDIS}. The observations of HR8799 were performed during five consecutive nights from 4 to 8 December 2014, using IRDIS in the broadband H filter ($1.48 - 1.77\ \mu$m) with an apodized Lyot mask \citep{Soummer_2005_ApodizedLyot, Carbillet_2011_ApodizedLyotCorono, Guerri_2011_ApodizedLyotCorono} of diameter 185~mas together with an undersized Lyot stop. A beam splitter located downstream the coronagraphic masks produces two identical parallel beams \citep{Beuzit_2008_SPHERE}, which results in two well separated images per acquisition, hereafter referred to as the \emph{left} and \emph{right} images. Each of the five observing sequences lasted for about half an hour, and consisted of 218 frames with a detector integration time (DIT) of 8 sec per frame. All observing sequences were obtained under fair seeing conditions (between $0\farcs8$ and $1\farcs5$), except on 7 December where the seeing was above $1\farcs5$. The sequences were acquired in pupil-stabilized mode to take advantage of the angular differential imaging \citep[ADI, ][]{Marois_2006a_ADI} technique. Due to the low elevation of HR8799 as seen from Cerro Paranal (maximum altitude of $44\degr$), the amount of parallactic angle rotation was however quite small, amounting to $8\fdg7$, $8\fdg5$, $8\fdg3$, $8\fdg1$ and $7\fdg8$ for the five nights, respectively. Four elongated diffraction spots, the so-called satellite spots, were created during the whole observing sequences by injecting a waffle pattern on the deformable mirror \citep{Langlois_2012_SPHERE} to help with the star centering procedure, as explained in the next section. 

\subsection{Data reduction}
\label{subsection:reduction}

The IRDIS raw frames were preprocessed using the SPHERE \texttt{EsoRex} pipeline. As a first step, master dark and flat frames were created from calibration data obtained for each night of observations. Then \texttt{EsoRex} identified the outlying pixels in the master dark frame by using a sigma clipping procedure and built a static bad-pixel map. Each frame was reduced by subtracting the corresponding master dark, dividing by the master flat and interpolating the pixels flagged in the bad-pixel map. At this stage we obtained two calibrated data cubes per night, one for each side of the IRDIS detector, resulting in ten data cubes. From each data cube we discarded bad frames by measuring the correlation of each frame with a reference frame that was tagged as good by visual inspection. Only the 85\% to 95\% most correlated frames were kept for post-processing, depending on the night. The night of 7 December was discarded due to its poor data quality, as already proposed by \citet{Apai_2016}.

\begin{figure}
	\centering
   	\includegraphics[height=4.6cm]{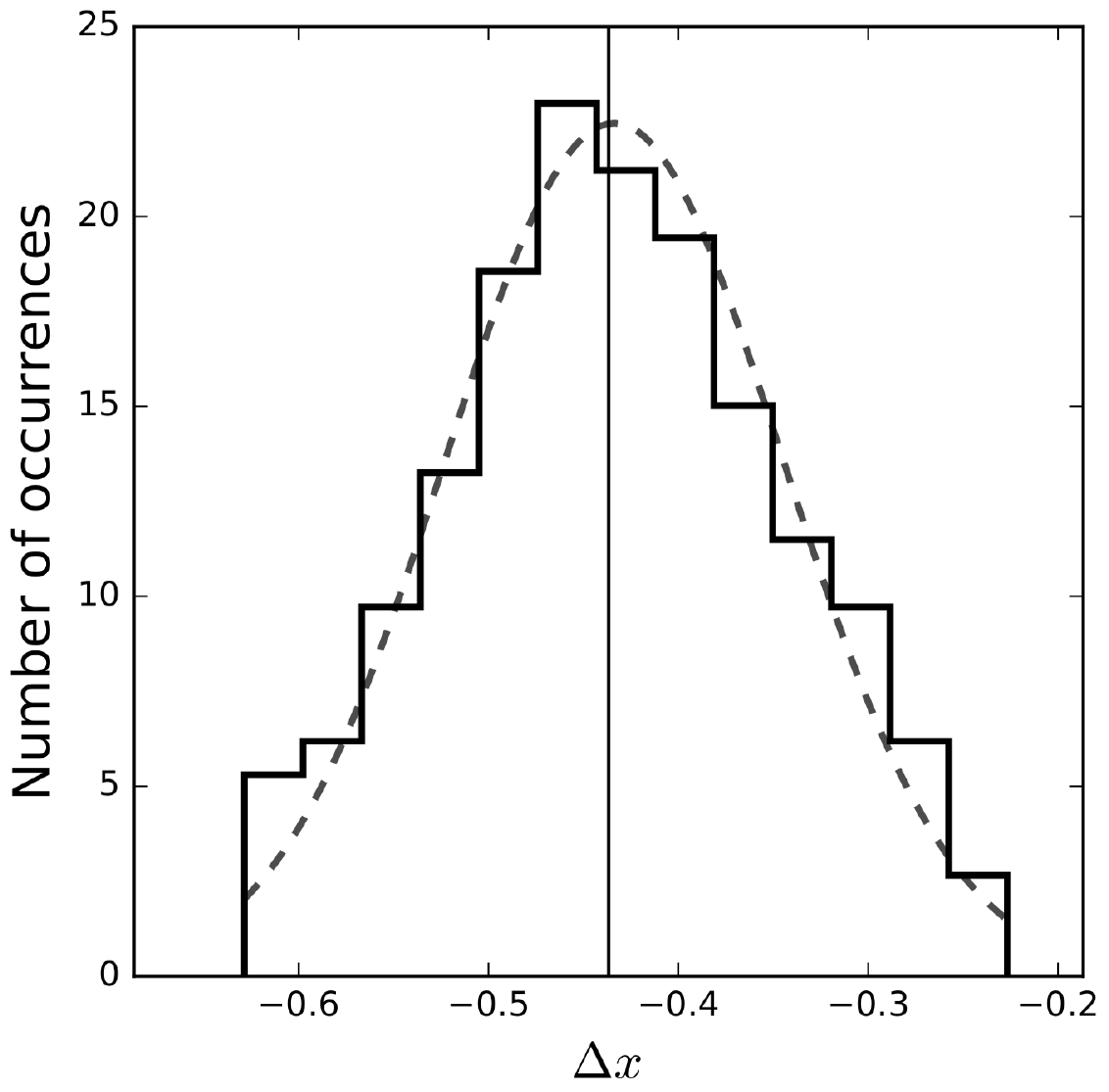}
	\includegraphics[height=4.6cm]{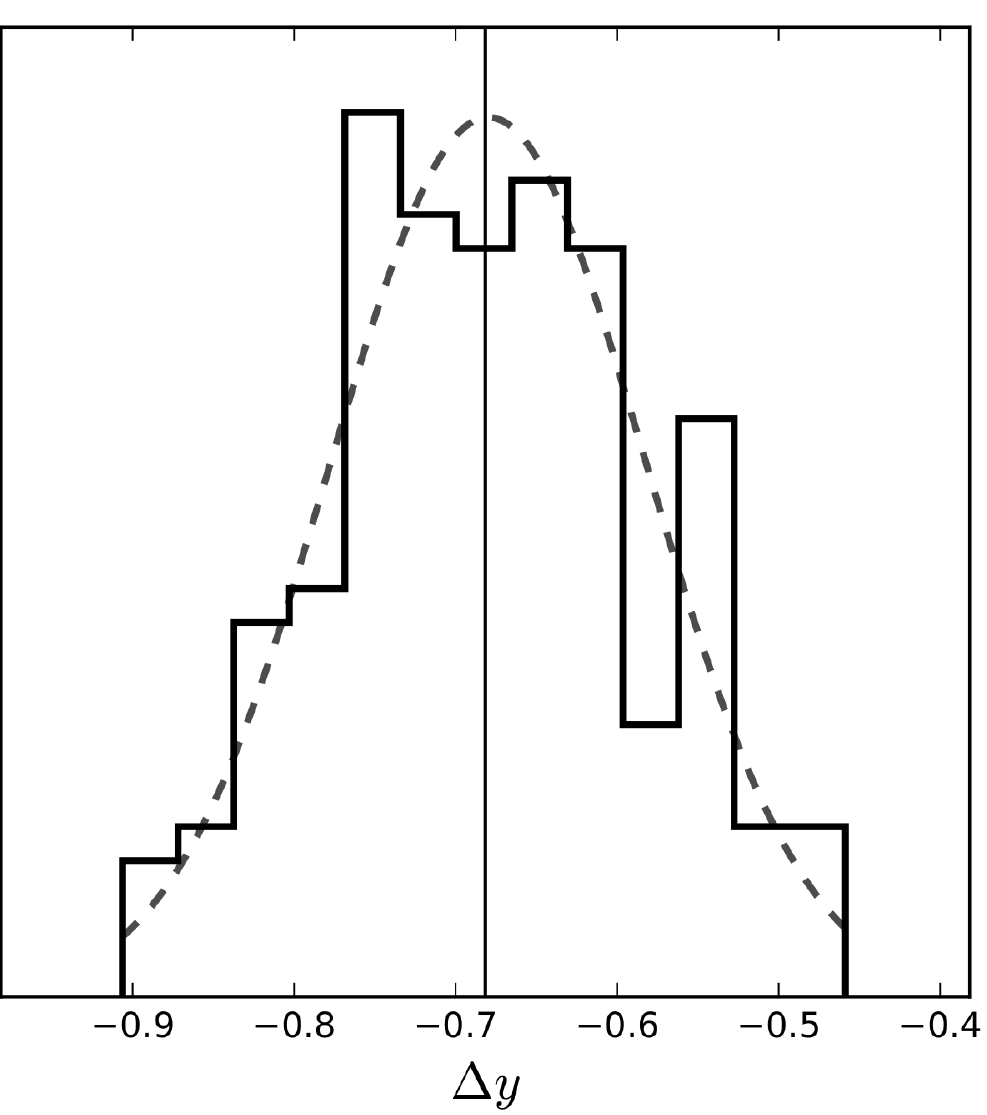}
     	\caption{Histogram of the horizontal and vertical offsets of the star with respect to the center of the frame in the 5 December (right) data cube. The vertical line represents the median of the histogram, and was used to globally re-center the data cube. The horizontal axis is in pixels, one pixel corresponding to 12.25 mas on sky.}
       	\label{figure:star_offsets}       
\end{figure}

We deliberately chose to skip the centering of the individual frames proposed by \texttt{EsoRex}. Instead, we used custom python routines\footnote{available in the VIP package \citep{Gomez_Gonzalez_2016_VIP}} to precisely measure the position of the star and the related uncertainty for each individual frame of all data cubes by exploiting the four satellite spots. Indeed, since the satellite spots have a high S/N and are designed to be symmetric with respect to the star, one can use them to infer the position of the star. In practice, due to their wavelength-dependent elongation and to residual atmospheric dispersion, the satellite spots are not perfectly symmetric with respect to the star \citep{Pathak16}. However, the symmetry is preserved at any given wavelength, and the spectrum-weighted astrometric position of the four satellite spots remains symmetric with respect to spectrum-weighted astrometric position of the star. To make sure to avoid the astrometric bias on the determination of the star position described by \citet{Pathak16}, the following strategy was adopted. For a given frame, we carefully fitted an asymmetric 2d Gaussian to each of the satellite spots to determine their respective centroid. Then, opposite centroids were connected by lines and the resulting intersection determined the estimated position $(x,y)$ of the star in detector coordinates. This was done for each frame to get the offset of the star from the center of the frame. For each data cube, a histogram of these offsets was built, and global offsets were obtained as the median of the vertical and horizontal offsets (see Fig.~\ref{figure:star_offsets}). All the frames were then shifted by the same amount for each cube to cancel the global offset, and cropped to a useful field-of-view of $511 \times 511$ pixels to reduce computation time in the post-processing. Our analysis suggests that a frame-by-frame recentering of the cube would not improve the final results, because the accuracy with which the stellar position can be determined in an individual frame is generally not smaller than the width of the histogram shown in Fig.~\ref{figure:star_offsets}. More details about the uncertainty on the position of the star are given in Sect.~\ref{subsection:star_position}.

The parallactic angles corresponding to the individual frames of each data cube were independently calculated frame by frame. The MJD time at the middle of each frame was derived from the information given by the \texttt{MJD-OBS} and \texttt{HIERARCH ESO DET FRAM UTC} header cards, which respectively give the time at the start and the end of the observing sequence, by dividing the total integration time equally into 218 parts. The parallactic angles were calculated using the spherical trigonometry formula given in \citet{Meeus_1998_AstroAlgo} based on the equatorial coordinates precessed to the epoch of the observations and corrected for nutation, aberrations, and refraction.

\subsection{Angular differential image processing}
\label{subsection:ADI}

We carried out the data post-processing with the open-source Vortex Imaging Processing\footnote{\texttt{https://github.com/vortex-exoplanet/VIP}} package \citep[\texttt{VIP},][]{Gomez_Gonzalez_2016_VIP} written in Python 2.7. Our post-processing is based on ADI techniques, which aim to reduce the quasi-static speckle noise and reveal the presence of off-axis sources by constructing and subtracting a reference on-axis point-spread function (PSF) from the individual frames of a data cube obtained in pupil tracking mode, where the star corresponds to the field rotation center \citep{Marois_2006a_ADI, Lafreniere_2007_LOCI}. Recently, \citet{Soummer_2012_KLIP} and \citet{Amara_2012_PYNPOINT} proposed to take advantage of PCA to make ADI post-processing more efficient. The PCA-ADI algorithm implemented in \texttt{VIP} is based on the approach presented in \citet{Amara_2012_PYNPOINT}, which can be summarized as follows:
\begin{itemize}
\item construct a set of orthogonal reference images, the so-called principal components (PCs), using singular value decomposition \citep[SVD, see e.g.][]{Press_2007_NumericalRecipies} of the data cube;
\item project all the frames of the cube onto a truncated set of  $n_{\text{PC}}$ ($< n_{\text{frame}}$) PCs, where $n_{\text{frame}}$ represents the total number of frames in a data cube;
\item reconstruct the data cube using a linear combination of PCs, and subtract the result from the original data cube to obtain a cube of residual frames;
\item rotate and collapse this cube of residuals to obtain the final image.
\end{itemize}

To optimize the determination of the astrometry, the S/N for each planet needs to be maximized. The S/N calculation implemented into \texttt{VIP} is based on a Student \emph{t}-test \citep[][]{Gosset_1908_Student} and follows the recommendation of \citet{Mawet_2014_Student} on small sample statistics \citep[see][for more details]{Gomez_Gonzalez_2016_LLSG}. The S/N of the planets in the final, post-processed image depends mainly on the number of PCs used when building the reconstructed cube. A small number of PCs leads to an incomplete representation of the speckle noise, while a large number of PCs tends to capture the signal of the planet in the reconstructed cube, which leads to a lower algorithmic throughput for the planetary signal after subtraction. An optimum number of PCs can generally be found to maximize the planet S/N \citep{Meshkat_2014}. For each data cube, we thus performed a grid search on the number of PCs to maximize the mean S/N in a region of one resolution element in diameter around each companion. The optimal $n_{\text{PC}}$ for each data cube is reported in Table~\ref{table:nightsAstrometry}. Let us note that the PCA implemented in \texttt{VIP} comes with several SVD libraries, such as the efficient \texttt{randomized SVD} \citep[][]{Halko_2011_randSVD}  and the well-known \texttt{LAPACK} \citep[see e.g.][]{Anderson_1990_LAPACK}. We refer to Gomez Gonzalez et al.\ (submitted) for a complete discussion of all the SVD libraries available in \texttt{VIP}.

\begin{figure}
	\centering
   	\includegraphics[width=9cm]{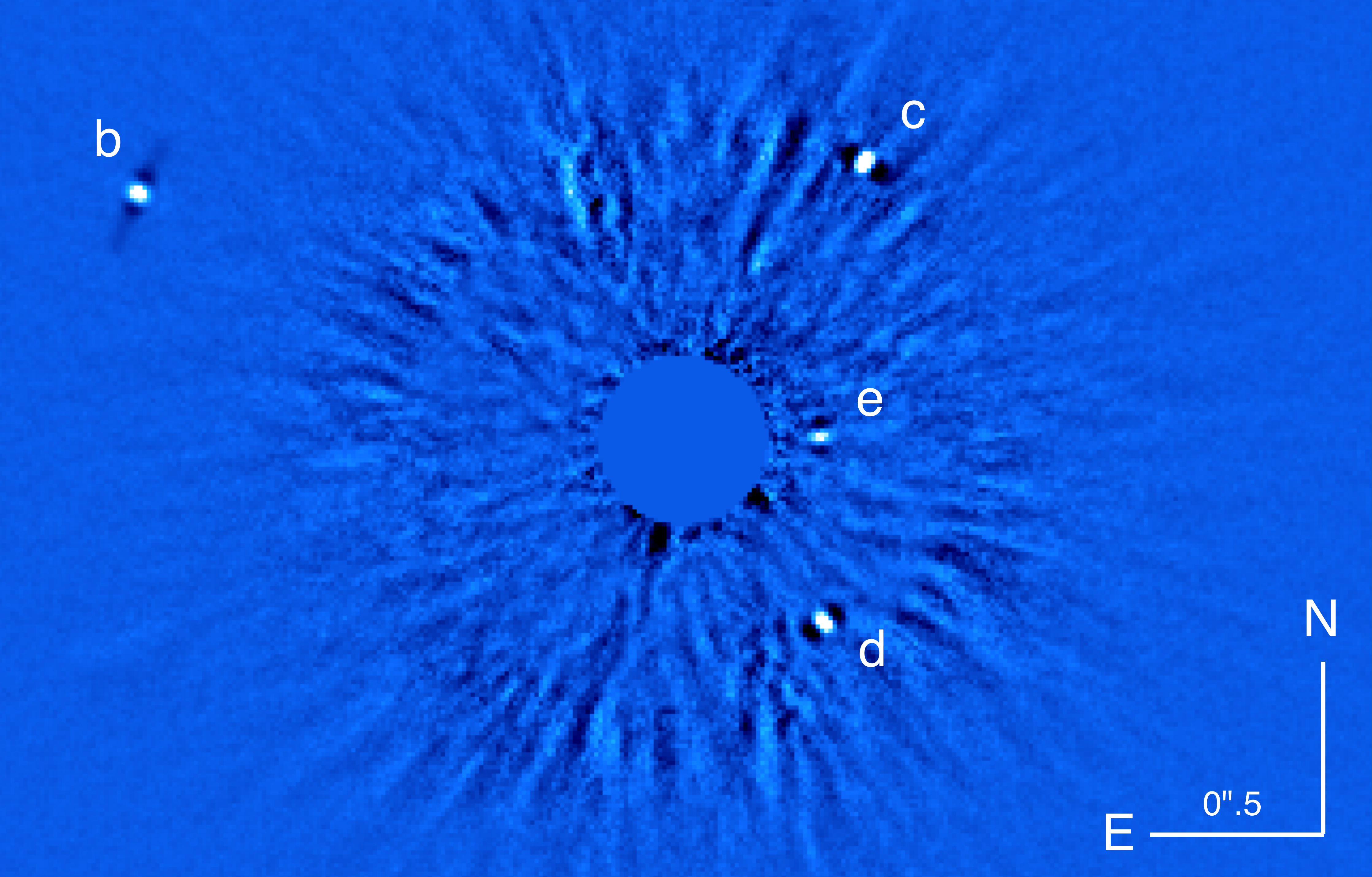}
     	\caption{Illustration of a full-frame PCA ADI post-processed SPHERE/IRDIS image of HR8799 acquired with broadband $H$ filter, left part, during the night of 4 December 2014. The central part was masked with a disk of radius of 20 pixels.} 
       	\label{figure:HR8799fullframe}       
\end{figure} 

Figure~\ref{figure:HR8799fullframe} illustrates a \texttt{VIP} post-processed image using full-frame PCA, where all the pixels of each frame are used at once to construct the reference images through SVD. The close region surrounding the host star is most affected by residual speckle noise and was masked with a disk of radius of 20 pixels to better reveal the planets in Fig. \ref{figure:HR8799fullframe}. Throughout the present analysis, we also performed annulus-wise PCA, which consists in performing PCA only for a thin annulus passing through a companion, with a typical width of a few resolution elements. Although full-frame PCA and annulus-wise PCA may lead to slightly different results, this choice does not significantly affect the final astrometry, which is dominated by other sources of error (see Sect.~\ref{section:RA}). Furthermore, annulus-wise PCA is significantly faster when performed on a single annulus, which is useful when dealing with large data cubes and/or when PCA is performed a large number of times (see Sect. \ref{subsection:position_and_stat_error}).

\section{Robust astrometry}
\label{section:RA}

The astrometric position of the HR8799bcde planets based on the December 2014 SPHERE/IRDIS data set has already been determined by \citet{Zurlo_2016_HR8799} and \citet{Apai_2016}. The \citet{Zurlo_2016_HR8799} final astrometry was obtained from the combination of four independent image-processing pipelines, by the quadratic sum of the error bar from each data reduction pipeline plus the standard deviation associated to the individual positions. \citet{Apai_2016} used their implementation of the KLIP algorithm to derive the planets positions by injecting artificial planets with negative count rates, and used a manual inspection of the image quality and of the subtraction residuals to estimate the error bars. Here, we propose to go beyond these approaches and to study in details the various contributions to the astrometric error budget, in an attempt to derive more reliable error bars. Our study is also meant to explore the ultimate astrometric accuracy of a state-of-the-art instrument such as SPHERE, and to identify possible ways to improve the astrometric accuracy in future studies.

What we call \emph{robust astrometry} consists in performing a proper evaluation of the statistical errors and systematic biases affecting the final astrometric estimation. The whole procedure consists of four steps: (i) the description and estimation of the instrumental calibration errors, (ii) the determination of the planets position with respect to the star and the related statistical error through Bayesian inference with MCMC sampling, (iii) the determination of the systematic error due to residual speckles, and (iv) the calculation of the error on the star position. Throughout the rest of this section, we provide details for each step of the process.

\subsection{Instrumental calibration and related errors}
\label{subsection:instruNoise}

To derive accurate astrometric measurements from IRDIS images, various astrometric calibrations need to be performed, namely the determination of the plate scale, the orientation of the north and the optical distortion. Firstly, the plate scale, given in arcsec per pixel, depends on the characteristics of all the optical elements composing the instrument. It allows to convert positions given in pixel into arcsec. Secondly, when observing in pupil-stabilized mode, the vertical axis of the detector does not necessarily point towards north. Two contributions need to be taken into account: (i) the pupil offset, which accounts for the zero point position of the derotator and is assumed to be constant between runs, and (ii) the so-called true north, which accounts for a variation in the detector orientation with respect to the sky due to thermal or mechanical stresses, and which needs to be estimated during each observing run. Thirdly, the distortion in SPHERE/IRDIS is mainly dominated by an anamorphic magnification between the horizontal and vertical axis of the detector. This effect is due to the presence of toric mirrors in the common path of the instrument \citep[see e.g.,][]{Zurlo_2016_HR8799}. 

Details of the observations used to derive those astrometric calibrations for IRDIS are described in \citet{Zurlo_2016_HR8799}. We refer to that paper for the details, but we still provide the reader with the practical information used in this study. The astrometric calibrations were obtained from IRDIS observations of the globular cluster $47$ Tuc acquired on 15 December 2014 with the same instrument setup and filter, and compared to the Hubble Space Telescope data of the same field, precessed to the same epoch and corrected for the differential proper motions of the individual stars. The values derived by \citet{Zurlo_2016_HR8799} for the plate scale and true north based on this data set have recently been revised by the SPHERE consortium, using their improved knowledge of the instrument. This revised estimation, described in \citet{Maire16}, leads to a plate scale of $12.251 \pm 0.009$ mas/pixel and a true north orientation of $-1\fdg709 \pm 0\fdg051$. These values are valid for both the left and right parts of the IRDIS detector. 
The pupil offset, based on commissioning and guaranteed-time data obtained on several astrometric fields, is equal to $135\fdg99 \pm 0\fdg11$. Finally, the IRDIS distortion measured on sky is dominated by an anamorphism of $0.60\% \pm 0.02\%$ between the horizontal and vertical directions \citep{Maire16}. Although the SPHERE calibration plan includes the daily measurement of distortion maps based on pinhole grids, we found that the quality of the astrometric estimations does not improve by using these maps. Prior to any post-processing, we thus simply rescaled each frame of each cube by a factor 1.006 along the $y$ axis. To take into account the uncertainty on this correction, an additional error of $0.02\%$ on the plate scale will be considered in the following analysis.

\subsection{Planet position and statistical error}
\label{subsection:position_and_stat_error}

\begin{figure}
	\centering
   	\includegraphics[width=4.45cm]{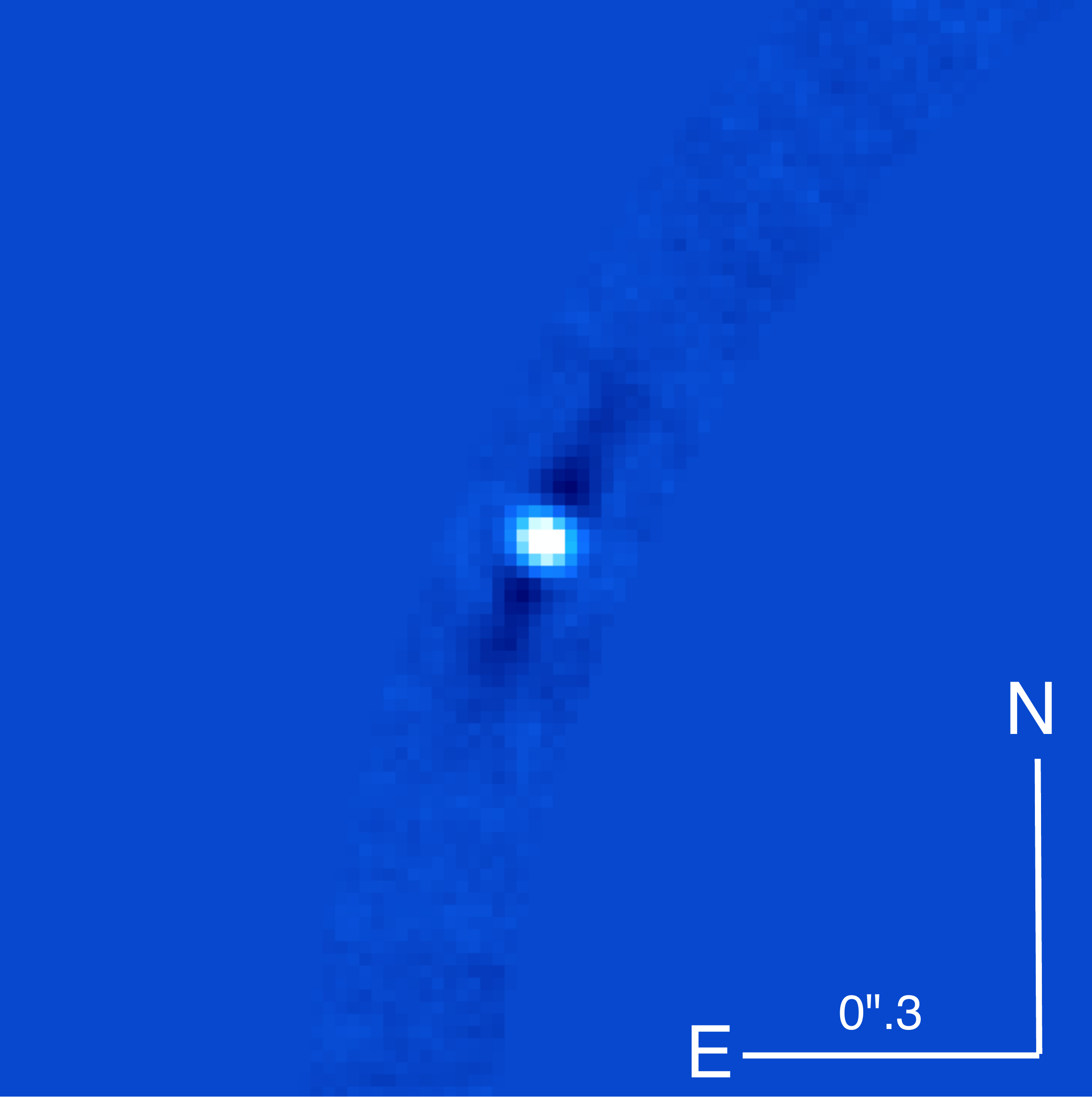}
   	\includegraphics[width=4.45cm]{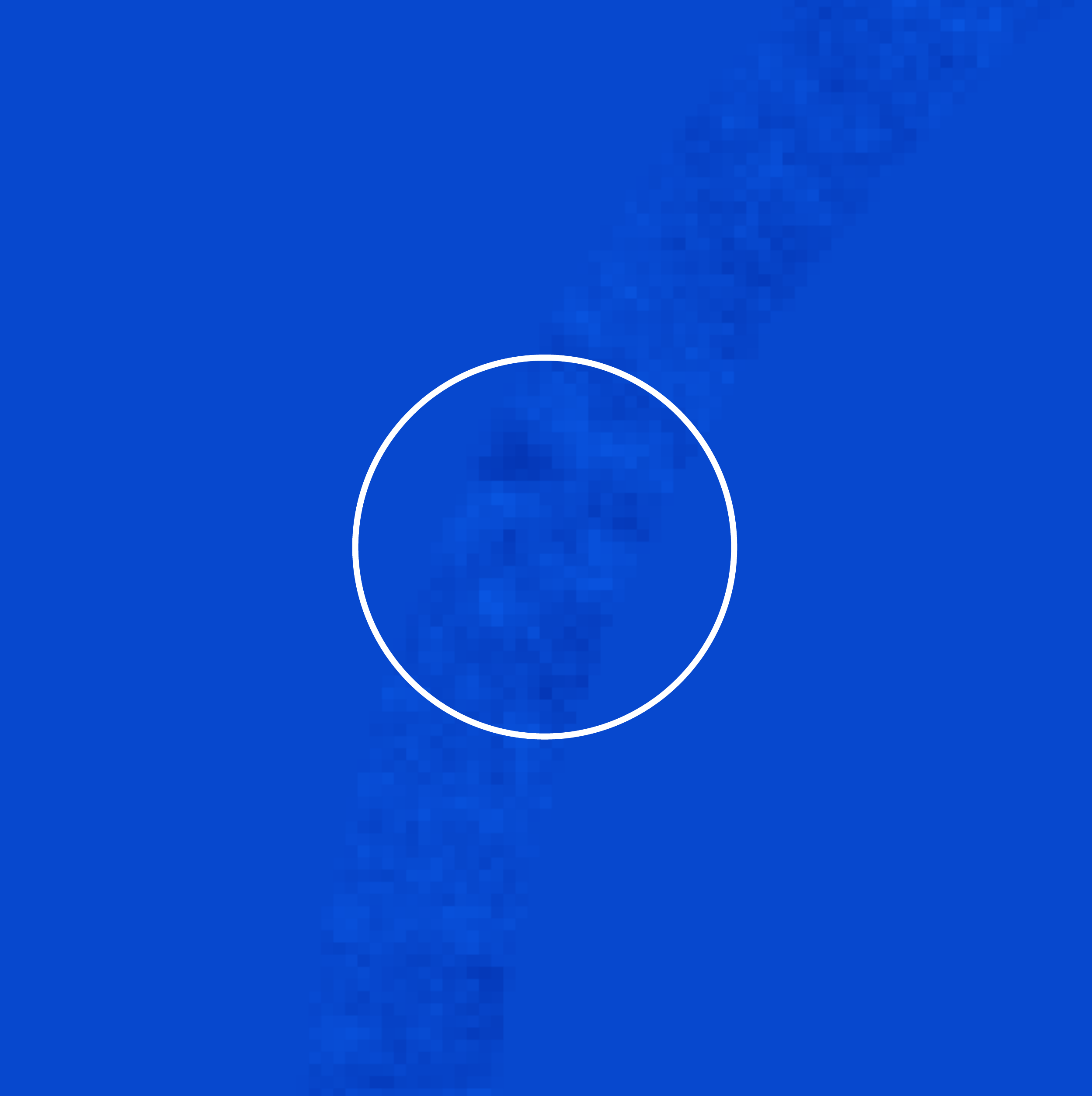}		
     	\caption{Illustration of the annulus-wise PCA post-processing and merit function evaluation used in the negative fake companion technique. \textit{Left}. No NEGFC was injected before annulus-wise PCA processing. \textit{Right}. A NEGFC was injected at the position and flux minimizing the merit function. The white circle illustrates the fixed circular aperture from which the pixel values $I_j$ have been extracted to evaluate the merit function. The same color scale was used for both images.}
       	\label{figure:chi2 pca annulus}       
\end{figure}

The next step in the robust astrometry process consists in determining for each data cube the position of the planets with respect to the host star and in estimating the statistical error related purely to the photon noise of the underlying thermal background and speckles through Bayesian inference based on MCMC simulations. This step does not describe the effect of the speckles themselves on the measured planet position, which will be discussed separately in Sect.~\ref{subsection:speckleNoise}. Our astrometric measurements are based on the negative fake companion technique \citep[NEGFC, see e.g.][]{Marois_2010_NegFC, Lagrange_2010_betapic}, which consists in injecting in the data cube a negative PSF template with the aim of canceling out the companion as well as possible in the final post-processed image based on a well-chosen merit function. The NEGFC technique is an iterative process, for which a step can be described as follows. For the chosen position/flux combination, a negative fake companion is injected in each frame of the data cube, and annular-wise PCA-ADI processing is performed on a single annulus passing through the considered companion. The intensities $I_j$ of $N$ pixels are then extracted within a circular region of a radius equal to a few resolution elements, centered on a first guess position defined at the start of the iterative process (which means that the position of the circular aperture is fixed and does not change during the process). Assuming that the noise affecting the $j$-th pixel value is given by $\sigma_j = \sqrt{I_j}$ (pure photon noise), we define the merit function as follows:
\begin{equation}
\chi^2 \propto \sum_{j=1}^{N} |I_j| \ .
\label{chi2}
\end{equation}
The position/flux of the NEGFC is then optimized to minimize the merit function in a three-step approach, as described in the following paragraphs. The resulting post-processed images, before and after injection of a NEGFC, are represented in Fig.~\ref{figure:chi2 pca annulus}. Because no off-axis PSF was acquired in December 2014 with the same observing setup as for the HR8799 observations, the adopted PSF template corresponds to unsaturated off-axis images of $\beta$ Pictoris obtained with SPHERE/IRDIS during science verification on 30 January 2015 (PI: A.-M. Lagrange) with the same observing mode as for HR8799 (same coronagraph, same broadband H filter, similar seeing $\sim 1"$). The influence of this choice will be discussed at the end of Sect.~\ref{subsection:position_and_stat_error}, together with a discussion of the effect of PSF chromatic dispersion on the measured planet position.

\paragraph{First guess estimation.}
\label{paragraph:first_guess}

With the optimal number of PCs in hand (see Table~\ref{table:nightsAstrometry}), we derive a first guess on the position of each companion in each PCA-ADI post-processed image by simply identifying the highest pixel value in the close vicinity of the companion. We then derive a first guess on the flux of the companion by injecting a NEGFC at that position and by evaluating the merit function for a grid of possible fluxes. Only the flux is optimized during this stage, while the companion position is fixed to our first guess. 

\paragraph{Nelder-Mead optimization.}
\label{paragraph:nelder-mead}

Although the first guess estimation results in a rough determination of the position/flux, and would constitute a valid initial set of parameters to start an MCMC-based Bayesian inference process (as presented in the next paragraph), it may turn out to be very time consuming due to the large number of merit function evaluations required to reach convergence in the MCMC and properly sample the posterior distributions. Thus, we propose to refine the first guess of the position/flux of the companions for the purpose of initializing the MCMC sampling close to the highly probable solution. To this aim, we use the first position/flux estimation as an initial guess for a Nelder-Mead simplex-based optimization \citep{Nelder_1965_Simplex} implemented into the \texttt{SciPy} Python library\footnote{\texttt{http://www.scipy.org}}. The adopted merit function is the one defined in Eq.~\ref{chi2}, and the position ($r, \theta$) of the NEGFC is now allowed to vary during the fit in addition to its flux. As expected, this leads to a significant improvement of the position/flux determination. The right panel of Fig.~\ref{figure:chi2 pca annulus} illustrates the result of an annulus-wise PCA-ADI post-processing performed on a single annulus passing through HR8799b, after injecting a NEGFC characterized by a position and flux minimizing the merit function.

The PCA-ADI algorithm that we first used in \texttt{VIP} relied on a \texttt{randomized SVD} library, which approximates the SVD of the data cube by using random projections and thereby provides increased computational efficiency (for details see Gomez Gonzalez et al., submitted). Although this randomized approach is very efficient, the random process induces random variations in the merit function that can be significant compared to the variations of the merit function between two steps, especially when approaching the minimum. This can prevent the optimization process from reaching the true minimum of the merit function, or even from converging. Therefore, we decided to use the more classical, yet slower, deterministic SVD approach proposed in the \texttt{LAPACK} library for our PCA-ADI processing in the simplex optimization, as well as for the rest of this study. This choice is all the more important when the companion is located in a region dominated by residual speckle noise.

\begin{table*}[!t]
\caption{Final HR8799bcde astrometric measurements with respect to the host star for nights of 4, 5, 6, and 8 December 2014, derived from SPHERE/IRDIS broadband H measurements (left and right parts), in terms of RA/DEC (columns 3-4) and in polar coordinates (columns 5-6). In addition, we list the derived optimal number of principal component $n_{\text{PC}}$ (column 2), as well as the statistical error bars (columns 7-8) and the speckle noise error bars (columns 9-10), both in polar coordinates.}
\label{table:nightsAstrometry}     
\centering                          
\begin{tabular}{cccccccccc}        
\hline\hline                		
Date	and side			& $n_{\text{PC}}$	&  $\Delta \text{RA}\ ['']$		& $\Delta \text{DEC}\ ['']$ 		& $\Delta r\ ['']$	& $\Delta \theta\ [\degr]$	& $\sigma_{\text{stat},r}\  ['']$ 	& $\sigma_{\text{stat},\theta}\  [\degr]$	& $\sigma_{\text{spec},r}\  ['']$		& $\sigma_{\text{spec},\theta}\  [\degr]$  \\			                         
\hline
HR8799b \\ 
2014-12-04 L		& $3$			& $1.5754$				& $0.7019$	& $1.7247$	& $65.985$			& $0.0003$					& $0.007$								& $0.0002$					& $0.007$		\\ 
2014-12-04 R		& $2$			& $1.5750$				& $0.7015$	& $1.7242$	& $65.994$			& $0.0003$					& $0.008$								& $0.0002$					& $0.007$		\\ 
2014-12-05 L		& $7$			& $1.5761$				& $0.7026$	& $1.7256$	& $65.975$			& $0.0003$					& $0.009$								& $0.0002$					& $0.009$		\\ 
2014-12-05 R		& $6$			& $1.5760$				& $0.7024$	& $1.7254$	& $65.977$			& $0.0003$					& $0.008$								& $0.0002$					& $0.008$		\\ 
2014-12-06 L		& $6$			& $1.5730$				& $0.7008$ 	& $1.7221$	& $65.985$			& $0.0004$					& $0.013$								& $0.0002$					& $0.009$		\\ 
2014-12-06 R		& $6$			& $1.5739$				& $0.7000$ 	& $1.7225$	& $66.023$			& $0.0004$					& $0.013$								& $0.0002$					& $0.009$		\\ 
2014-12-08 L		& $4$			& $1.5743$				& $0.7016$	& $1.7236$	& $65.980$			& $0.0003$					& $0.013$								& $0.0003$					& $0.011$		\\ 
2014-12-08 R		& $4$			& $1.5736$				& $0.7021$	& $1.7231$	& $65.956$			& $0.0003$					& $0.008$								& $0.0003$					& $0.010$		\\            
\hline
HR8799c \\ 
2014-12-04 L		& $5$			& $-0.5116$				& $0.7971$	& $0.9471$	& $327.307$			& $0.0002$					& $0.014$								& $0.0006$					& $0.048$		\\ 
2014-12-04 R		& $6$			& $-0.5127$				& $0.7984$	& $0.9488$	& $327.293$			& $0.0002$					& $0.010$								& $0.0006$					& $0.044$		\\ 
2014-12-05 L		& $13$			& $-0.5089$				& $0.7992$	& $0.9475$	& $327.512$			& $0.0003$					& $0.012$								& $0.0008$					& $0.059$		\\ 
2014-12-05 R		& $14$			& $-0.5103$				& $0.8003$	& $0.9492$	& $327.479$			& $0.0004$					& $0.015$								& $0.0008$					& $0.053$		\\ 
2014-12-06 L		& $15$			& $-0.5113$				& $0.7979$	& $0.9477$	& $327.351$			& $0.0005$					& $0.020$								& $0.0006$					& $0.047$		\\ 
2014-12-06 R		& $18$			& $-0.5118$				& $0.7986$	& $0.9485$	& $327.342$			& $0.0005$					& $0.013$								& $0.0007$					& $0.052$		\\ 
2014-12-08 L		& $20$			& $-0.5104$				& $0.7987$	& $0.9479$	& $327.421$			& $0.0004$					& $0.026$								& $0.0010$					& $0.077$		\\              
2014-12-08 R		& $7$			& $-0.5128$				& $0.7986$	& $0.9491$	& $327.291$			& $0.0003$					& $0.016$								& $0.0012$					& $0.088$		\\ 
\hline
HR8799d \\ 
2014-12-04 L		& $5$			& $-0.3990$				& $-0.5250$	& $0.6594$	& $217.233$			& $0.0012$					& $0.024$								& $0.0012$					& $0.093$		\\ 
2014-12-04 R		& $5$			& $-0.3994$				& $-0.5244$	& $0.6592$	& $217.292$			& $0.0004$					& $0.027$								& $0.0011$					& $0.092$		\\ 
2014-12-05 L		& $21$			& $-0.4008$				& $-0.5233$	& $0.6592$	& $217.448$			& $0.0006$					& $0.035$								& $0.0013$					& $0.085$		\\ 
2014-12-05 R		& $21$			& $-0.3999$				& $-0.5221$	& $0.6576$	& $217.454$			& $0.0005$					& $0.039$								& $0.0013$					& $0.075$		\\ 
2014-12-06 L		& $21$			& $-0.4008$				& $-0.5233$	& $0.6592$	& $217.446$			& $0.0005$					& $0.022$								& $0.0010$					& $0.077$		\\ 
2014-12-06 R		& $20$			& $-0.3999$				& $-0.5230$	& $0.6584$	& $217.397$			& $0.0005$					& $0.017$								& $0.0010$					& $0.080$		\\ 
2014-12-08 L		& $18$			& $-0.3982$				& $-0.5208$	& $0.6556$	& $217.405$			& $0.0004$					& $0.030$								& $0.0029$					& $0.136$		\\ 
2014-12-08 R		& $46$			& $-0.4007$				& $-0.5208$	& $0.6571$	& $217.575$			& $0.0005$					& $0.029$								& $0.0027$					& $0.123$		\\ 
\hline
HR8799e \\ 
2014-12-04 L		& $9$			& $-0.3859$				& $0.0117$	& $0.3861$	& $271.735$			& $0.0010$					& $0.103$								& $0.0022$					& $0.202$		\\ 
2014-12-04 R		& $16$			& $-0.3852$				& $0.0099$	& $0.3854$	& $271.468$			& $0.0013$					& $0.077$								& $0.0039$					& $0.292$		\\ 
2014-12-05 L		& $10$			& $-0.3829$				& $0.0121$	& $0.3831$	& $271.803$			& $0.0006$					& $0.044$								& $0.0029$					& $0.196$		\\ 
2014-12-05 R		& $12$			& $-0.3841$				& $0.0125$	& $0.3843$	& $271.859$			& $0.0005$					& $0.055$								& $0.0024$					& $0.167$		\\ 
2014-12-06 L		& $8$			& $-0.3858$				& $0.0097$	& $0.3859$	& $271.436$			& $0.0006$					& $0.034$								& $0.0022$					& $0.182$		\\ 
2014-12-06 R		& $23$			& $-0.3865$				& $0.0113$	& $0.3867$	& $271.668$			& $0.0006$					& $0.048$								& $0.0019$					& $0.159$		\\ 
2014-12-08 L		& $15$			& $-0.3843$				& $0.0139$	& $0.3846$	& $271.072$			& $0.0016$					& $0.186$								& $0.0049$					& $0.357$		\\     
2014-12-08 R		& $11$			& $-0.3862$				& $0.0159$	& $0.3865$	& $272.360$			& $0.0008$					& $0.145$								& $0.0082$					& $0.534$		\\     

\hline            
\end{tabular}
\end{table*}

\paragraph{MCMC and final positions.}
\label{paragraph:MCMC}

\begin{figure}
	\centering
   	\includegraphics[width=9cm]{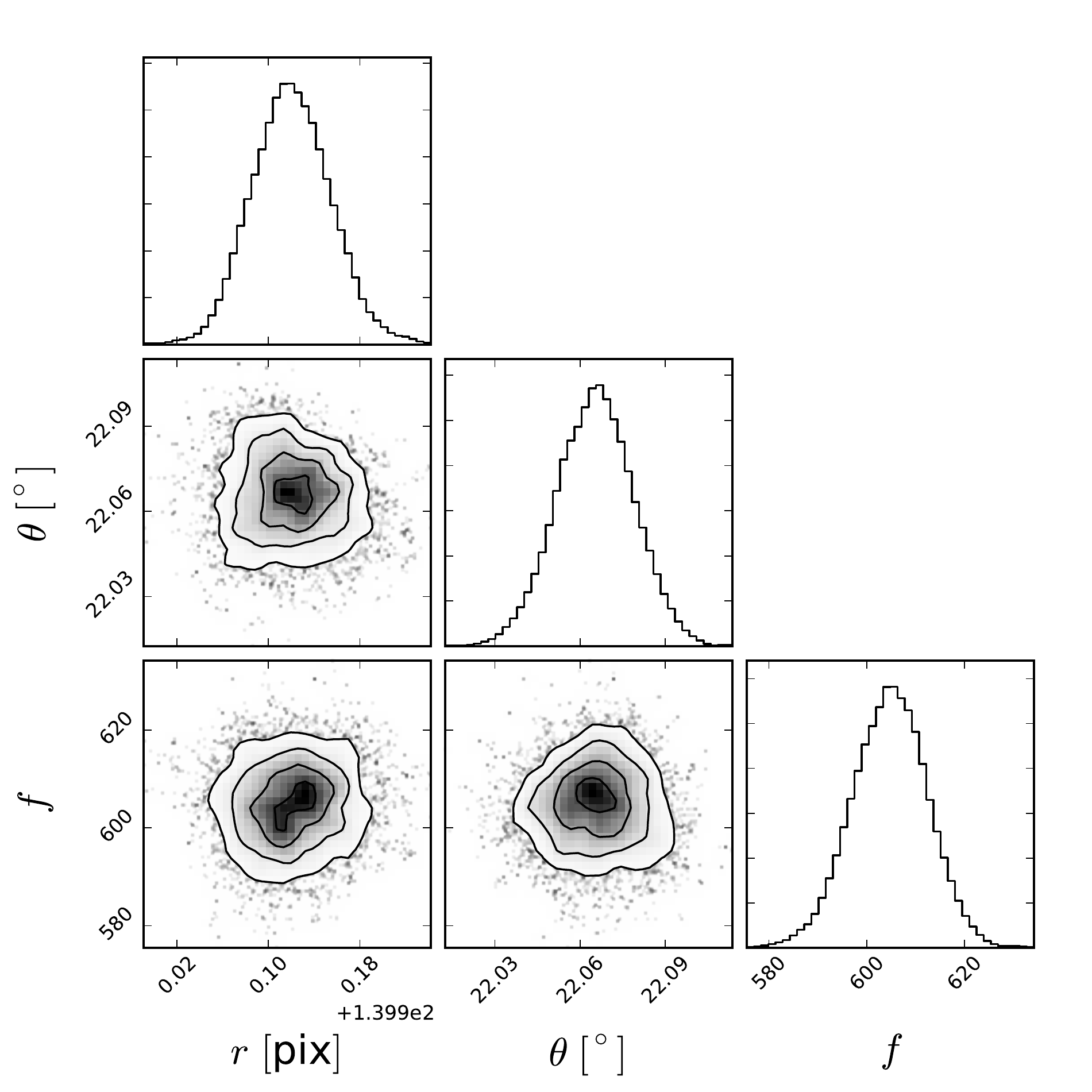}
     	\caption{Illustration of a typical corner plot obtained from the MCMC simulations using the NEGFC technique. The target companion is HR8799b observed during the night of 6 December 2014. The radial distance $r$ (in pixel) and azimuth $\theta$ (in degree) are detector coordinates with respect to the host star. The diagonal panels illustrate the posterior PDFs while off-axis ones illustrate the correlation between them.}
       	\label{figure:astrometryMCMC}        
\end{figure} 

Because the merit function used in the Nelder-Mead optimization is not strictly convex, it is not guaranteed that the optimization will converge at the exact position of the planet, as it could potentially get stuck in a local minimum. Although this behaviour was generally not observed  \citep[as shown in][]{Morzinski_2015}, we decided to use the NEGFC technique coupled with an MCMC approach to obtain the final flux and position of the HR8799 planets, expressed in polar coordinates with respect to the host star. Let us recall briefly that the MCMC approach aims to sample the posterior probability density function (PDF), i.e. the probability of the position/flux parameters given the data cube and the prior knowledge \citep[see e.g.][]{Hogg_2010_MCMC}. The \texttt{VIP} module dedicated to NEGFC technique embeds the \texttt{emcee} package \citep{emcee}, which implements an affine-invariant ensemble sampler for MCMC proposed by \citet{Goodman_MCMC_2010}. Such an ensemble is composed of \emph{walkers}, which can be considered as Metropolis-Hastings chains. The main difference between walkers and Metropolis-Hastings chains lies in the fact that the proposal distribution for a given walker depends, at a given step, on the position of all other walkers in the ensemble. Conversely, the proposal distributions involved in the Metropolis-Hastings algorithm are independent. Besides being more efficient in terms of the number of calls to the cost function, one major advantage of \texttt{emcee} is that it relies on only two calibration parameters, in comparison with the $\sim N^2$ parameters required for a Metropolis-Hastings algorithm in an $N$-dimensional parameter space to properly sample the PDF and speed up the process \citep[for more details, see][and references therein]{emcee, Goodman_MCMC_2010}. 

\begin{figure*}
	\centering
   	\includegraphics[width=8.9cm]{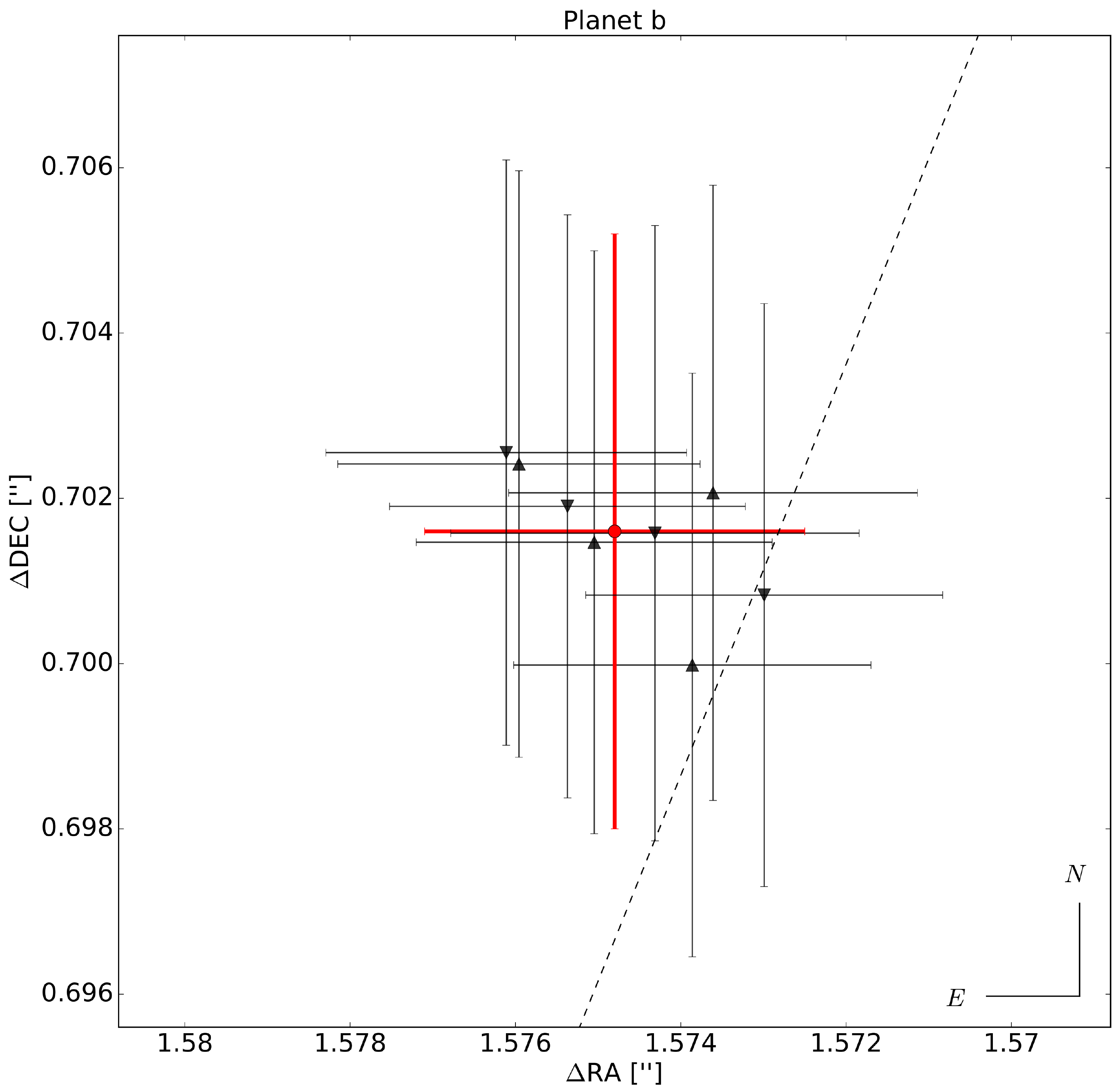} \hspace*{2mm}
	\includegraphics[width=9cm]{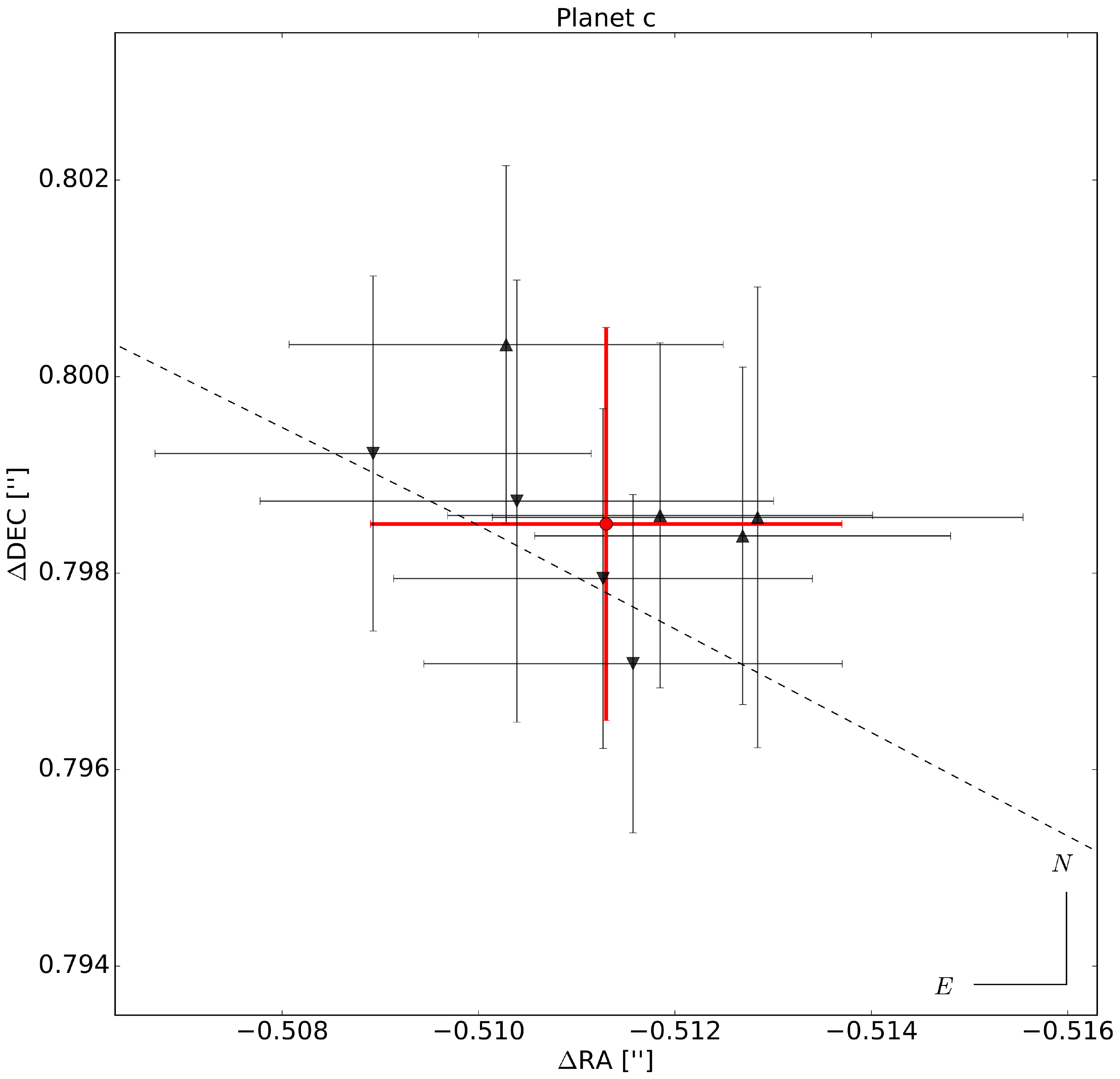} \\
	\includegraphics[width=9cm]{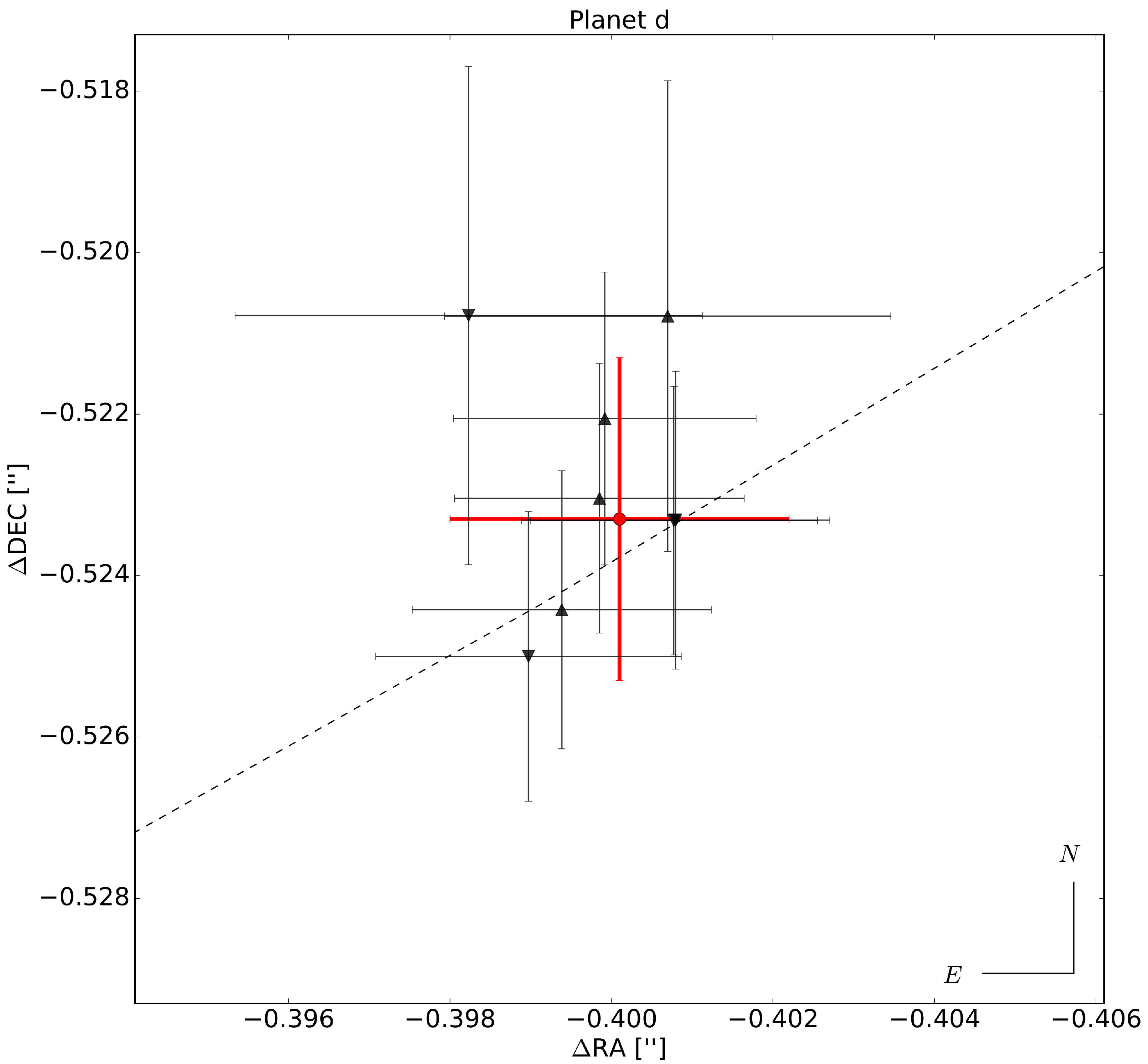} \hspace*{2mm}
	\includegraphics[width=8.8cm]{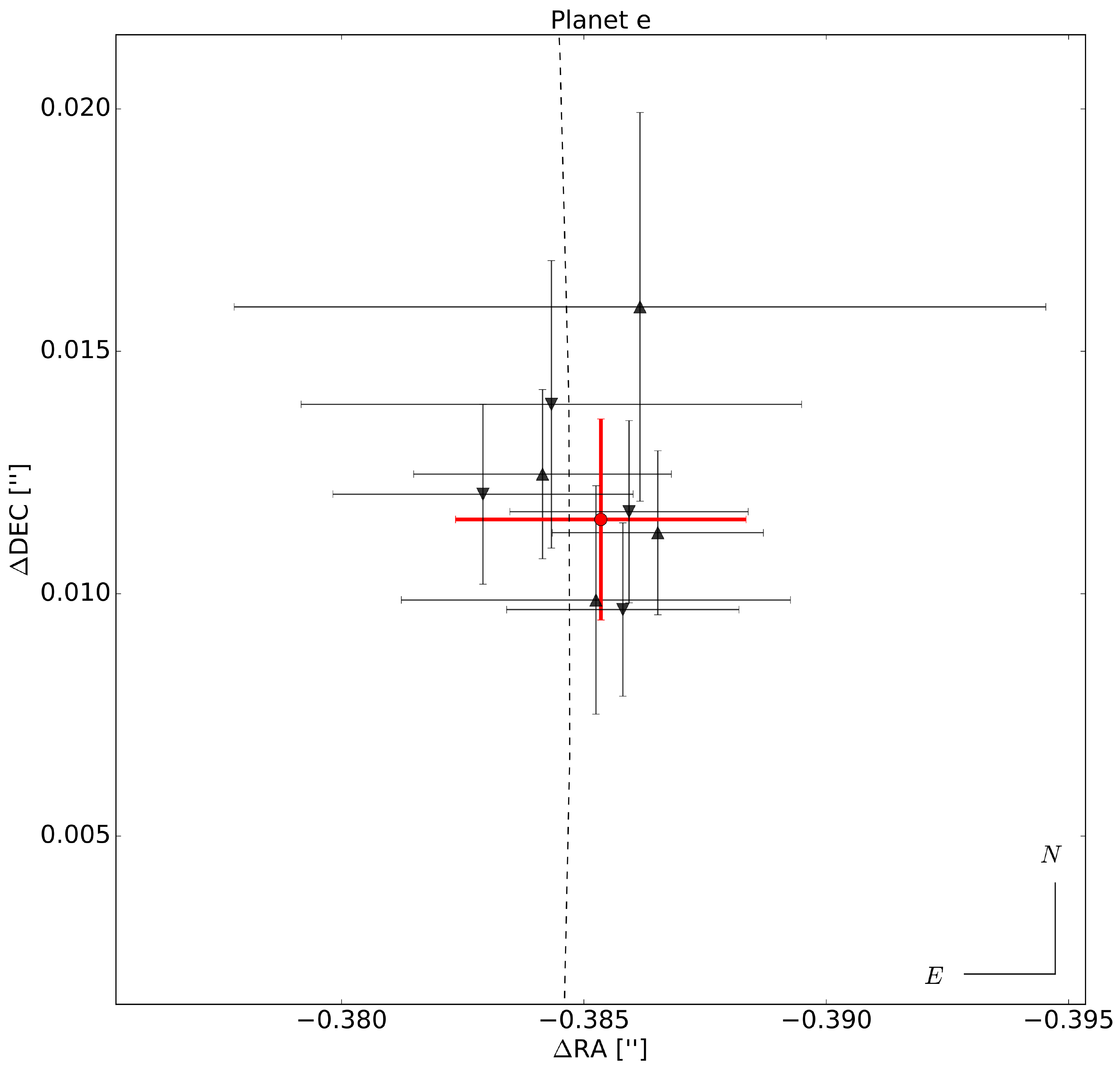}
     	\caption{Astrometry for HR8799bcde observed during the nights of 4, 5, 6, and 8 December 2014. The positions obtained from the left (resp. right) data cubes are represented with downward (resp.\ upward) black triangles. The error bars on the individual data points take into account all the contributions discussed in Sects.~\ref{subsection:instruNoise}, \ref{subsection:position_and_stat_error}, \ref{subsection:speckleNoise}, and \ref{subsection:star_position}. The red dots correspond to the final astrometric measurements for each planet, together with the final error bar discussed in Sect.~\ref{subsection:final_astrometry}. The dashed lines represent the best orbital solutions for each planet in terms of reduced $\chi^2$ reported in Table~\ref{table:ConfidenceIntervals}.}
       	\label{figure:nightAstrometryZoom}       
\end{figure*}

For each data cube and each companion, we carried out MCMC simulations to sample posterior PDFs related to the planet polar coordinates ($r,\theta$) with respect to the host star and the planet flux $f$. For each MCMC simulation, we used 200 walkers firstly initialized in a small ball around the solution obtained from the Nelder-Mead optimization. The chain was sufficiently close to convergence\footnote{None of the MCMC convergence tests available in the literature can conclude formally on the chain convergence (see Sect.~\ref{subsection:PyAstrOFit_package}.} to allow Bayesian inference after typically 200 steps. More details about convergence statistical tests are given in Sect.~\ref{subsection:PyAstrOFit_package}, where we describe the \texttt{PyAstrOFit} Python package. In Fig. \ref{figure:astrometryMCMC}, the so-called corner plot illustrates the posterior PDFs and the correlation between the parameters $(r, \theta, f)$ for HR8799b observed during the night of 6 December 2014. Similar results were obtained for other planets and observing nights. Although a flux estimation for each planet is obtained, we focus our analysis only on the astrometry in this paper. 

Taking into account the plate scale, the true north and pupil offset orientation (see Sect.~\ref{subsection:instruNoise}), we have projected the HR8799bcde highly probable sets of polar coordinates onto the north and east directions. As a result, the eight HR8799bcde final positions for the four nights (left and right parts) are reported in Table~\ref{table:nightsAstrometry} and displayed in Fig.~\ref{figure:nightAstrometryZoom}. These positions will be used in Sect.~\ref{subsection:final_astrometry} to deduce the final HR8799 astrometry for epoch 2014.93. In addition to obtaining the highly probable position/flux for a given companion, the MCMC simulations give a robust estimation of the statistical error on the astrometry (i.e., related purely to photon noise). This error, reported in columns 7 and 8 of Table \ref{table:nightsAstrometry}, generally constitutes a minor contribution to the error budget, as discussed in the next sections.

\begin{figure*}
	\centering
   	\includegraphics[width=6cm]{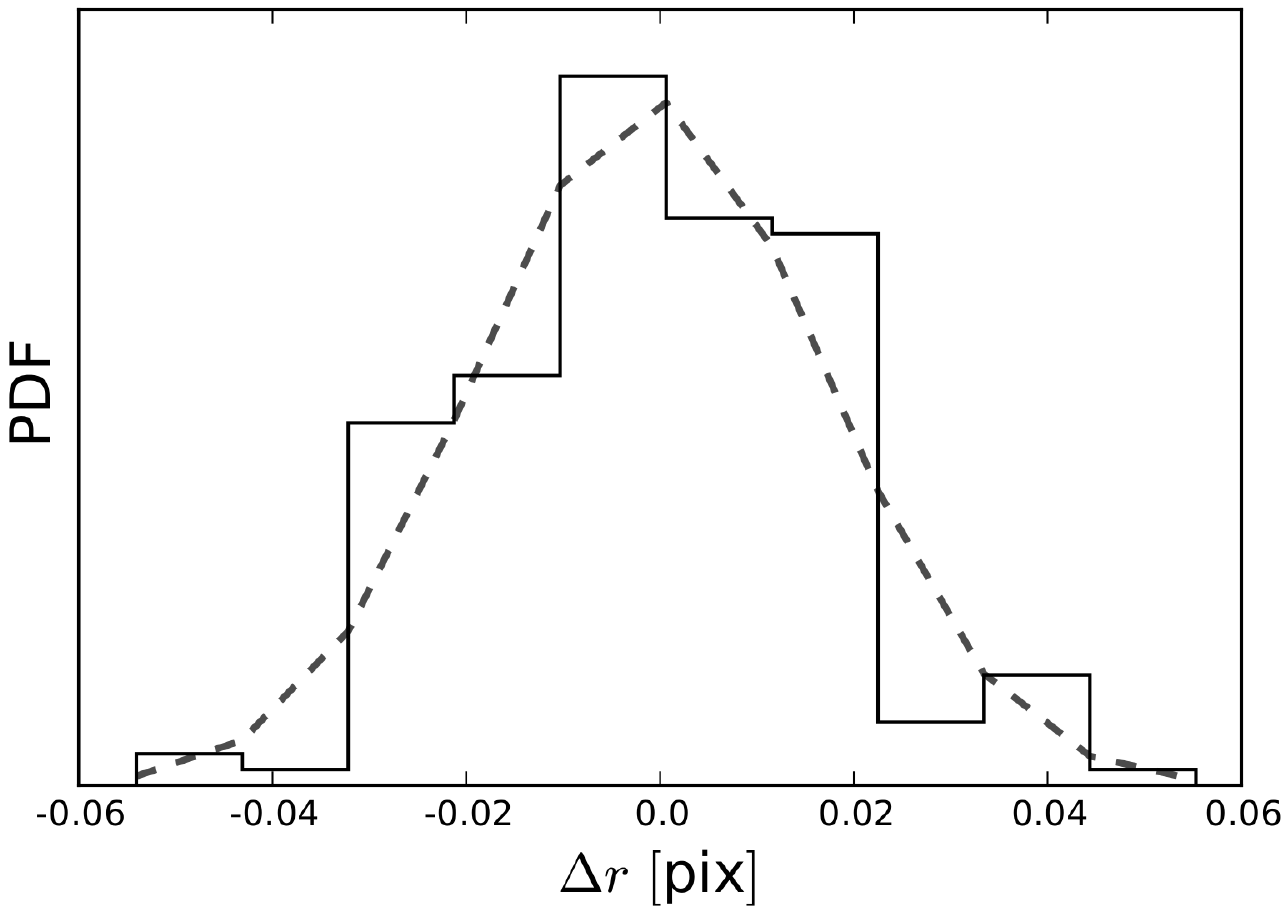}
	\includegraphics[width=5.85cm]{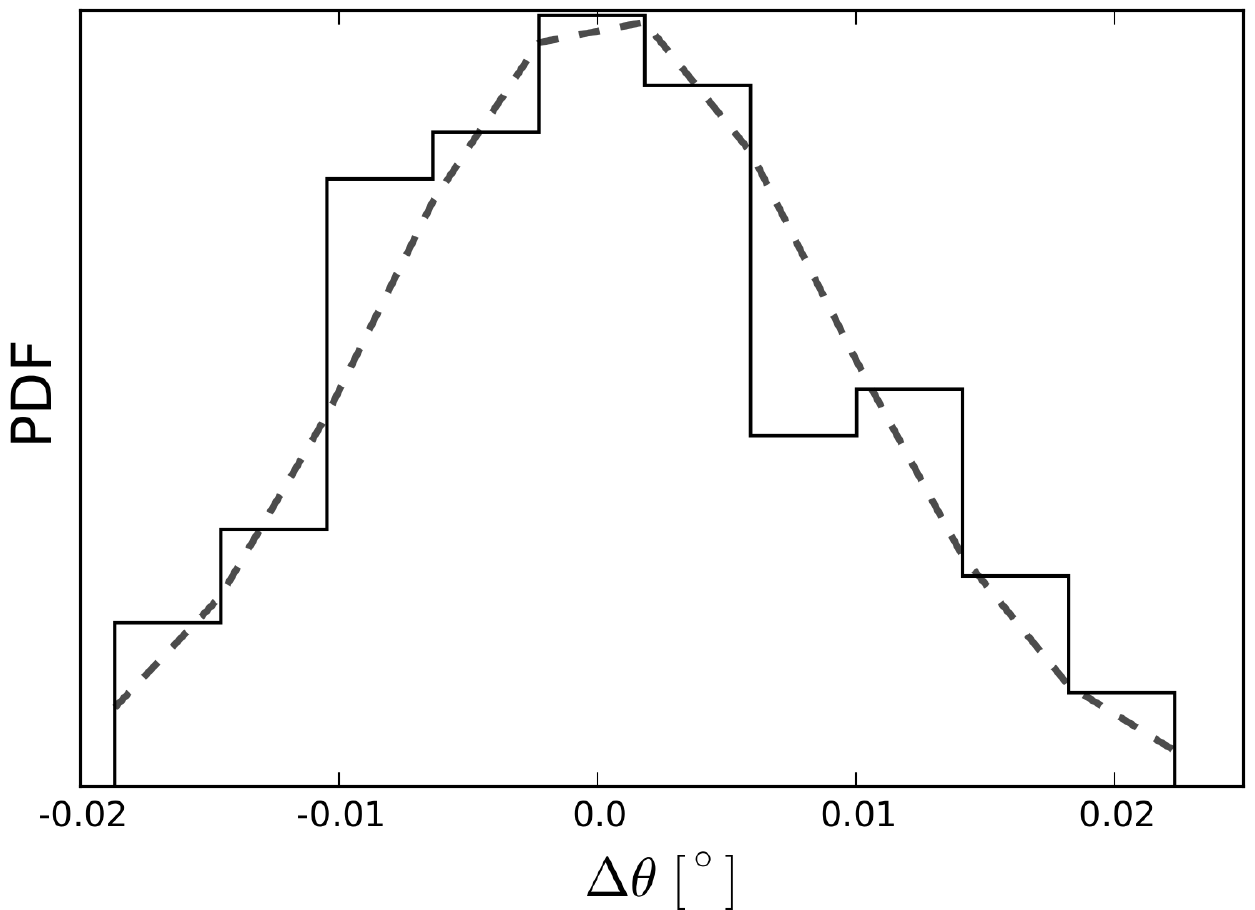}
	\includegraphics[width=6cm]{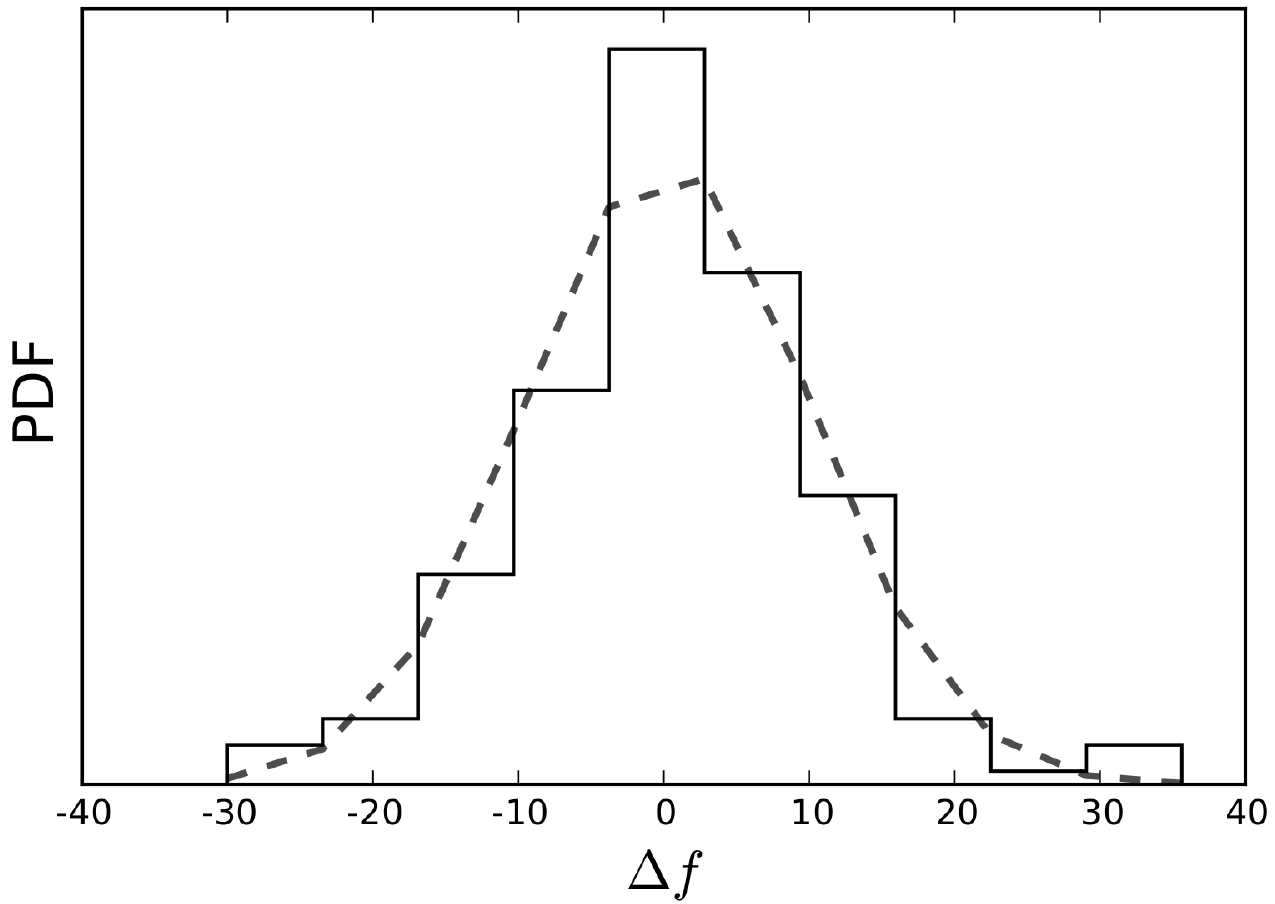}
     	\caption{Speckle noise estimation for HR8799b observed on 6 December 2014. The histograms illustrate the offsets between the true position/flux of a fake companion and its position/flux obtained from the NEGFC technique. The dashed lines correspond to the 1D gaussian	fit from which we determine the speckle noise.}
       	\label{figure:speckleHist}       
\end{figure*} 

\paragraph{Influence of the template PSF.}
\label{paragraph:PSF}

Since a non-saturated, off-axis PSF was not obtained in the same observing mode during the nights where HR8799 was observed, we chose as a PSF template for our NEGFC analysis the closest off-axis PSF in time obtained with the same observing mode under similar weather conditions, which turned out to be an off-axis PSF of beta Pictoris obtained on 30 January 2015. The fact that both the instrument and the atmospheric conditions may have changed within the interval leads to a possible bias in our measurement of the planets position, which could vary from night to night. To evaluate this bias, we have taken a series of twelve off-axis PSFs observed in the same mode under good atmospheric conditions, obtained in 2015 in the context of the SHARDDS survey (J.~Milli, personal communication). For each planet and each observing night in our HR8799 data set, we successively used the twelve off-axis PSFs as templates for the NEGFC technique, and derived the planets astrometry using the method described above. The dispersion of the astrometric measurements gives us an estimation of the bias that can be introduced by using a non-contemporaneous PSF. The observed dispersion does not depend much on the planet nor on the observing night, and has an overall standard deviation of 0.6~mas. This error bar will be added quadratically to the other error sources in Sect.~\ref{subsection:final_astrometry}.

\paragraph{Influence of residual dispersion.}
\label{paragraph:ADC}

Another source of imperfection in the recovery of the planets astrometry for broadband observations is the residual atmospheric dispersion after correction by the atmospheric dispersion correctors (ADC) included in the SPHERE optical path. While the small angular separation between the star and planets ensures the residual dispersion to be almost perfectly equal for {all of them, their different spectra can result in a chromatic offset between their measured positions. The residual dispersion after correction by the SPHERE ADC has been shown to be smaller than 1.2~mas rms for zenith angles as large the maximum of $54\degr$ encountered in the present data set \citep{Hibon16}. Taking into account the H-band spectrum of the star and of the four planets \citep{Bonnefoy16}, we estimate that the maximum astrometric offset between the star and planets due to residual dispersion cannot be larger than 0.25~mas in the worst case where residual dispersion shows a linear trend across the H band. This contribution is negligible in our final astrometric error budget.

\subsection{Systematic error due to residual speckles}
\label{subsection:speckleNoise}

Performing PCA-ADI removes a large fraction of the quasi-static speckle noise and significantly improves the S/N of the companions. Although highly effective, this process is not perfect and some level of residual speckle noise remains in the post-processed images. Such noise has a major impact on photometric and astrometric measurements \citep{Guyon_2012_HighPrecision}, and needs to be taken into account in the error budget. Since speckle noise is known to have a radial dependence, we propose to estimate its impact by injecting fake companions in the data cube at the radial distance of the real planets but for a wide range of angular positions, and by testing the ability of the Nelder-Mead optimization to find their position and flux through the NEGFC technique. The first step in this process is to create an ``empty'' data cube by injecting four NEGFCs characterized by the highly probable positions/fluxes derived from the previous MCMC simulations. In the empty cube, we inject a fake companion characterized by a flux $f_{\text{true}}$ and a radial distance $r_{\text{true}}$, both corresponding to the highly probable solution, but at an arbitrarily chosen angular coordinate $\theta_{\text{true}, i}$. Using the NEGFC technique coupled with the Nelder-Mead optimization, we determine the position/flux $(r_i, \theta_i, f_i)$ of the fake companion. We then compute the offsets $\Delta r_i = r_{\text{true}} - r_i$, $\Delta \theta_i = \theta_{\text{true},i} - \theta_i$, $\Delta f_i = f_{\text{true}} - f_i$ between the known position/flux characterizing the fake companion and the solution obtained from the optimization process. The same process is repeated for a series of 360 azimuths equally spaced between $0\degr$ and $360\degr$. These 360 realizations are used to build three normalized histograms,  respectively for $\Delta r$, $\Delta \theta$ and $\Delta f$. The histograms for $\Delta r$ and $\Delta \theta$ are then fitted with a Gaussian function, and the standard deviations $\sigma_{\text{spec},r}$ and $\sigma_{\text{spec},\theta}$  of the Gaussian functions are used as an estimation of the speckle noise affecting the radial and azimuthal coordinates. A similar approach was already used by e.g. \citet{Maire_2015_HR8799}. We illustrate in Fig.~\ref{figure:speckleHist} the three histograms for HR8799b observed on 6 December 2012. The results obtained for all the planets and data cubes are reported in column 9 and 10 in Table \ref{table:nightsAstrometry}. It appears clear that the error induced by speckle noise increases for decreasing angular separations of the companion with respect to the host star.  Indeed, the brightness of the residual speckles increases closer to the star. We also note that speckle noise is always larger than statistical noise, except for HR8799b.

Another possible way to evaluate speckle noise is to measure the influence of the number of PCs used in the PCA post-processing on the position/flux determination, as proposed e.g.\ by \citet{Pueyo_2015_HR8799}. Indeed, the residual speckle pattern changes as a function of the number of PCs. To check the consistency of this method with the one proposed above, we determined the position/flux of each companion in each data cube using the NEGFC technique with the Nelder-Mead optimization using a number of PCs ranging from 5 to 90 (for a number of PCs $> 90$, the companion self-subtraction becomes too important to get a high S/N). We then constructed three normalized histograms, respectively for $r$, $\theta$ and $f$. As expected, the standard deviations of these histograms are similar to those deduced above. 

Finally, we note that the residual speckle noise estimated here is in good agreement with the semi-empirical estimation of the astrometric accuracy based on the planet S/N proposed in the case of pure photon noise by \citet[][Eq.~A1]{Guyon_2012_HighPrecision}, if we extrapolate this relation to the speckle-dominated regime in the following way, as already proposed by \citet{Mawet_2015_HR3549}: $\sigma_{1{\rm D}}[\lambda/D] = 1/(\pi S/N)$. Using such a semi-empirical formula therefore looks like a possible way to get a quick estimation of the astrometric error bar related to speckle noise, although we recommend to go through the analysis presented in this section to obtain a robust estimation.

\subsection{Error on the star position}
\label{subsection:star_position}

Inside SPHERE, a dedicated differential tip-tilt sensor is used to obtain an image of the PSF just upstream of the coronagraph, and is used as an input for closed-loop control of the star position with respect to the coronagraph, thereby ensuring a stable star centering \citep{Fusco_2006_SPHERE, Baudoz_2010_SPHERE}. Based on laboratory measurements, the expected accuracy of the star centering is supposed to be around $0.5$~mas on sky \citep{Baudoz_2010_SPHERE}. As mentioned in Sect.~\ref{subsection:reduction}, no individual frame centering was applied to the data cubes in our analysis, but rather a global centering of all frames in each individual cube using the same $x,y$ offsets.

Here, we independently estimate the uncertainty on the mean star position for each data cube. The evaluation of this uncertainty is based on the histogram of the $x$ and $y$ offsets measured for all individual frames by the centroid plus intersection method described in Sect.~\ref{subsection:reduction}. The mean position of the star in a given data cube can be obtained by a Gaussian fit of the two histograms, as illustrated in Fig.~\ref{figure:star_offsets}. Based on this figure, we will assume in the following discussion that the histograms follow a Gaussian distribution, so that the accuracy on the determination of the mean stellar position in a given cube is given by the standard deviation of the best-fit Gaussian divided by the square root of the number of realizations. The standard deviations of the best-fit Gaussians are given in Table~\ref{table:starPosition} in terms of RA and DEC, by projecting the $(\sigma_{\star,x}, \sigma_{\star,y})$-error ellipses expressed in detector coordinates onto the north and east directions. We note that the derived stellar jitter estimation is slightly larger than predicted in \citet{Baudoz_2010_SPHERE}, with values varying from $0.76$~mas to $1.44$~mas depending on the night (i.e., around 0.1 pixel in detector coordinates). Based on these values, and taking into account the $\sim200$ frames present in each data cube, the error bar on the mean stellar position in any given cube amounts to less than 0.1~mas, and is therefore completely negligible in our final noise budget.

\begin{table}
\caption{Estimation of the stellar jitter in the eight data cubes.}
\label{table:starPosition}
\centering
\begin{tabular}{ccc}
\hline\hline
Date and side	& $\sigma_{\star, \text{RA}}$ [$''$]   	& $\sigma_{\star, \text{DEC}}$ [$''$] 	\\
\hline
2014-12-04 L	&$0.00076$	&$0.00078$ \\
2014-12-04 R	&$0.00076$	&$0.00078$ \\
2014-12-05 L	&$0.00084$	&$0.00081$ \\
2014-12-05 R	&$0.00087$	&$0.00085$ \\
2014-12-06 L	&$0.00077$	&$0.00079$ \\
2014-12-06 R	&$0.00078$	&$0.00080$ \\
2014-12-08 L	&$0.00143$	&$0.00139$ \\
2014-12-08 R	&$0.00144$	&$0.00139$ \\	
\hline			
\end{tabular}
\end{table}

However, this contribution represents only the purely statistical error on the determination of the star position. We also need to take into account possible systematic biases on the determination of the star position based on the satellite spots. To this end, we have obtained a data set on a relatively bright star, using the waffle mode of the DM but without coronagraph. The star was mildly saturated at its center to increase the S/N on the satellite spots. We determined the center of the star based on a truncated Moffat profile, to reject the saturated part of the PSF, and compared this estimation with the prediction based on the satellite spots. We checked that the two estimations match with an accuracy better than 0.1 pixel, which represents our best estimation of an upper limit on a possible bias. This also confirms that the method proposed in Sect.~\ref{subsection:reduction} to determine the stellar position from the satellite spots does not lead to major astrometric bias, even in the presence of residual atmospheric dispersion. Here, we will conservatively assume that a bias of 0.1~pixel (1.2~mas) affects our determination of the mean star position in all cubes. 

\subsection{Final astrometry}
\label{subsection:final_astrometry}

Particular care must be taken when combining the results and error bars of several astrometric measurements, especially in the presence of correlated errors. How the various error bars add up needs a specific discussion. Firstly, we note that our experimental determination of the error bar related to residual speckles inherently takes into account the contribution of photon noise. Indeed, the empirical intensity of the speckles includes the contribution of the photon noise associated to all sources of signal at any given location (stellar residuals, planet, sky emission and thermal background). This is backed up by the fact that the error bar associated to speckle noise generally dominates the error bar associated to photon noise. Only in the case of planet b are they of the same order of magnitude, which reflects the fact that residual speckles are very faint compared to residual background noise at that angular distance from the star.

Secondly, we make the conservative assumption that the errors related to speckle noise are fully correlated, not only between the left and right data cubes obtained on a same night, but also between all nights. The assumption of full correlation between the left and right data cubes is justified by the fact that the signals recorded by the two parts of the detector are almost identical (to within photon noise and some minor differential aberrations that amount to a few nm rms at most), and is backed up by the fact that the estimated error bars are almost identical for the left and right sides for most of the nights and planets (see Table~\ref{table:nightsAstrometry}). The assumption that speckle noise is fully correlated from night to night is more debatable. It is indeed expected that speckle noise will be partly correlated between successive nights, because residual speckles are often associated to non-common path aberrations in the instrument that can vary on very long timescales. To be on the conservative side, we will assume a full correlation of speckle noise in all data sets. The error bar on the final astrometry regarding speckle noise should then be computed as the median of all speckle noise-related error bars. We note however that the estimations of the speckle noise-related error bars significantly vary from one night to the other (see Table~\ref{table:nightsAstrometry}), which suggests that this noise is at least partly uncorrelated, and that our final error bars will be pessimistic.

\begin{table*}
\caption{The final HR8799bcde astrometric measurements with respect to the star for epoch 2014.93.}
\label{table:FinalAstrometry}     
\centering                          
\begin{tabular}{ccccc}        
\hline\hline                
Planet 		& $\Delta r$ [$''$] & $\Delta \theta$ [$\degr$] & $\Delta \text{RA}$ [$''$] & $\Delta \text{DEC}$ [$''$] \\                         
\hline
HR8799b 		& $1.7241 \pm 0.0019$& $65.99 \pm 0.13$ & $1.5748 \pm 0.0023$ 	& $0.7016 \pm 0.0036$ \\   
HR8799c 		& $0.9481 \pm 0.0017$& $327.37 \pm 0.16$ & $-0.5113 \pm 0.0024$ 	& $0.7985 \pm 0.0020$ \\  	
HR8799d 		& $0.6587 \pm 0.0019$& $217.40 \pm 0.19$ & $-0.4001 \pm 0.0021$ 	& $-0.5233 \pm 0.0020$ \\  	
HR8799e 		& $0.3855 \pm 0.0030$ & $271.71 \pm 0.31$ & $-0.3853 \pm 0.0030$ 	& $0.0115 \pm 0.0021$ \\           
\hline                         
\end{tabular}
\end{table*}

Thirdly, we proposed in the previous section that the final error bar related to the determination of the star position is dominated by a systematic bias that can amount up to 1.2~mas, and that the variability of the PSF shape can induce a bias of up to 0.6~mas. These biases will be added quadratically to our final astrometric error bar for all planets. The same applies to instrumental calibration errors, which are supposed to affect all data cubes in the exact same way. Indeed, appropriate observations of astrometric fields were not performed on each of the five HR8799 observing nights. We therefore had to rely on an astrometric calibration carried out by the SPHERE consortium one week later (see Sect.~\ref{subsection:instruNoise}), which was used as a reference for all five nights. Although we could not check the stability of the calibration over a few nights, we note that the latest IRDIS astrometric calibrations by the SPHERE consortium show that the time variations of plate scale and true north are mostly within their estimated error bars, based on two years of astrometric fields observations, while the pupil offset and anamorphic factor are mostly constant \citep{Maire16}. This suggests that our final estimation of the astrometric error bar should not include any unaccounted bias related to the variability of the IRDIS astrometric calibration. That being said, we still recommend that, in future observing programs dedicated to precise astrometric measurements, observations of standard astrometric fields be obtained during each individual night to ensure a high astrometric robustness.

Based on these assumptions, the computation of the final astrometry and related error bars proceeds as follows for each planet:
\begin{itemize}
\item define the final astrometry of the four planets as the weighted mean of the eight individual positions (left and right parts of the detector for the four nights), using as a weight the inverse of the variance of speckle noise;
\item estimate the final error bar related to speckle noise as the median of the individual error bars on the eight astrometric measurements;
\item add quadratically the contribution of speckle noise, the upper limit on the stellar centering bias, and the contribution of instrumental calibration errors to obtain the final astrometric error bars.
\end{itemize}
All these calculations are performed in polar coordinates, reflecting the fact that error bars generally have different behaviors along the radial and azimuthal directions. The last step is based on the following formulae:
\begin{align}
\sigma_{\text{tot},r}^2 &= \text{PLSC}^2 (\sigma_{r,\text{spec}}^2+\sigma_{r,\star}^2 + \sigma_{r,\text{PSF}}^2 + \sigma_{r,\text{AF}}^2 r^2) + \sigma_{\text{PLSC}}^2 r^2 \ , \\
\sigma_{\text{tot},\theta}^2 &= \sigma_{\theta,\text{spec}}^2 + \sigma_{\theta,\star}^2+ \sigma_{\theta,\text{PSF}}^2 + \sigma_{\theta,\text{AF}}^2 +\sigma_{\text{PO}}^2 + \sigma_{\text{TN}}^2 \ ,
\end{align}
where $r$ is the radial distance in pixels, $\sigma_{r,\text{spec}}$ and $\sigma_{\theta,\text{spec}}$ the final radial (pixels) and azimuthal (degrees) error bars related to speckle noise, $\sigma_{r,\star}$ and $\sigma_{\theta,\star}$ the radial (pixels) and azimuthal (degrees) stellar centering biases, $\sigma_{r,\text{PSF}}$ and $\sigma_{\theta,\text{PSF}}$ the radial (pixels) and azimuthal (degrees) error bars related to the imperfection of the PSF template in the NEGFC analysis, $\sigma_{r,\text{AF}}$ and $\sigma_{\theta,\text{AF}}$ the radial and azimuthal errors on the anamorphic factor expressed in percent, and where PLSC refers to the plate scale in $\arcsec/$pixel, PO to the pupil offset and TN to the true north, both in degrees. The final astrometries and related error bars are given for the four planets in Table~\ref{table:FinalAstrometry} and are illustrated in Fig.~\ref{figure:nightAstrometryZoom}. Table~\ref{table:FinalAstrometry} includes a projection of the error bars onto the RA and DEC directions, to comply with the usage. However, we suggest that expressing the error bars in polar coordinates is more appropriate, because polar coordinates usually correspond to the major and minor axes of the error ellipse. Another, even more appropriate way to proceed would be to specify the error ellipse by its three parameters (two axes and position angle). In the present case, the error bars are sufficiently symmetric to proceed with RA/DEC error bars, even though we note that the HR8799b error bars are significantly asymmetric, the angular error bar being twice as large as the radial one. This is mostly due to the large uncertainty on the pupil offset ($0\fdg11$, see Sect.~\ref{subsection:instruNoise}), which severely affects planets located far from the star.

\begin{table}
\caption{Comparison between the final error bars ($\sigma_{\text{tot}}$) listed in Table \ref{table:FinalAstrometry} and the standard deviation of the eight positions per planet displayed in Fig.~\ref{figure:nightAstrometryZoom} (see also Table \ref{table:nightsAstrometry}).}
\label{table:ComparisonEBvsSTD}     
\centering                          
\begin{tabular}{ccccc}        
\hline\hline                
& $\sigma_{\text{tot},\Delta\text{RA}}$ & $\sigma(\text{$\Delta$RA})$ & $\sigma_{\text{tot},\Delta\text{DEC}}$ & $\sigma(\text{$\Delta$DEC})$ \\
Planet  & [mas] & [mas] & [mas] & [mas] \\
\hline
HR8799b 		& $2.3$ & $1.1$ & $3.6$ & $0.8$ \\   
HR8799c 		& $2.4$ & $1.2$ & $2.0$ & $0.9$ \\  	
HR8799d 		& $2.1$ & $0.9$ & $2.0$ & $1.4$ \\  	
HR8799e 		& $3.0$ & $1.2$ & $2.1$ & $1.9$ \\           
\hline                         
\end{tabular}
\end{table}

To check the consistency of our error bars, we compared the statistical distribution of the eight individual data points obtained for each planet to the individual error bars on the eight data points. Table~\ref{table:ComparisonEBvsSTD} shows that the final error bars are generally about twice larger than the dispersion of the individual data points. This is related to the fact that the major error sources (speckle noise, stellar position bias, instrumental calibration) are supposed to be fully correlated between individual measurements, so that the final error bar has a similar size as the individual ones. This suggests that an improvement by up to a factor two in astrometric accuracy could be achieved by improving the astrometric calibration. That said, the individual error bars are in relatively good adequacy with the dispersion of the data points (see Fig.~\ref{figure:nightAstrometryZoom}), although we note a significant asymmetry in the distribution of the data points towards the NE-SW direction. This asymmetry looks quite consistent between the four planets, and we therefore suggest that it comes from a time variability in the bias on the stellar position measurement (the only error source that is naturally expressed in RA/DEC), which could be related to variations in the PSF shape and/or in the diffraction pattern created by the DM on a night-to-night timescale. This variation remains within the expected amplitude of about 0.1 pixel for the star position bias.

For planet b, the main contribution to the error budget comes from the imperfect astrometric calibration and from the uncertainty on the star position, while speckle noise is negligible. This is consistent with the fact that HR8799b lies in a region that is not significantly affected by residual speckles (see Fig.~\ref{figure:HR8799fullframe}). For planet c, although speckle noise significantly increases, the noise budget remains dominated by the astrometric calibration and stellar position uncertainties. The dominance of stellar centering noise in the astrometric error budget of these two planets is backed up by the fact that the dispersion in the individual astrometric measurements for planets b and c has a similar amplitude and shape (see Fig.~\ref{figure:nightAstrometryZoom}), as expected for a global centering error. For the two inner most planets (d and e), speckle noise progressively becomes the dominant contributor to the error budget, and once again this is consistent with Fig.~\ref{figure:nightAstrometryZoom}, where the dispersion of the astrometric data points increases significantly, especially for planet e. We finally note that our astrometric measurements are in general agreement with the astrometric measurements derived in \citet{Zurlo_2016_HR8799} and \citet{Apai_2016} to within error bars, but that our error bars are two to three times smaller, thanks to a careful evaluation of all systematic error sources. For the orbital architecture analysis presented in the next section, we will thus only use our data reduction for the IRDIS data set of December 2014.

\section{Orbital fitting analysis}
\label{section:PyAstrOFit}

\subsection{The \texttt{PyAstrOFit} Python package}
\label{subsection:PyAstrOFit_package}

To perform our analysis of the HR8799bcde orbital architecture, we have adopted the Bayesian framework. With the aim of making our results reproducible as well as allowing anyone to straightforwardly perform similar analysis, we introduce the \texttt{PyAstrOFit} package\footnote{\texttt{https://github.com/vortex-exoplanet/PyAstrOFit}} implemented in Python 2.7, which is fully dedicated to orbital fitting using the MCMC approach. The code is open source and has been used to carry out our analysis and to produce all the figures presented in this section. \texttt{PyAstrOFit} is composed of several modules but the core of the package relies on three main modules, referred to as \texttt{Orbit}, \texttt{Sampler} and \texttt{Inference}: 
\begin{itemize}
\item The \texttt{Orbit} module is used to instantiate an orbit object, which includes the required data to model or represent any bound orbit. Unbound orbits are not considered because they would require the use of universal Keplerian variables and Stumpff functions \citep{Beust_2016_MCMC}.
\item The \texttt{Sampler} module constitutes the core of the Markov chains construction. It embeds the \texttt{emcee} package \citep{emcee}, which implements the affine-invariant ensemble sampler for MCMC proposed by \citet{Goodman_MCMC_2010}.
\item The \texttt{Inference} module is dedicated to Bayesian inference. Its main purpose is to represent both the posterior PDFs and the correlations between parameters from the Markov chain, to determine the confidence intervals, and to derive a set of allowable orbits or the best solution in terms of reduced $\chi^2$.
\end{itemize} 
The \texttt{PyAstrOFit} sampler comes with several \emph{convergence} diagnostic tools. In practice, one can never be sure that a chain has actually converged, but there exists several tests to evaluate whether the chain appears to be close to convergence (or more precisely, far from non-convergence\footnote{In the rest of the text, we will generally use ``convergence'' as a shortcut for ``far from non-convergence''.}):
\begin{itemize}
\item The \emph{acceptance rate} \citep[][]{MacKay_2003_InformationTheory}, which corresponds to the fraction of accepted to proposed candidates, can be monitored: if the acceptance rate is too high, the chain is probably not mixing well, while a low acceptance rate indicates that too many proposed candidate are rejected (which is symptomatic of a walker stuck in a given position). 
\item The \emph{Gelman-Rubin} $\hat{R}$ statistical test \citep[][]{Gelman_1992_MCMC, Ford_2006_MCMC, Gelman_2014_BayesianDataAnalysis} compares, for each parameter, the variance estimated from nonoverlapping parts of the chain to the variance of their estimates of the mean. A large $\hat{R}$ value may arise from slow chain mixing or multimodality \citep[][]{Cowles_1996_MCMC}. Conversely, a $\hat{R}$ value close to $1$ indicates that the Markov chain is close to convergence.
\item The \emph{lag $\rho_k$ autocorrelation} corresponds to the correlation between every draw and its $k$th lag. A relatively high $\rho_{k=K}$ value for a given $K$ indicates a high degree of correlation between the draws, a slow mixing and a chain far from ence.
\item The \emph{integrated autocorrelation time} $\tau$ \citep[see e.g.][]{emcee, Christen_2010_MCMC, Goodman_MCMC_2010}, also called \emph{inefficiency factor}, aims to give an estimate of the number of posterior PDF evaluations required to draw an independent sample. The smaller $\tau$, the better.
\end{itemize}

\begin{figure*}
	\centering
   	\includegraphics[width=18cm]{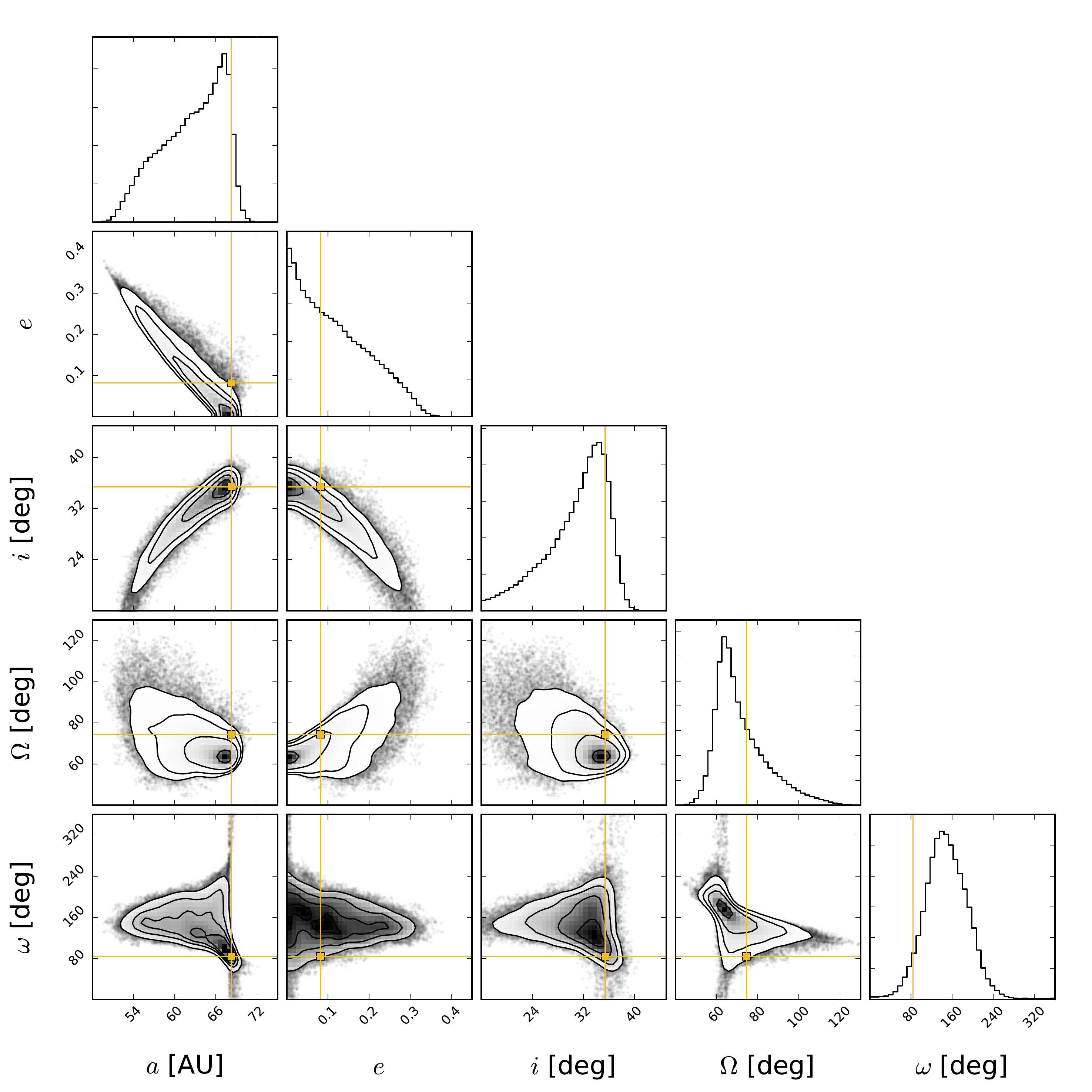}
     	\caption{Results of the MCMC simulations for HR8799b, displayed as a corner plot for the Keplerian elements $a$, $e$, $i$, $\Omega$ and $\omega$. The diagonal panels illustrate the posterior PDFs while the off-axis ones illustrate the correlation between the parameters. The yellow lines and crosses correspond to the best solution in terms of reduced $\chi^2$.}
       	\label{image:corner_b}       
\end{figure*}

\begin{figure*}
	\centering
   	\includegraphics[width=18cm]{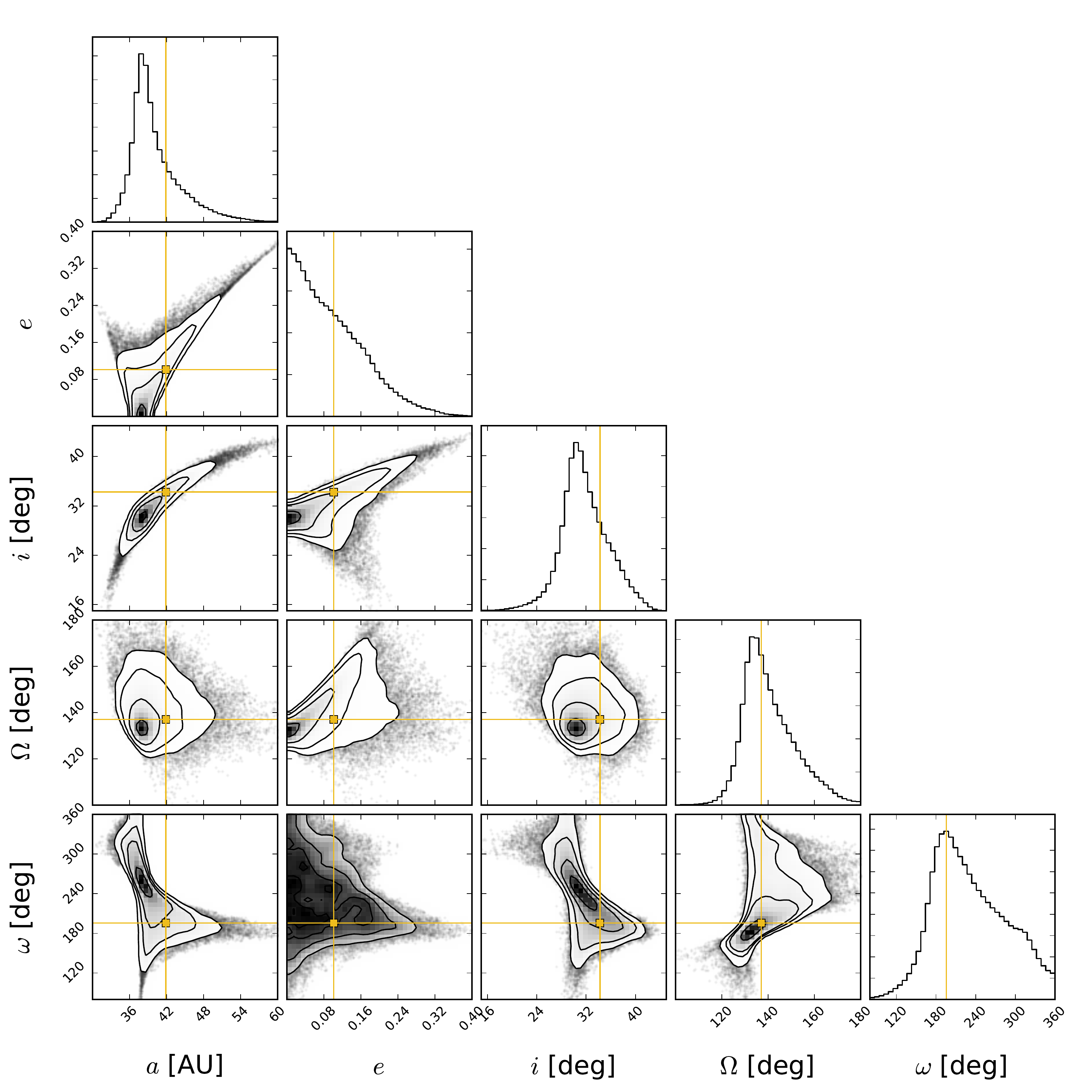}
     	\caption{Same as Fig.~\ref{image:corner_b}, for HR8799c.}
       	\label{image:corner_c}       
\end{figure*}

\begin{figure*}
	\centering
   	\includegraphics[width=18cm]{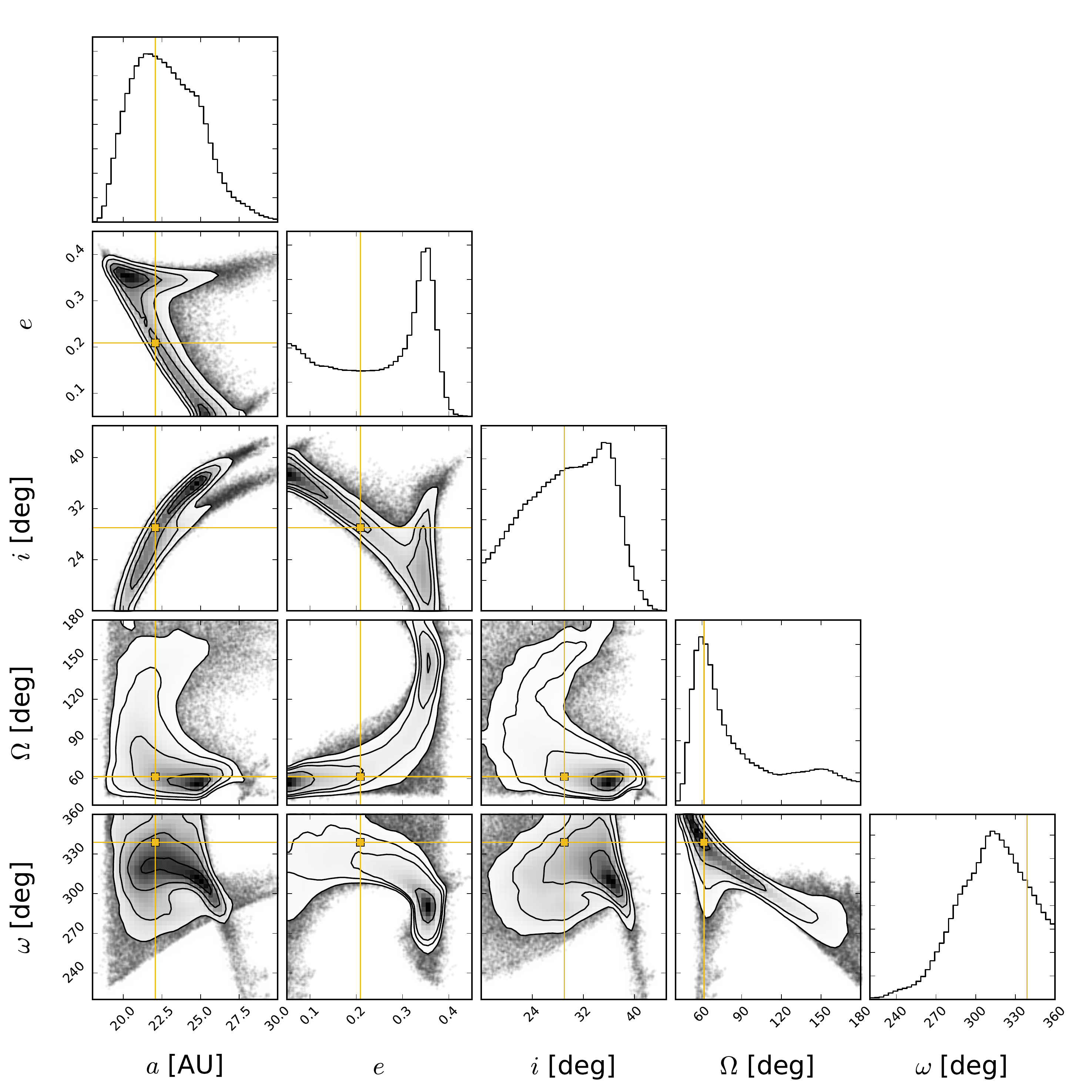}
     	\caption{Same as Fig.~\ref{image:corner_b}, for HR8799d.}
       	\label{image:corner_d}       
\end{figure*}

\begin{figure*}
	\centering
   	\includegraphics[width=18cm]{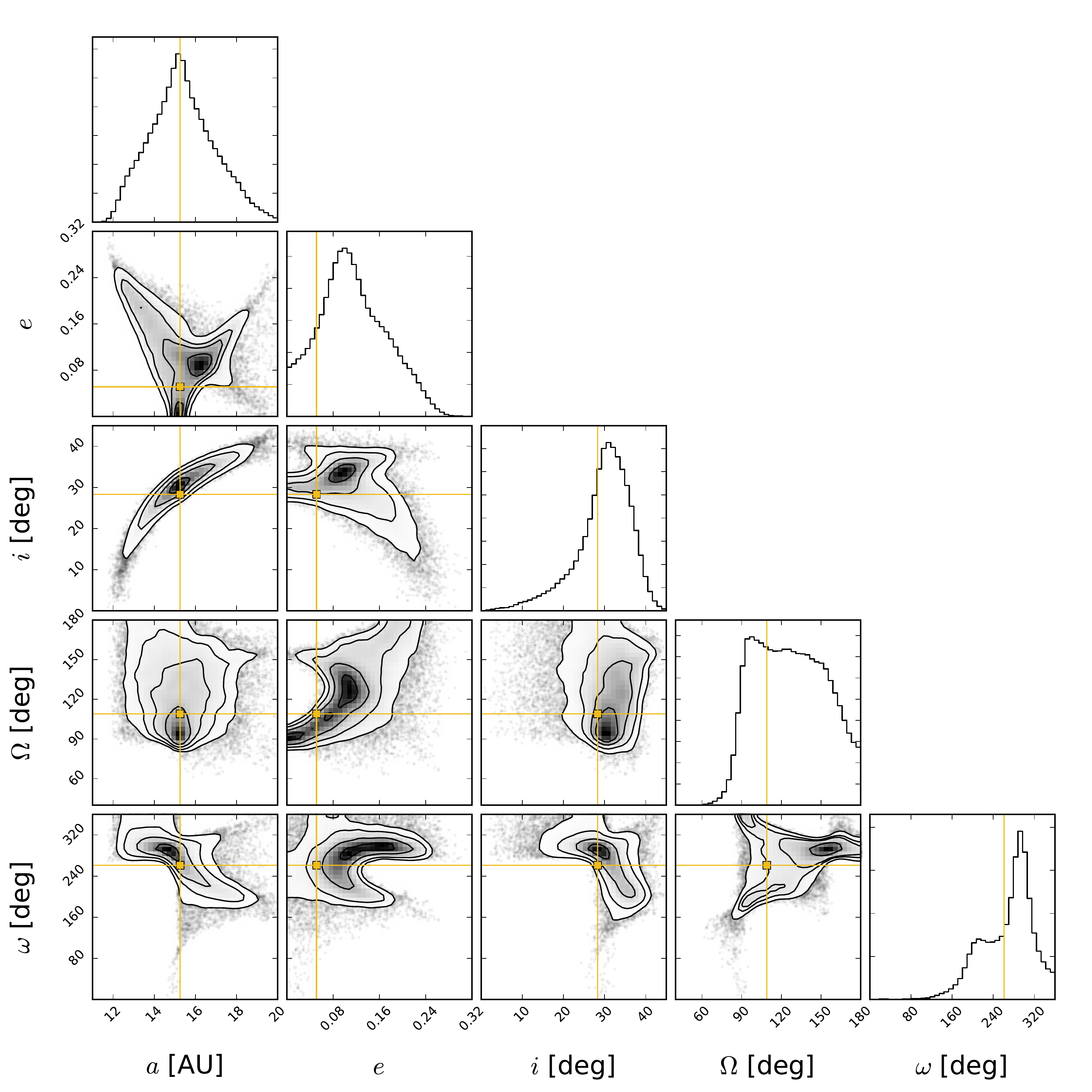}
     	\caption{Same as Fig.~\ref{image:corner_b}, for HR8799e.}
       	\label{image:corner_e}       
\end{figure*}

Following \citet{Ford_2006_MCMC} and \citet{Chauvin_2012_betaPic}, the sampling can be done on three different state vectors noted $\vec{x} = (a, e, i , \omega, \Omega, t_{p})$, $\vec{x}' = (\log{P}, e, \cos{i} , \Omega + \omega, \Omega - \omega, t_{p})$ where $P$ represents the orbital period, or $\vec{u}(\vec{x})$ defined at Eq. (A.1) in \citet{Chauvin_2012_betaPic}. As suggested by \citet{Ford_2006_MCMC}, adopting a uniform prior distribution for $\vec{x}'$ may help to improve the convergence of the chain. Indeed, a uniform prior distribution for $\cos{i}$ in the interval $[-1, 1]$ implies a prior distribution proportional to $\sin{i}$ in the interval $[-90\degr, 90\degr]$ for the inclination. As a consequence, it implies that orbits characterized by $i \simeq 0\degr$ (face-on) are considered intrinsically less probable than the ones characterized by $i \simeq 90\degr$ (edge-on). 

Which statistical test one should adopt as a convergence criterion is a question with no trivial solution. Nonetheless, some recommandations can be found in the literature \citep[see e.g.][]{Cowles_1996_MCMC}. For instance, \cite{Gelman_2014_BayesianDataAnalysis} theoretically demonstrated that the optimal Markov chain mixing to sample normally distributed posterior PDFs is characterized by an acceptance rate equal to $\sim 0.44$ when adopting the Metropolis-Hastings algorithm. It is widely agreed that an acceptance rate between $0.2$ and $0.5$ constitutes a appropriate value to ensure a good Markov chain mixing. The criteria adopted in this analysis are defined in the next section when we address the HR8799 orbit fitting, and are generally similar to those used by \citet{Pueyo_2015_HR8799} in their study of the HR8799 orbital parameters. When all the statistical tests meet the criteria, we consider that the part of the chain satisfying them has converged. The \texttt{Inference} module can then be used to draw independent samples from the chain to construct the final posterior PDFs for each Keplerian parameter. The confidence intervals are defined in terms of highest density regions \citep[HDR,][]{Hyndman96}, also referred to as highest posterior density intervals. For a given confidence level $1-\alpha$ ($0 \leq \alpha \leq 1$), the idea is to take a horizontal line and shift it up until the area under the regions of the PDF located above this line represents a fraction $1-\alpha$ of the total area under the PDF. The projection to the $x$ axis of this area defines the $100(1-\alpha)$\% HDR. To infer our confidence intervals, we choose to use the $68.3$\% HDRs. In the ideal case where the PDF $f$ is unimodal, the HDR corresponds to the smallest of all intervals $[a,b]$ that satisfy $\text{Pr}(a \leq x \leq b) = 1-\alpha$, which happens such that $f(a) = f(b)$.

The \texttt{Inference} module also comes with various tools to display the results such as corner plots to illustrate the 
PDFs and their corresponding correlation, walk plots to illustrate the mixing of the chain, or the illustration of allowable orbits together with the data. More information about all the \texttt{PyAstrOFit} possibilities and tutorials dedicated to each
module can be found on the \texttt{GitHub} repository.

\subsection{HR8799 orbital fitting with \texttt{PyAstrOFit}}
\label{subsection:PyAstrOFit_HR8799}

Here, we revisit the MCMC-based Bayesian analysis described in \citet{Pueyo_2015_HR8799} using more robust convergence criteria before using our chains for inference, and using an extended data set by adding not only the SPHERE astrometric data presented in this work but also the latest astrometric measurements from the literature (see below). We update those results to what is described in this section. As a consequence, this analysis supersedes the one in \citet{Pueyo_2015_HR8799}.

For our orbital analysis, we have assumed the distance of the system and the mass of the host star to be respectively 39.4~pc \citep{vanLeeuwen_2007_HR8799} and  $1.51 M_{\sun}$ \citep{Baines_2012_HR8799}. The possibility of inferring these parameters from the orbital fitting modules has not been implemented into \texttt{PyAstrOFit} yet, but could be the subject of a future update. The astrometric positions used for the orbit fitting come from the works of \citet{Marois_2008_HR8799}, \citet{Lafreniere_2009_HR8799}, \citet{Fukagawa_2009_HR8799}, \citet{Metchev_2009_HR8799}, \citet{Hinz_2010_HR8799}, \citet{Currie_2011_HR8799}, \citet{Currie_2012_HR8799}, \citet{Currie_2014_HR8799}, \citet{Bergfors_2011_HR8799}, \citet{Galicher_2011_HR8799}, \citet{Soummer_2011_HR8799}, \citet{Esposito_2013_HR8799}, \citet{Maire_2015_HR8799}, \citet{Zurlo_2016_HR8799}, and \citet{Konopacky16} along with the positions derived in the present work. A compilation of all the astrometric measurements for HR8799bcde is provided in Appendix (Table \ref{table:Compilation}). All these data are included into the \texttt{PyAstrOFit} source code, and can easily be queried by an interested user. 

The presence of systematic errors, whose careful calibration is described in Sect.~\ref{section:RA}, turned out to be a significant nuisance when trying to reconcile contemporaneous astrometric measurements from various instruments. For instance the difference in astrometry for HR8799b between \citet{Currie_2014_HR8799} and \citet{Pueyo_2015_HR8799} could be explained by an offset in the true north position between Palomar and Keck (this offset has a lesser impact for the planets at smaller separations). Given the relative paucity of data in the astrometric calibrator (based on a single astrometric binary) presented in \citet{Pueyo_2015_HR8799}, compared to the long history of high precision Keck astrometry \citep[see e.g.,][]{Yelda10,Service16}, we chose to include only the \citet{Currie_2014_HR8799} points in our analysis. Another example would be the discrepancy for the position of HR8799d between these two papers, which can be easily traced back to the presence of a bright residual speckle near planet d in the Palomar/P1640 data. Because there is no CPU efficient method to carry out the negative injection in IFS data without ADI, \citet{Pueyo_2015_HR8799} could not use the method described in Sect.~\ref{subsection:instruNoise}, and only used a method similar to that presented in Sect.~\ref{subsection:speckleNoise} (at other azimuth angles, where there was no bright speckle). As a consequence the \citet{Pueyo_2015_HR8799} uncertainties on HR8799d are most likely under-reported, and we instead use the contemporaneous estimate of \citet{Currie_2014_HR8799}. These two examples illustrate the complexity of precision astrometry in high contrast imaging and the importance of carrying out all the steps of robust astrometry as described in the present paper. Because the HR8799 system has been observed by multiple instruments since 2012, and because we had access to the P1640 data, we could conduct these instrument to instrument sanity checks and choose the most robust published astrometry. Before 2012, the measurements are more sparse and we thus decided to include all of them in the orbit fitting in the absence of further information.

The affine invariant sampler implemented in \texttt{emcee} comes with only two hyperparameters to be tuned: the number of walkers and an adjustable scale parameter $a > 1$, which has a direct impact on the acceptance rate of each walker \citep[see][]{Goodman_MCMC_2010}. All our simulations were performed with 1200 walkers. For a given Keplerian parameter, the set composed of the first element of each walker constitutes the initial distribution, which depends on how we decide to initialize the walkers. The equilibrium distribution, which we expect to be close to the posterior distribution when the chain has converged, should not depend on this initial distribution, see e.g. \citet{Meyn_Tweedie_2009}. As discussed in \citet{emcee}, starting the simulation with an initial distribution close to the expected posterior distribution speeds up the convergence. This can be done by initializing the walkers in a small $N$-dimensional ball in the parameter space around a highly probable solution. However, such an approach  not only requires a priori knowledge of the main posterior distribution peak, but can also jeopardize the chain convergence if the posterior distribution is multi-modal. Various alternatives are proposed in \citet{emcee}, and we have tested some of them. In particular, we can start the walkers uniformly over a given range in the parameter space. All our tests have led to identical results for all the Keplerian elements of each planet. 

We have started the chain construction with a minimum of 1000 steps per walker before beginning any convergence tests. During the MCMC run, the acceptance rate was monitored and the hyperparameter $a$ (initially set to $2$) was dynamically tuned in order to ensure an acceptance rate between 0.2 and 0.5 for at least 75\% of the walkers. The walkers for which acceptance rate was outside $[0.2, 0.5]$ were discarded and not used for Bayesian inference. We have considered that a chain has converged when the Gelman-Rubin statistical test $\hat{R} < 1.01$ was satisfied three times in a row for all Keplerian parameters. The convergence was reached after typically 40,000 steps per walker, for a total computing time equal to 3 hours using a computer equipped with 28 processing units.

\begin{table*}
\caption{Confidence intervals ($1\sigma$) and best solutions in term of $\chi^2_{\text{min}}$ for all the Keplerian elements, as well as for the orbital period $P$.}
\label{table:ConfidenceIntervals}
\centering
\begin{tabular}{lcccc}
\hline\hline
 & HR8799b & HR8799c & HR8799d & HR8799e\\
\hline
$a$ [AU]					&$[59.3, 68.7]$ & $[33.4, 42.6]$		& $[18.58, 23.08]$	& $[13.5, 16.7]$\\
$a_{\chi^2_{\text{min}}}$   & $68.22$			& $35.80$				& $22.44$			& $15.25$\\
$P$ [yr]					&$[456.4, 569.5]$	& $[157.2, 226,4]$		& $[65.2, 90.3]$	& $[40.4, 55.6]$\\
$P_{\chi^2_{\text{min}}}$   & $395.87$		    & $174.41$			    & $86.55$			& $48.49$\\
$e$						    &$[0.0, 0.155]$	& $[0.0, 0.169]$		& $[0.050, 0.410]$	& $[0.003, 0.129]$\\
$e_{\chi^2_{\text{min}}}$   & $0.081$			& $0.062$				& $0.173$			& $0.051$\\
$i$ [deg]					&$[27.0, 37.3]$		& $[26.6, 38.2]$		& $[19.5, 37.9]$	& $[22.5, 34.8]$\\
$i_{\chi^2_{\text{min}}}$   & $35.43$			& $27.43$				& $31.93$			& $28.31$\\
$\Omega$	 [deg]			&$[55.7, 79.8]$		& $[125.1, 153.0]$		& $[45.0, 180.0]$	& $[84.6, 158.4]$\\
$\Omega_{\chi^2_{\text{min}}}$	& $74.45$		& $136.1$				& $51.2$			& $108.9$\\
$\omega$	[deg]				& $[100.6, 193.9]$	& $[154.8, 342.0]$		& $[286.9, 360.0]$	& $[183.6, 349.2]$ \\
$\omega_{\chi^2_{\text{min}}}$	& $83.85$		& $336.07$			& $350.8$			& $260.9$\\
$t_p$ [JD]					& [1683.54, 1772.04]	 & [1809.24, 1935.29]	& [1965.6, 1987.3]	& [1952.0, 1997.75] \\
$t_{p,\chi^2_{\text{min}}}$        	& $1723.13$		& $1907.85$			& $1974.28$		& $1979.21$\\
\hline		
\end{tabular}
\end{table*}

For each planet, the corner plots are depicted in Figs.~\ref{image:corner_b} to \ref{image:corner_e}. They illustrate the resulting posterior PDFs and the correlation between the Keplerian parameters $a$, $e$, $i$, $\Omega$ and $\omega$. The yellow lines represent the best solution in terms of reduced $\chi^2$. These solutions do not necessarily coincide with the peaks of the posterior PDFs. Following the procedure described in Section \ref{subsection:PyAstrOFit_package}, the confidence intervals were inferred from the posterior PDFs and are summarized in Table \ref{table:ConfidenceIntervals} together with the best solution in terms of reduced $\chi^2$. In addition, Figs. \ref{figure:full_highly_probable_orbits} and \ref{figure:zoom_highly_probable_orbits} given in the Appendix illustrate a set of $1000$ allowable orbits characterized by $\chi^2 < \chi^2_{\text{min}} + 0.1$ and for which all the Keplerian parameters are in the confidence intervals reported in Table \ref{table:ConfidenceIntervals}.

\subsection{Discussion}
\label{subsection:discussion}

\begin{figure}[!ht]
	\centering
	\includegraphics[width=7cm]{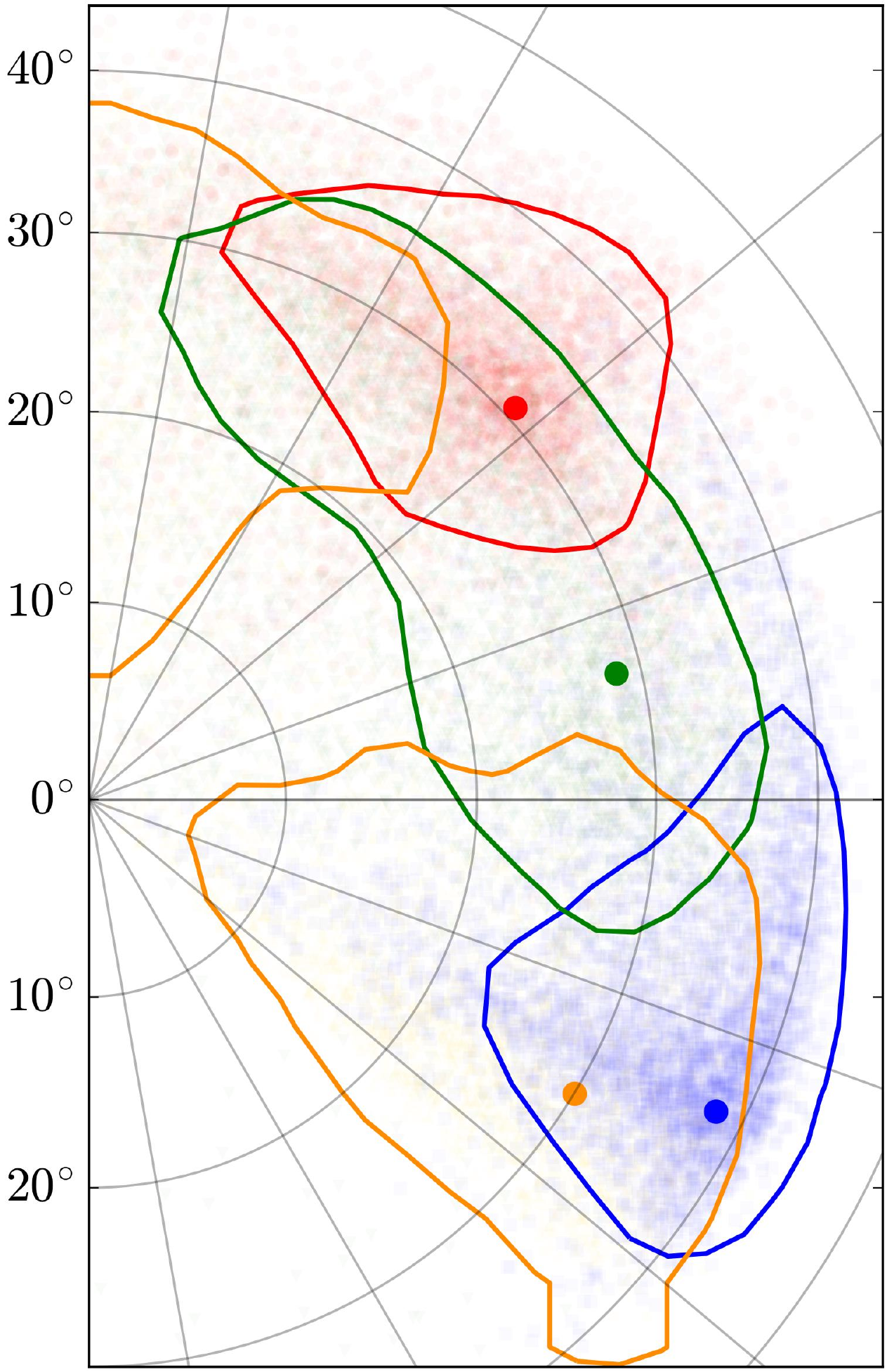}
     	\caption{Coplanarity test for HR8799bcde. The scatter plot illustrates the $\hat{\vec{n}}$ vector coordinates on the sky plane (see text) respectively for the HR8799b (blue), HR8799c (red), HR8799d (orange) and HR8799e (green) allowable orbits. All the points on a given circle refer to orbits characterized by the same inclination. Similarly, all points on a given radial branch refer to orbits characterized by the same longitude of ascending node. The angle between two spokes corresponds to $20\degr$. The solid lines represent the 68\% confidence interval around the most probable $\hat{\vec{n}}$ vector coordinates, which are depicted by thick points.}
       	\label{figure:coplanarity}       
\end{figure}

One of the most striking results of our MCMC analysis concerns the eccentricity of planet d, which shows a clear peak around $e_{\text{d,peak}} \sim \boldsymbol{0.35}$, and rejects the circular orbit hypothesis outside its $1\sigma$ confidence interval (although a significant set of solutions characterized by $e_\text{d} < 0.2$ cannot be ruled out). Actually, all four planets show possible signs of non-circular orbits at various levels, as the eccentricity posterior PDFs are generally rather broad, or in some cases (planets d and e) not monotonically decreasing.

Another striking result concerns the orientation of the orbits. The inclination of all the planets is similar and lies between about $\boldsymbol{20\degr}$ and $\boldsymbol{38\degr}$. It clearly rules out a family of solutions previously proposed in the literature \citep[see e.g.][]{Marois_2010b_HR8799, Currie_2011_HR8799}, for which the planets have a face-on orbit ($i=0\degr$). The longitude of the ascending node $\Omega$ generally shows a large confidence interval for all planets, but we can readily note a significant difference (at the $> 2\sigma$ level) between the $\Omega$ derived for planets b and c. Comparing the individual inclinations and longitudes of ascending nodes has however a limited usefulness, and we therefore propose to compare the three-dimensional relative orientations of the orbits for all four planets, to test the coplanarity of the system. This can be done by projecting onto the sky plane the normalized vector $\hat{\vec{n}}$ orthogonal to the orbital plane, defined by:
\begin{equation}
\label{equation:n}
\hat{\vec{n}} = \left[\sin{(i)} \cos{(\Omega - \pi/2)}, \sin{(i)} \sin{(\Omega - \pi/2)}, \cos{(i)} \right] \ .
\end{equation}
The scatter plot represented in Fig. \ref{figure:coplanarity} illustrates the vector $\hat{\vec{n}}$ coordinates on the sky plane for the allowable orbits of the four planets. The pole of the polar grid locates the projected vector that points towards Earth. All the points on a given arc of a circle refer to orbits characterized by the same inclination. Similarly, all points on a given spoke refer to orbits characterized by the same longitude of ascending node. Planets d and e have very wide distributions of orientations, which are compatible with any other individual planet. We can even note that the 68\% confidence interval is disjoint for planet d, which echoes the bimodal PDFs seen in Fig.~\ref{image:corner_d}. However, there is a clear discrepancy between the orbital planes of planets b and c, for which the orbital planes show a mutual inclination of $35\degr$, and the 68\% confidence intervals are largely disjoint. This suggests that the system might not be coplanar, at a significance level of about $2\sigma$. Taken together with the evidence for non-zero eccentricity for planet d, this represents new empirical constraints that may be hard to reconcile with the mean-motion resonance scenarios currently proposed in the literature \citep{Fabrycky_2010_HR8799,Gozdziewski_2014_HR8799}. Long-lived, non-resonant orbital architectures do not seem to predict these peculiar features either \citep{Gotberg16}.

\begin{table}
\caption{Relative probabilities (in $\%$) associated with different mean-motion resonances between consecutive pairs of planet.}
\label{table:mmr}
\centering
\begin{tabular}{cccc}
\hline\hline
Period ratio & $P_b/P_c$ & $P_c/P_d$ & $P_d/P_e$ \\ 
\hline
$1.5 (\pm 0.075)$ & 7.9 & 2.0 & 17.1 \\
$2.0 (\pm 0.100)$ & 16.3 & 12.1 & 13.5 \\
$2.5 (\pm 0.125)$ & 10.3 & 17.0 & 3.7 \\
$3.0 (\pm 0.150)$ & 2.1 & 8.9 & 0.6 \\
\hline		
\end{tabular}
\end{table}

\begin{figure*}
	\centering
   	\includegraphics[width=6.04cm]{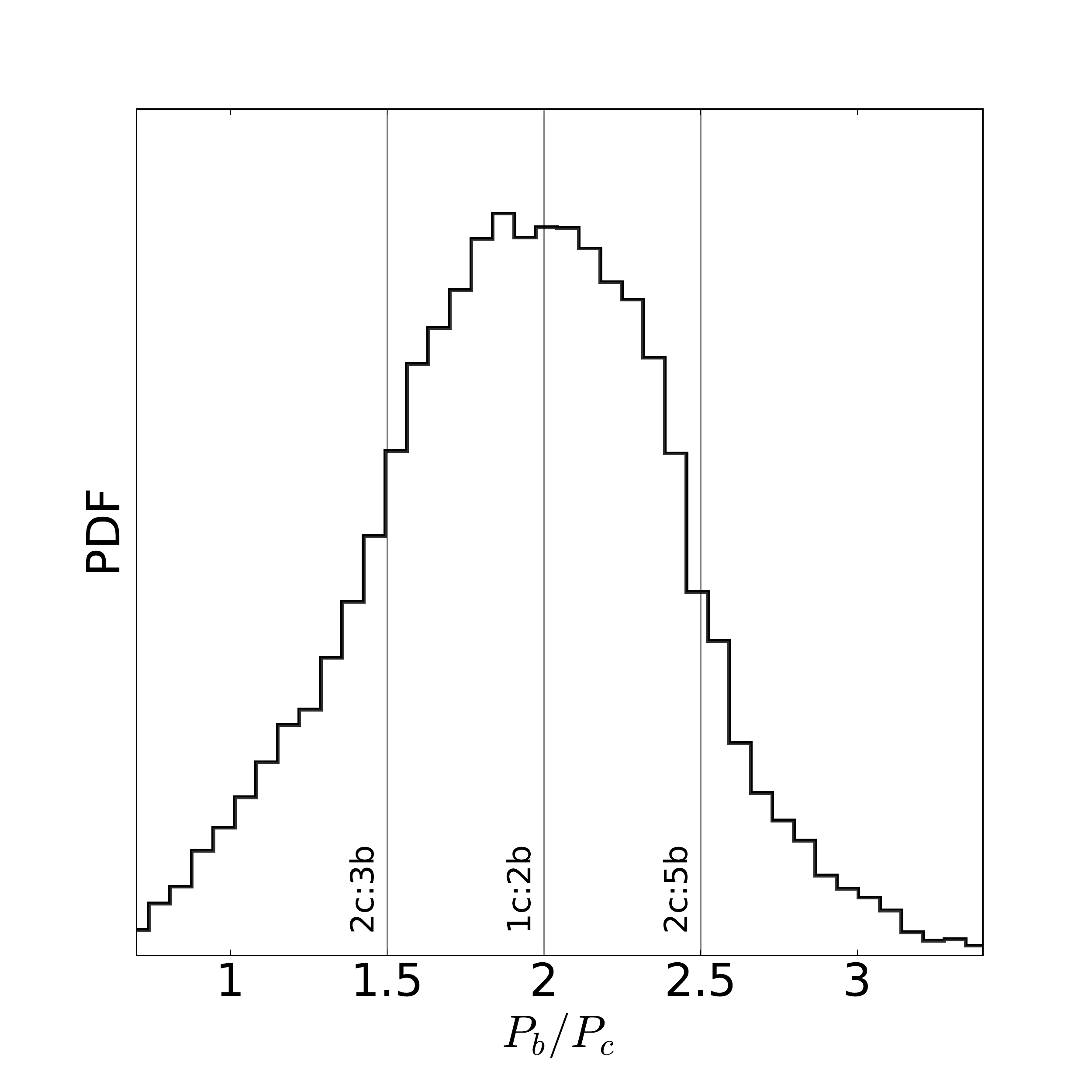}
	\includegraphics[width=6.03cm]{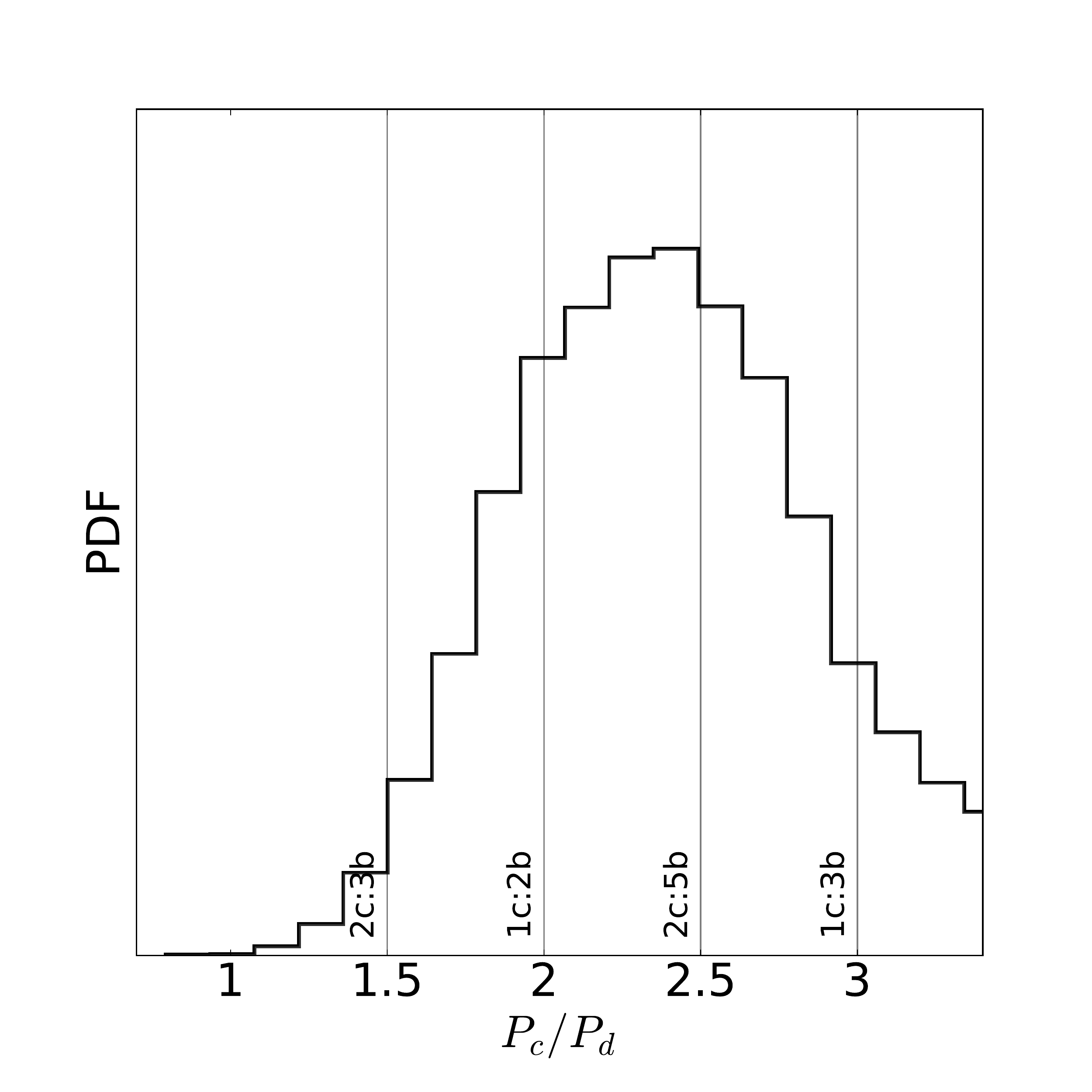}
	\includegraphics[width=6.04cm]{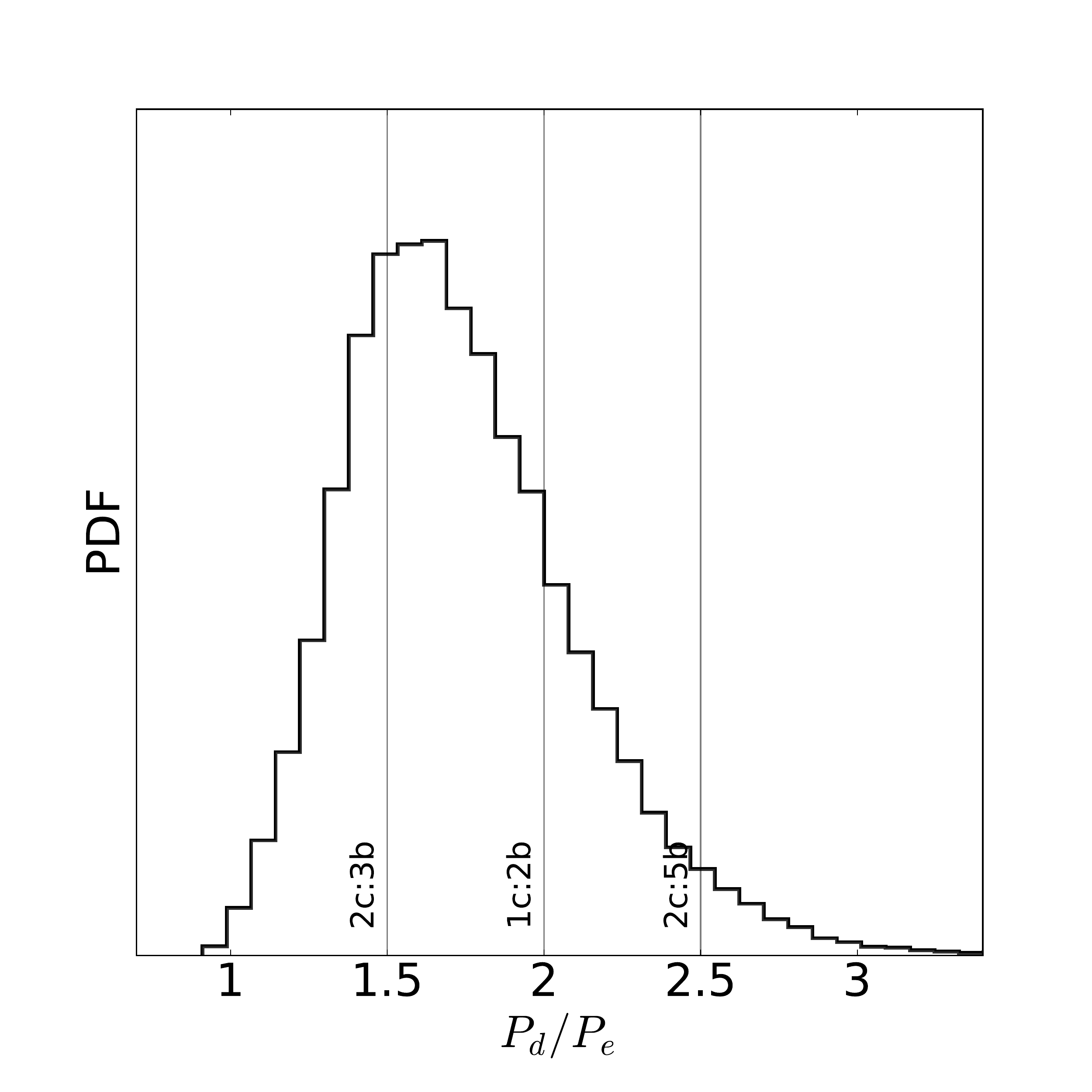}
     	\caption{Histograms of the period ratios $P_b/P_c$ (left panel), $P_c/P_d$ (middle panel) and $P_d/P_e$ (right panel) derived from the semi-major axis posterior PDFs given in Figs. \ref{image:corner_b} to \ref{image:corner_e} and adopting $1.51 M_{\sun}$ for the host star mass. These period ratios allow to identify which mean-motion resonances are compatible with the allowable orbits derived from our MCMC analysis. It appears clear that the most compatible mean-motion resonance is 1e:2d:4c:8b.}
       	\label{figure:mmr}       
\end{figure*} 

A thorough analysis of the system dynamics and stability, taking into account the most recent positions, would be of great 
interest but is beyond the scope of this paper. Yet, an interesting clue in this context is the compatibility of the planet periods with mean-motion resonances  \citep{Fabrycky_2010_HR8799,Gozdziewski_2014_HR8799}. We therefore propose to simply identify mean-motion resonances compatible with our results. To this aim, we illustrate in Fig. \ref{figure:mmr} the distribution of the ratios of periods, respectively $P_b/P_c$,  $P_c/P_d$ and $P_d/P_e$, obtained by dividing the respective PDFs. These PDFs were obtained by applying the Kepler's third law to the semi-major axis PDFs illustrated in Figs.~\ref{image:corner_b} to \ref{image:corner_e}, assuming a mass of $1.51 M_\sun$ \citep{Baines_2012_HR8799} for the star-planets system. The highly probable period ratios and the associated confidence intervals are respectively $P_b/P_c = 1.899_{-0.475}^{+0.458}$, $P_c/P_d = 2.339_{-0.572}^{+0.541}$ and $P_d/P_e = 1.578_{-0.360}^{+0.342}$. It appears clear that the most compatible mean-motion resonance between planets b and c is 1c:2b. For the other couple of planets, the most probable mean-motion resonances are respectively 2d:5c and 2e:3d, while the 1c:2d and 1d:2e solutions also have a high probability. In Table~\ref{table:mmr}, we derive the probability of various mean-motion resonances by determining the fraction of orbits with ratios of periods in a range of $10 \%$ around the selected resonance, e.g., comprised in the $[1.9, 2.1]$ interval for the 1:2 resonance, as proposed in \citet{Currie_2012_HR8799}. The 1e:2d:4c:8b configuration has been identified in the literature as a stable resonant configuration for the system \citep[see e.g.][]{Soummer_2011_HR8799, Currie_2012_HR8799, Gozdziewski_2014_HR8799}. The probabilities reported in Table \ref{table:mmr} show that this configuration is likely, with probabilities respectively equal to $16.25\%$, $12.07\%$ and $13.52\%$ for the b--c, c--d and d--e pairs.

Comparing our results with other works is not an easy task, because of the various assumptions made in the literature. We can still note that the confidence intervals reported in Table \ref{table:ConfidenceIntervals} include a significant fraction of the solutions derived in  \citet{Gozdziewski_2014_HR8799}, \citet{Currie_2012_HR8799}, \citet{Maire_2015_HR8799} and \citet{Zurlo_2016_HR8799}. In the latter two papers, only circular orbits characterized by well-defined mean-motion resonances are considered for the orbit fitting. This may explain the slight discrepancies between the results obtained from the two different approaches. In particular, the semi-major axis $a$ derived in \citet{Maire_2015_HR8799} and \citet{Zurlo_2016_HR8799} correspond systematically to the upper bound of our confidence intervals, which is linked to almost circular orbits. The only works that can be directly compared with ours are the studies of \citet{Pueyo_2015_HR8799} and \citet{Konopacky16}, which include PDFs for all the orbital elements resulting from an MCMC posterior sampling or from Monte Carlo simulations. We will refrain from giving a direct comparison of our study with the work of \citet{Pueyo_2015_HR8799} due to the discrepancies identified in Sect.~\ref{subsection:PyAstrOFit_HR8799}, and rather shortly discuss the compatibility of our results with those of \citet{Konopacky16}.

 Comparing of our PDFs with those presented by \citet{Konopacky16} shows a broad consistency between the two analyses, although their constraints on the orbital parameters are generally broader than ours, especially regarding the semi-major axis and inclination. This is most probably due to the shorter time baseline used in their analysis. Some discrepancies can be noted, though. One of them concerns the argument of the periastron for the four planets, which are generally not consistent. This is not unexpected since this orbital parameter is particularly difficult to constrain based on such small phase coverage for orbits with low eccentricity. Besides the argument of periastron, we can note some intriguing differences. The main one relates to coplanarity: our MCMC analysis does not support their conclusion that the system is most probably coplanar. More precisely, \citet{Konopacky16} do not favour the solutions for planet c with a longitude of ascending node $\Omega_c$ in the range $[125,\,153]$ deg derived in our study, while a broad peak can also be noticed around $\Omega_c=130$~deg in their PDF. They argue that a higher number of low-eccentricity solutions do favour an $\Omega_c$ near $\sim$50~deg. This is at odds with our analysis, as we do not find any peak around 50~deg in the PDF of $\Omega_c$, while the large majority of our solutions for planet c have an eccentricity lower than 0.2. Another difference concerns the inclination found for the orbital planes of planets d and e, which peaks around 45~deg in their analysis, while our results are around 30~deg for both planets. The discrepancy between these results reaches a significance level of about $2\sigma$. It is difficult to assess whether these discrepancies could be (partly) related to instrument-specific biases and/or to the different approaches used to sample the PDFs, or if they only result from the increased time baseline and enlarged data set in our study.


\section{Conclusions}
\label{section:Conclusions}

In the first part of this work, we have presented an independent data reduction of SPHERE/IRDIS images of the planetary system HR8799, acquired in the broadband H filter using coronagraphic imaging during the December 2014 science verification run. To achieve a robust determination of the astrometric position of the planets with respect to the star, we took advantage of the angular differential imaging post-processing algorithm implemented in the \texttt{VIP} pipeline \citep{Gomez_Gonzalez_2016_VIP}, based on principal component analysis, and performed a detailed analysis of the various contributions to the astrometric error budget. The resulting astrometric positions agree within $1 \sigma$ with previous estimations based on the same data set \citep{Zurlo_2016_HR8799,Apai_2016}, with error bars two to three times smaller thanks to a careful estimation of systematic errors. The main contribution to the astrometric error depends on the angular distance from the star: the error budget is dominated by the uncertainty on the stellar position ($\sim 1$~mas) and instrumental calibration errors for planet b, while residual speckle noise increases for smaller angular separations and becomes dominant for planet e. We note that these revised error bars match the early expectations of SPHERE in terms of astrometric accuracy ($\sim 2$ mas), and suggest that the astrometric accuracy could even be further improved (especially for planets located outside the speckle-noise dominated regime) by a more careful IRDIS astrometric calibration and by improving upon our estimation of the bias on the star center determination using dedicated calibration programs. In practice, nothing seems to prevent SPHERE/IRDIS from reaching a 1~mas astrometric accuracy in the future based on a careful calibration plan.

In the second part of the paper, we presented the open-source \texttt{PyAstrOFit} package written in Python 2.7, fully dedicated to orbital fitting within the Bayesian framework. Thanks to \texttt{PyAstrOFit}, we performed the orbital motion analysis and derived posterior PDFs for the six Keplerian elements of each planet. While planets b, c and e are characterized by small eccentricities, 
the eccentricity of planet d clearly peaks at $e_\text{d} = 0.35$, yet without ruling out solutions with smaller eccentricities. The combination of the posterior PDFs for the inclination and the longitude of the ascending node allowed us to evaluate the distribution of the projection on the sky of the vector perpendicular to the highly probable orbits. Our results seem to rule out the coplanarity between the four planets, in particular for planets b and c, whose orbital planes are shown to be inclined by $35\degr$ with respect to each other. 
If confirmed by future astrometric monitoring, the significant eccentricity of planet d and the non-coplanarity of planets b and c would become stringent constraints for dynamical models of the HR8799 planetary system. We note in particular that an eccentricity larger than 0.3 is generally not compatible with the predictions based on systems locked in multiple mean-motion resonances \citep{Gozdziewski_2014_HR8799}, nor with long-lived non-resonant systems \citep{Gotberg16}. New dynamical simulations based on the latest astrometric measurements will be particularly useful to constrain the origin and fate of this system. We also recommend future studies to give up on the circular and coplanar assumptions, which are not backed up by our detailed MCMC simulations, although we should not overlook the fact that orbital fitting can be strongly affected by unaccounted astrometric biases or underestimation of astrometric errors.

\begin{acknowledgements}
The research leading to these results has received funding from the European Research Council under the European Union's Seventh Framework Programme (ERC Grant Agreement n. 337569), and from the French Community of Belgium through an ARC grant for Concerted Research Action.
\end{acknowledgements}

%
%


\bibliographystyle{aa}
\bibliography{hr8799_sphere_rev2}

\begin{appendix} 


\section{Archival astrometric data}

Table~\ref{table:Compilation} compiles all astrometric measurements available in the literature for the HR8799 system.

\def\sbCOMP{.8}
\begin{table*}[h]
\tiny
 \caption[]{Compilation of astrometric measurements for HR8799bcde available in the literature and used in our orbital analysis. When more than one paper reports on the astrometry for a given data set, the values listed in the table refer to the latest astrometric measurement published in the literature. The positions and the associated error bars, the observation dates and the corresponding references can be easily retrieved using \texttt{PyAstrOFit}, through the command \texttt{PyAstrOFit.Planet}\_\texttt{data.get}\_\texttt{planet}\_\texttt{data('hr8799b')}, for instance. The list of all available planets in \texttt{PyAstrOFit} can be obtained 
through the command \texttt{PyAstrOFit.Planet}\_\texttt{data.which}\_\texttt{planet()}.}
\begin{tabular}{lccccc}
 \hline \hline
Epoch & HR8799b & HR8799c & HR8799d & HR8799e & References \\
 & $\Delta$RA $['']$, $\Delta$DEC $['']$& $\Delta$RA $['']$, $\Delta$DEC $['']$ & $\Delta$RA $['']$, $\Delta$DEC $['']$ & $\Delta$RA $['']$, $\Delta$DEC $['']$ \\ 
\hline
\scalebox{\sbCOMP}{1998.83}		& \scalebox{\sbCOMP}{$1.4110 \pm 0.0090$, $0.9860 \pm 0.0090$}			& $-$	
							& $-$																			& $-$	
							& \scalebox{\sbCOMP}{2}\\ 
							
\scalebox{\sbCOMP}{1998.83}		& \scalebox{\sbCOMP}{$1.4180 \pm 0.0220$, $1.0040 \pm 0.0200$}			& \scalebox{\sbCOMP}{$-0.8370 \pm 0.0260$, $0.4830 \pm 0.0230$}		
							& \scalebox{\sbCOMP}{$0.1330 \pm 0.0350$, $-0.5330 \pm 0.0340$}			& $-$	
							& \scalebox{\sbCOMP}{10}\\ 
							
\scalebox{\sbCOMP}{2002.54}		& \scalebox{\sbCOMP}{$1.4810 \pm 0.0230$, $0.9190 \pm 0.0170$}			& $-$	
							& $-$																			& $-$	
							& \scalebox{\sbCOMP}{3}\\ 
							
\scalebox{\sbCOMP}{2004.53}		& \scalebox{\sbCOMP}{$1.4710 \pm 0.0060$, $0.8840 \pm 0.0060$}			& \scalebox{\sbCOMP}{$-0.7390 \pm 0.0060$, $0.6120 \pm 0.0060$}		
							& $-$							& $-$ 	
							& \scalebox{\sbCOMP}{1,17}\\ 
							
\scalebox{\sbCOMP}{2005.54}		& \scalebox{\sbCOMP}{$1.4960 \pm 0.0050$, $0.8560 \pm 0.0050$}			& \scalebox{\sbCOMP}{$-0.7130 \pm 0.0050$, $0.6300 \pm 0.0050$}		
							& \scalebox{\sbCOMP}{$-0.0870 \pm 0.0100$, $-0.5780 \pm 0.0100$}							& $-$	
							& \scalebox{\sbCOMP}{11}\\ 

\scalebox{\sbCOMP}{2007.58}		& \scalebox{\sbCOMP}{$1.5040 \pm 0.0030$, $0.8370 \pm 0.0030$}			& \scalebox{\sbCOMP}{$-0.6830 \pm 0.0040$, $0.6710 \pm 0.0040$}		
							& \scalebox{\sbCOMP}{$-0.1790 \pm 0.0050$, $-0.5880 \pm 0.0050$}				& $-$ 	
							& \scalebox{\sbCOMP}{5, 17}\\ 

\scalebox{\sbCOMP}{2007.81}	& \scalebox{\sbCOMP}{$1.500 \pm 0.0070$, $0.8360 \pm 0.0070$}			& \scalebox{\sbCOMP}{$-0.6780 \pm 0.0070$, $0.6760 \pm 0.0070$}		
							& \scalebox{\sbCOMP}{$-0.1750 \pm 0.0100$, $-0.5890 \pm 0.0100$}		& $-$	
							& \scalebox{\sbCOMP}{1, 17}\\ 

\scalebox{\sbCOMP}{2008.52}		& \scalebox{\sbCOMP}{$1.5270 \pm 0.0040$, $0.7990 \pm 0.0040$}			& \scalebox{\sbCOMP}{$-0.6580 \pm 0.0040$, $0.7010 \pm 0.0040$}		
							& \scalebox{\sbCOMP}{$-0.2080 \pm 0.0040$, $-0.5820 \pm 0.0040$}							&  $-$	
							& \scalebox{\sbCOMP}{1}\\ 
							
\scalebox{\sbCOMP}{2008.61}		& \scalebox{\sbCOMP}{$1.5270 \pm 0.0020$, $0.8010 \pm 0.0020$}			& \scalebox{\sbCOMP}{$-0.6570 \pm 0.0020$, $0.7060 \pm 0.0020$}		
							& \scalebox{\sbCOMP}{$-0.2160 \pm 0.0020$, $-0.5820 \pm 0.0020$}							&  $-$	
							& \scalebox{\sbCOMP}{1}\\ 

\scalebox{\sbCOMP}{2008.71}	& \scalebox{\sbCOMP}{$1.5160 \pm 0.0040$, $0.8180 \pm 0.0040$}			& \scalebox{\sbCOMP}{$-0.6630 \pm 0.0030$, $0.6930 \pm 0.0030$}		
							& \scalebox{\sbCOMP}{$-0.2020 \pm 0.0040$, $-0.5880 \pm 0.0040$}		& $-$ 	
							& \scalebox{\sbCOMP}{1, 17}\\ 

\scalebox{\sbCOMP}{2008.89}		& \scalebox{\sbCOMP}{$1.5320 \pm 0.0200$, $0.7960 \pm 0.0200$}			& \scalebox{\sbCOMP}{$-0.6540 \pm 0.0200$, $0.7000 \pm 0.0200$}		
							& \scalebox{\sbCOMP}{$-0.2170 \pm 0.0200$, $-0.6080 \pm 0.0200$}							&  $-$	
							& \scalebox{\sbCOMP}{6, 7}\\ 
							
\scalebox{\sbCOMP}{2009.02}		& $-$																			& \scalebox{\sbCOMP}{$-0.6120 \pm 0.0300$, $0.6650 \pm 0.0300$}	
							& $-$																			& $-$	
							& \scalebox{\sbCOMP}{6}\\ 

\scalebox{\sbCOMP}{2009.58}	& \scalebox{\sbCOMP}{$1.5260 \pm 0.0040$, $0.7970 \pm 0.0040$}			& \scalebox{\sbCOMP}{$-0.6390 \pm 0.0040$, $0.7120 \pm 0.0040$}		
							& \scalebox{\sbCOMP}{$-0.2370 \pm 0.0030$, $-0.5770 \pm 0.0030$}        & \scalebox{\sbCOMP}{$-0.3060 \pm 0.0070$, $-0.2110 \pm 0.0070$} 	
							& \scalebox{\sbCOMP}{4, 17}\\ 
							
\scalebox{\sbCOMP}{2009.58}	& \scalebox{\sbCOMP}{$1.5310 \pm 0.0070$, $0.7940 \pm 0.0070$}			& \scalebox{\sbCOMP}{$-0.6350 \pm 0.0090$, $0.7220 \pm 0.0090$}		
							& \scalebox{\sbCOMP}{$-0.2500 \pm 0.0070$, $-0.5700 \pm 0.0070$}		& \scalebox{\sbCOMP}{$-0.3180 \pm 0.0100$, $-0.1950 \pm 0.0100$} 	
							& \scalebox{\sbCOMP}{4, 17}\\ 

\scalebox{\sbCOMP}{2009.62}		& \scalebox{\sbCOMP}{$1.5360 \pm 0.0100$, $0.7850 \pm 0.0100$}			& $-$		
							& $-$																			&  $-$	
							& \scalebox{\sbCOMP}{7}\\ 

\scalebox{\sbCOMP}{2009.70}		& \scalebox{\sbCOMP}{$1.5380 \pm 0.0300$, $0.7770 \pm 0.0300$}			& \scalebox{\sbCOMP}{$-0.6340 \pm 0.0300$, $0.6970 \pm 0.0300$}		
							& \scalebox{\sbCOMP}{$-0.2820 \pm 0.0300$, $-0.5900 \pm 0.0300$}																			&  $-$	
							& \scalebox{\sbCOMP}{6, 7}\\ 
							
\scalebox{\sbCOMP}{2009.76}		& \scalebox{\sbCOMP}{$1.5350 \pm 0.0200$, $0.8160 \pm 0.0200$}			& \scalebox{\sbCOMP}{$-0.6360 \pm 0.0400$, $0.6920 \pm 0.0400$}		
							& \scalebox{\sbCOMP}{$-0.2700 \pm 0.0700$, $-0.6000 \pm 0.0700$}							& $-$	
							& \scalebox{\sbCOMP}{8}\\ 

\scalebox{\sbCOMP}{2009.77}		& \scalebox{\sbCOMP}{$1.5320 \pm 0.0070$, $0.7830 \pm 0.0070$}			& \scalebox{\sbCOMP}{$-0.6270 \pm 0.0070$, $0.7160 \pm 0.0070$}		
							& \scalebox{\sbCOMP}{$-0.2410 \pm 0.0070$, $-0.5860 \pm 0.0070$}							&  \scalebox{\sbCOMP}{$-0.3060 \pm 0.0070$, $-0.2170 \pm 0.0070$}	
							& \scalebox{\sbCOMP}{7}\\ 

\scalebox{\sbCOMP}{2009.83}	& \scalebox{\sbCOMP}{$1.5240 \pm 0.0100$, $0.7950 \pm 0.0100$}			& \scalebox{\sbCOMP}{$-0.6360 \pm 0.0090$, $0.7200 \pm 0.0090$}		
							& \scalebox{\sbCOMP}{$-0.2510 \pm 0.0070$, $-0.5730 \pm 0.0070$}		& \scalebox{\sbCOMP}{$-0.3100 \pm 0.0090$, $-0.1870 \pm 0.0090$} 	
							& \scalebox{\sbCOMP}{4, 17}\\ 

\scalebox{\sbCOMP}{2009.84}		& \scalebox{\sbCOMP}{$1.5400 \pm 0.0190$, $0.8000 \pm 0.0190$}			& \scalebox{\sbCOMP}{$-0.6300 \pm 0.0130$, $0.7200 \pm 0.0130$}		
							& \scalebox{\sbCOMP}{$-0.2400 \pm 0.0140$, $-0.5800 \pm 0.0140$}							& $-$	
							& \scalebox{\sbCOMP}{9}\\ 

\scalebox{\sbCOMP}{2010.53}	& \scalebox{\sbCOMP}{$1.5320 \pm 0.0050$, $0.7830 \pm 0.0050$}			& \scalebox{\sbCOMP}{$-0.6190 \pm 0.0040$, $0.7280 \pm 0.0040$}		
							& \scalebox{\sbCOMP}{$-0.2650 \pm 0.0040$, $-0.5760 \pm 0.0040$}		& \scalebox{\sbCOMP}{$-0.3230 \pm 0.0060$, $-0.1660 \pm 0.0060$} 	
							& \scalebox{\sbCOMP}{4, 17}\\ 

\scalebox{\sbCOMP}{2010.55}		& \scalebox{\sbCOMP}{$1.5470 \pm 0.0060$, $0.7570 \pm 0.0090$}			& \scalebox{\sbCOMP}{$-0.6060 \pm 0.0060$, $0.7250 \pm 0.0060$}		
							& \scalebox{\sbCOMP}{$-0.2690 \pm 0.0060$, $-0.5800 \pm 0.0060$}							& \scalebox{\sbCOMP}{$-0.3290 \pm 0.0060$, $-0.1780 \pm 0.0060$}		
							& \scalebox{\sbCOMP}{4, 13}\\ 

\scalebox{\sbCOMP}{2010.83}	& \scalebox{\sbCOMP}{$1.5350 \pm 0.0150$, $0.7660 \pm 0.0150$}			& \scalebox{\sbCOMP}{$-0.6070 \pm 0.0120$, $0.7440 \pm 0.0120$}		
							& \scalebox{\sbCOMP}{$-0.2960 \pm 0.0130$, $-0.5610 \pm 0.0130$}		& \scalebox{\sbCOMP}{$-0.3410 \pm 0.0160$, $-0.1430 \pm 0.0160$} 	
							& \scalebox{\sbCOMP}{4, 11, 17}\\ 
							
\scalebox{\sbCOMP}{2011.55}	& \scalebox{\sbCOMP}{$1.5410 \pm 0.0050$, $0.7620 \pm 0.0050$}			& \scalebox{\sbCOMP}{$-0.5950 \pm 0.0040$, $0.7470 \pm 0.0040$}		
							& \scalebox{\sbCOMP}{$-0.3030 \pm 0.0050$, $-0.5620 \pm 0.0050$}		& \scalebox{\sbCOMP}{$-0.3520 \pm 0.0080$, $-0.1300 \pm 0.0080$} 	
							& \scalebox{\sbCOMP}{17}\\ 
							 
\scalebox{\sbCOMP}{2011.79}		& \scalebox{\sbCOMP}{$1.5790 \pm 0.0110$, $0.7340 \pm 0.0110$}			& \scalebox{\sbCOMP}{$-0.5610 \pm 0.0100$, $0.7520 \pm 0.0100$}		
							& \scalebox{\sbCOMP}{$-0.2990 \pm 0.0100$, $-0.5630 \pm 0.0100$}							& \scalebox{\sbCOMP}{$-0.3260 \pm 0.0110$, $-0.1190 \pm 0.0110$} 	
							& \scalebox{\sbCOMP}{12}\\ 
 
\scalebox{\sbCOMP}{2011.86}		& \scalebox{\sbCOMP}{$1.5460 \pm 0.0110$, $0.7250 \pm 0.0110$}			& \scalebox{\sbCOMP}{$-0.5780 \pm 0.0100$, $0.7670 \pm 0.0100$}		
							& \scalebox{\sbCOMP}{$-0.3200 \pm 0.0100$, $-0.5490 \pm 0.0100$}							& \scalebox{\sbCOMP}{$-0.3820 \pm 0.0110$, $-0.1270 \pm 0.0110$} 	
							& \scalebox{\sbCOMP}{12}\\ 
							
\scalebox{\sbCOMP}{2012.45\tablefootmark{a}}	& \scalebox{\sbCOMP}{$1.5630 \pm 0.0050$, $0.7060 \pm 0.0050$}			& \scalebox{\sbCOMP}{$-0.5580 \pm 0.0040$, $0.7650 \pm 0.0040$}		
							& \scalebox{\sbCOMP}{$-0.3230 \pm 0.0060$, $-0.5290 \pm 0.0060$}							& \scalebox{\sbCOMP}{$-0.3660 \pm 0.0060$, $-0.0900 \pm 0.0060$} 	
							& \scalebox{\sbCOMP}{14}\\ 
							
\scalebox{\sbCOMP}{2012.55}	& \scalebox{\sbCOMP}{$1.5450 \pm 0.0050$, $0.7470 \pm 0.0050$}			& \scalebox{\sbCOMP}{$-0.5780 \pm 0.0050$, $0.7610 \pm 0.0050$}		
							& \scalebox{\sbCOMP}{$-0.3390 \pm 0.0050$, $-0.5550 \pm 0.0050$}		& \scalebox{\sbCOMP}{$-0.3730 \pm 0.0080$, $-0.0840 \pm 0.0080$} 	
							& \scalebox{\sbCOMP}{17}\\ 
							
\scalebox{\sbCOMP}{2012.82}	& \scalebox{\sbCOMP}{$1.5490 \pm 0.0040$, $0.7430 \pm 0.0040$}			& \scalebox{\sbCOMP}{$-0.5720 \pm 0.0030$, $0.7680 \pm 0.0030$}		
							& \scalebox{\sbCOMP}{$-0.3460 \pm 0.0040$, $-0.5480 \pm 0.0040$}		& \scalebox{\sbCOMP}{$-0.3700 \pm 0.0090$, $-0.0760 \pm 0.0090$} 	
							& \scalebox{\sbCOMP}{17}\\ 
							
\scalebox{\sbCOMP}{2012.83}		& \scalebox{\sbCOMP}{$1.5580 \pm 0.0060$, $0.7290 \pm 0.0090$}			& \scalebox{\sbCOMP}{$-0.5570 \pm 0.0060$, $0.7630 \pm 0.0060$}		
							& \scalebox{\sbCOMP}{$-0.3430 \pm 0.0060$, $-0.5550 \pm 0.0060$}							& \scalebox{\sbCOMP}{$-0.3710 \pm 0.0060$, $-0.0800 \pm 0.0060$}		
							& \scalebox{\sbCOMP}{13}\\ 
							
\scalebox{\sbCOMP}{2013.79}	& \scalebox{\sbCOMP}{$1.5450 \pm 0.0220$, $0.7240 \pm 0.0220$}			& \scalebox{\sbCOMP}{$-0.5420 \pm 0.0220$, $0.7840 \pm 0.0220$}		
							& \scalebox{\sbCOMP}{$-0.3820 \pm 0.0160$, $-0.5220 \pm 0.0160$}		& \scalebox{\sbCOMP}{$-0.3730 \pm 0.0130$, $-0.0170 \pm 0.0130$} 	
							& \scalebox{\sbCOMP}{17}\\ 
 
\scalebox{\sbCOMP}{2013.81}		& \scalebox{\sbCOMP}{$1.5624 \pm 0.0085$, $0.7133 \pm 0.0130$}			& \scalebox{\sbCOMP}{$-0.5383 \pm 0.0060$, $0.7838 \pm 0.0131$}		
							& \scalebox{\sbCOMP}{$-0.3771 \pm 0.0070$, $-0.5380 \pm 0.0111$}							& \scalebox{\sbCOMP}{$-0.3938 \pm 0.0105$, $-0.0357 \pm 0.0168$} 	
							& \scalebox{\sbCOMP}{15}\\ 
							
\scalebox{\sbCOMP}{2014.53}		& \scalebox{\sbCOMP}{$1.5700 \pm 0.0060$, $0.7070 \pm 0.0060$}			& \scalebox{\sbCOMP}{$-0.5220 \pm 0.0040$, $0.7910 \pm 0.0040$}		
							& \scalebox{\sbCOMP}{$-0.3900 \pm 0.0050$, $-0.5300 \pm 0.0060$}							& \scalebox{\sbCOMP}{$-0.3860 \pm 0.0090$, $-0.0080 \pm 0.0090$} 	
							& \scalebox{\sbCOMP}{16}\\ 
							
\scalebox{\sbCOMP}{2014.54}	& \scalebox{\sbCOMP}{$1.5600 \pm 0.0130$, $0.7250 \pm 0.0130$}			& \scalebox{\sbCOMP}{$-0.5400 \pm 0.0130$, $0.7990 \pm 0.0130$}		
							& \scalebox{\sbCOMP}{$-0.4000 \pm 0.0110$, $-0.5340 \pm 0.0110$}		& \scalebox{\sbCOMP}{$-0.3870 \pm 0.0110$, $0.0030 \pm 0.0110$} 	
							& \scalebox{\sbCOMP}{17}\\ 
							
\scalebox{\sbCOMP}{2014.62}		& $-$																			& $-$	
							& \scalebox{\sbCOMP}{$-0.3910 \pm 0.0040$, $-0.5290 \pm 0.0040$}							& \scalebox{\sbCOMP}{$-0.3840 \pm 0.0020$, $-0.0050 \pm 0.0020$} 	
							& \scalebox{\sbCOMP}{16}\\ 

\scalebox{\sbCOMP}{2014.93}		& \scalebox{\sbCOMP}{$1.5748 \pm 0.0023$, $0.7016 \pm 0.0036$}			& \scalebox{\sbCOMP}{$-0.5113 \pm 0.0024$, $0.7985 \pm 0.0020$}		
							& \scalebox{\sbCOMP}{$-0.4001 \pm 0.0021$, $-0.5233 \pm 0.0020$}							& \scalebox{\sbCOMP}{$-0.3853 \pm 0.0030$, $0.0115 \pm 0.0021$} 	
							& \scalebox{\sbCOMP}{18}\\  
\hline
\end{tabular}
\tablebib{
(1) \citet{Marois_2008_HR8799}, (2) \citet{Lafreniere_2009_HR8799}, (3) \citet{Fukagawa_2009_HR8799}, (4) \citet{Marois_2010b_HR8799}, (5) \citet{Metchev_2009_HR8799}, (6) \citet{Hinz_2010_HR8799}, (7) \citet{Currie_2011_HR8799}, 
(8) \citet{Bergfors_2011_HR8799}, (9) \citet{Galicher_2011_HR8799}, (10) \citet{Soummer_2011_HR8799}, (11) \citet{Currie_2012_HR8799}, (12) \citet{Esposito_2013_HR8799}, (13) \citet{Currie_2014_HR8799}, (14) \citet{Pueyo_2015_HR8799}, 
(15) \citet{Maire_2015_HR8799}, (16) \citet{Zurlo_2016_HR8799}, (17) \citet{Konopacky16}, (18) this work.\\
\tablefoottext{a}{The epoch reported in \citet{Pueyo_2015_HR8799} has been discarded from our orbital fitting, see Sect.~\ref{subsection:PyAstrOFit_HR8799} for details.}
}
\label{table:Compilation}
\end{table*}


\section{Illustration of the orbital solutions}

Figures~\ref{figure:full_highly_probable_orbits} and \ref{figure:zoom_highly_probable_orbits} give an illustration of the highly probable orbits resulting from our MCMC simulations.

\def\zoom{9cm}
\begin{figure*}
	\centering
   	\includegraphics[width=\zoom]{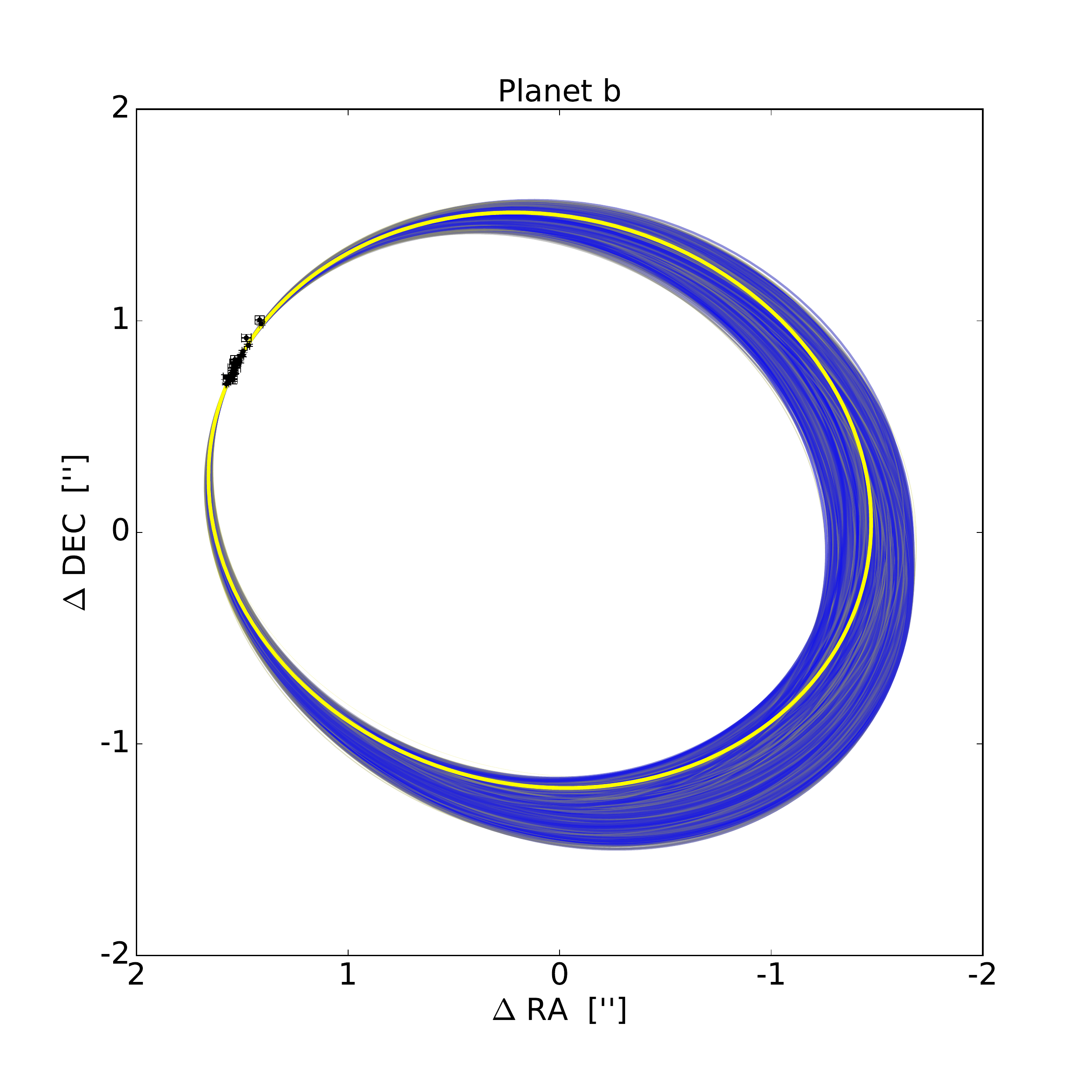}
	\includegraphics[width=\zoom]{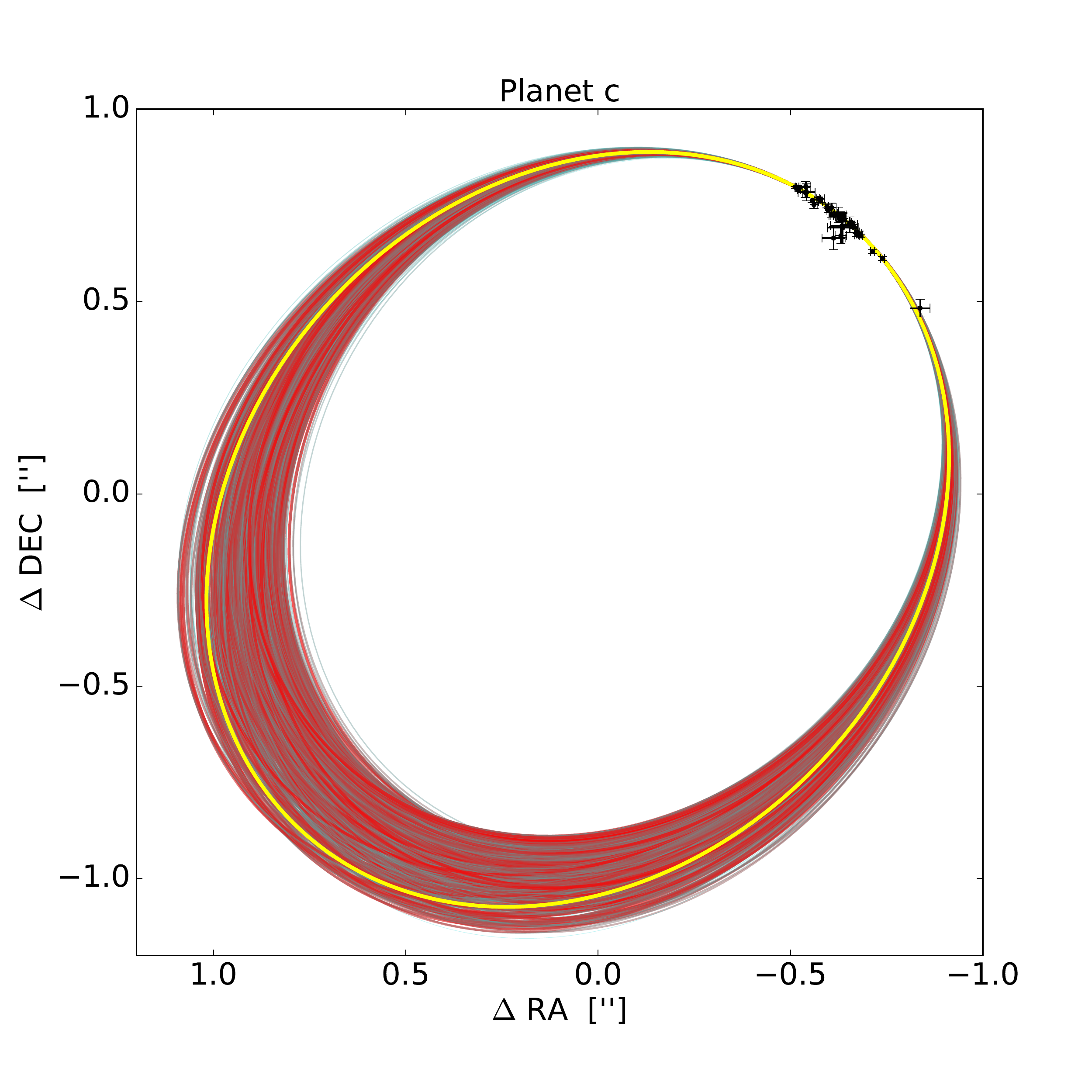}
	\includegraphics[width=\zoom]{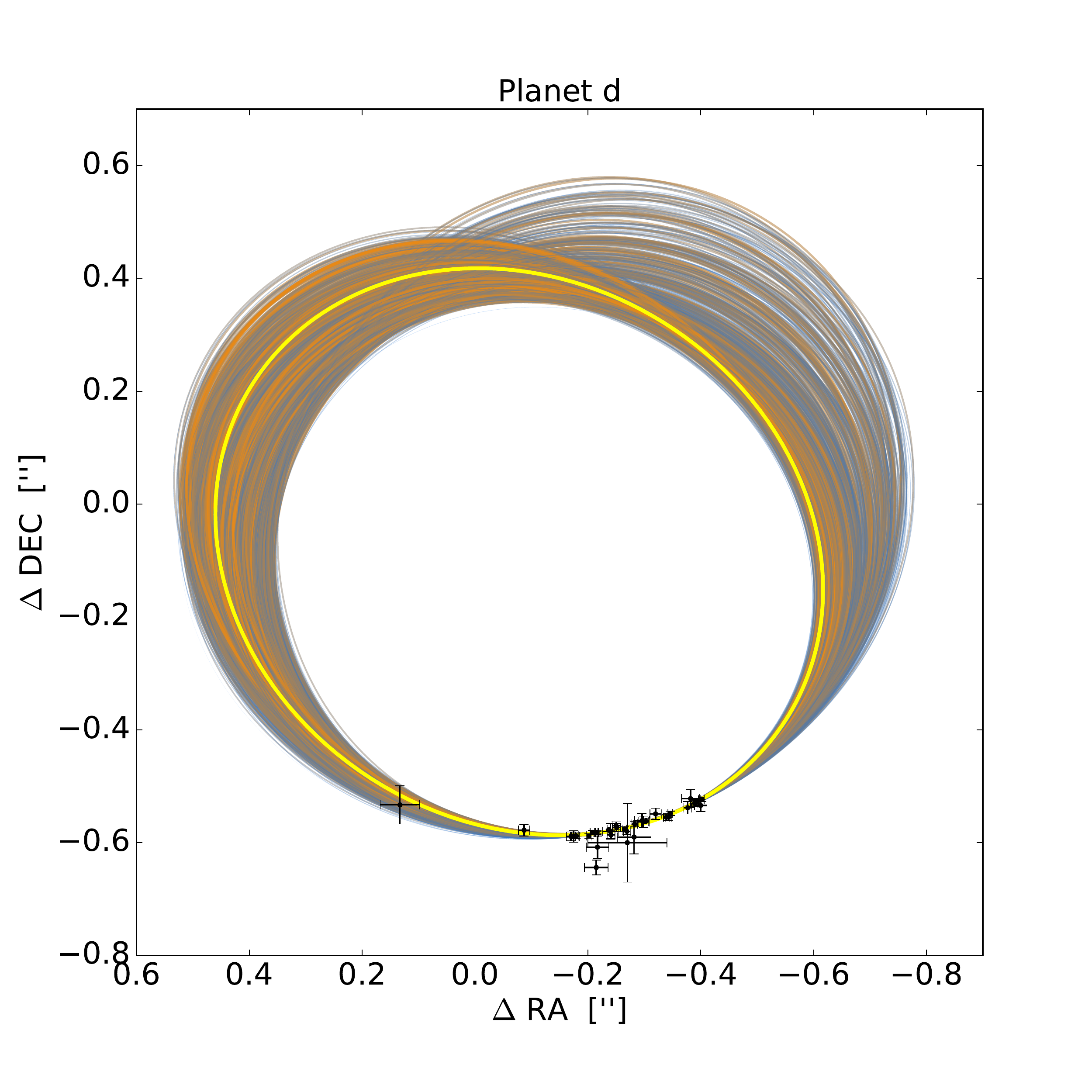}
	\includegraphics[width=\zoom]{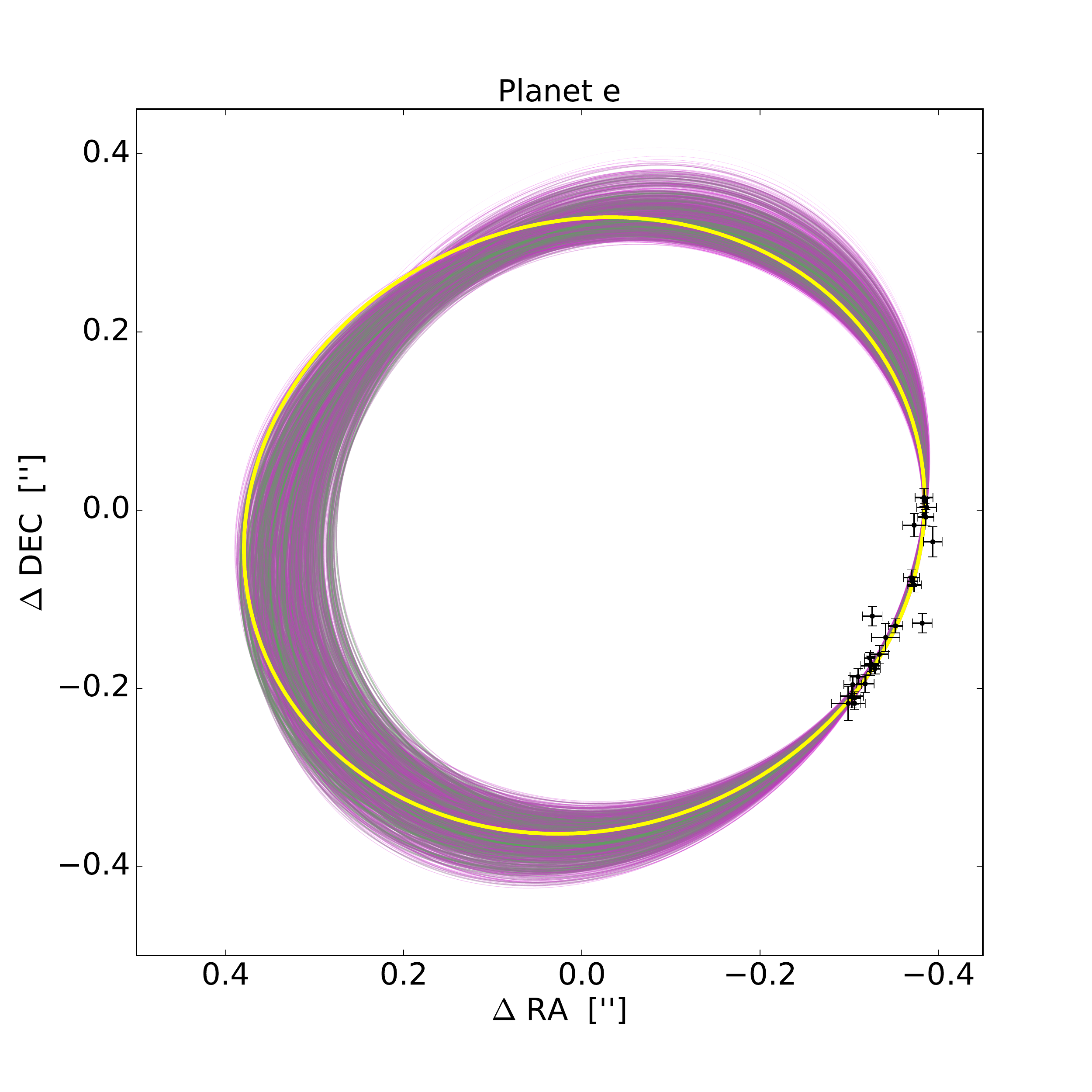}
     	\caption{Illustration of the on-sky projection of 1000 allowable orbits for HR8799bcde, all characterized by $\chi^2 < \chi^2_{\text{min}} + 0.1$ and by Keplerian parameters in the confidence intervals reported in Table \ref{table:ConfidenceIntervals}. The yellow orbit corresponds to the $\chi^2_{\text{min}}$ solution reported in Table \ref{table:ConfidenceIntervals}.}
       	\label{figure:full_highly_probable_orbits}       
\end{figure*} 

\def\zoom{9cm}
\begin{figure*}
	\centering
   	\includegraphics[width=\zoom]{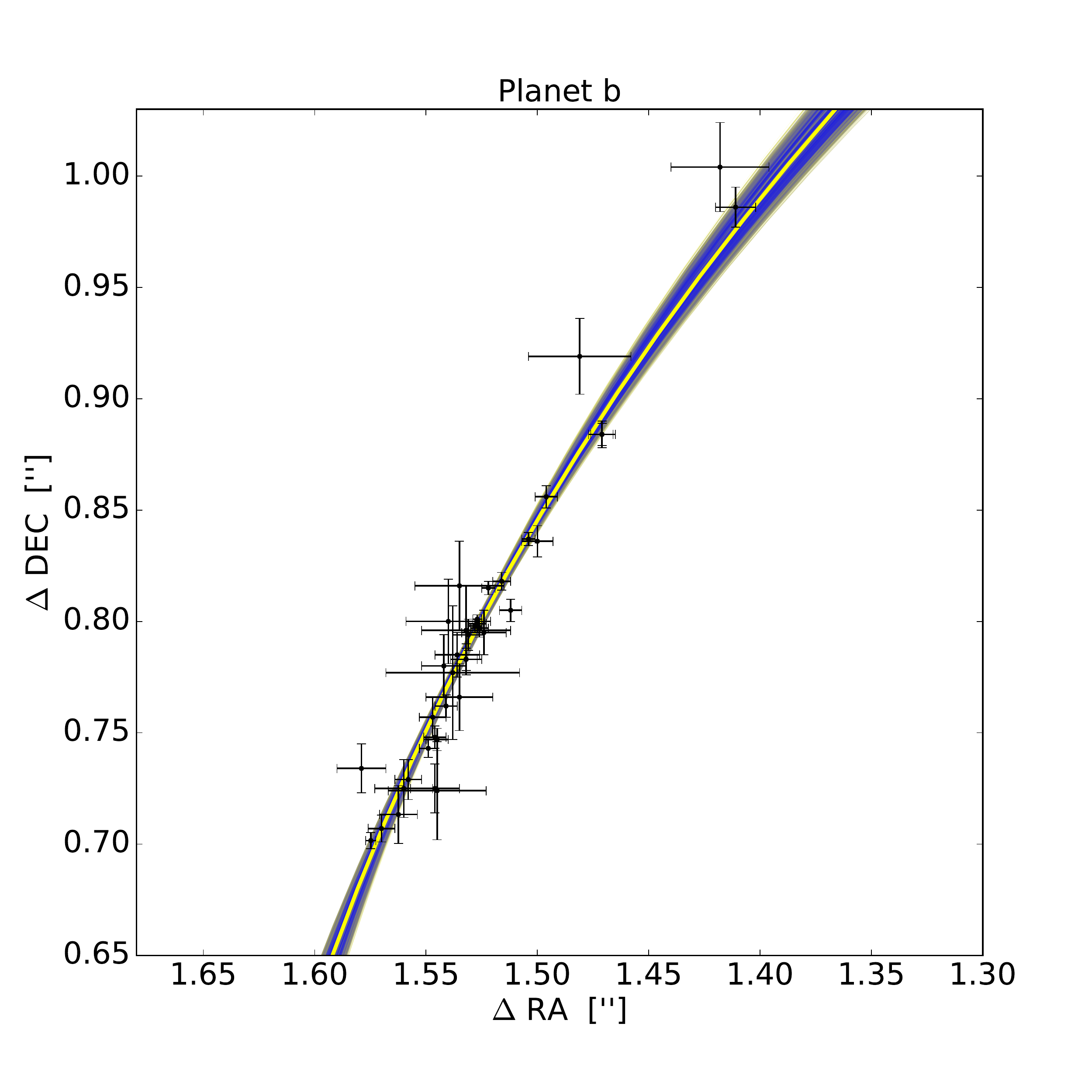}
   	\includegraphics[width=\zoom]{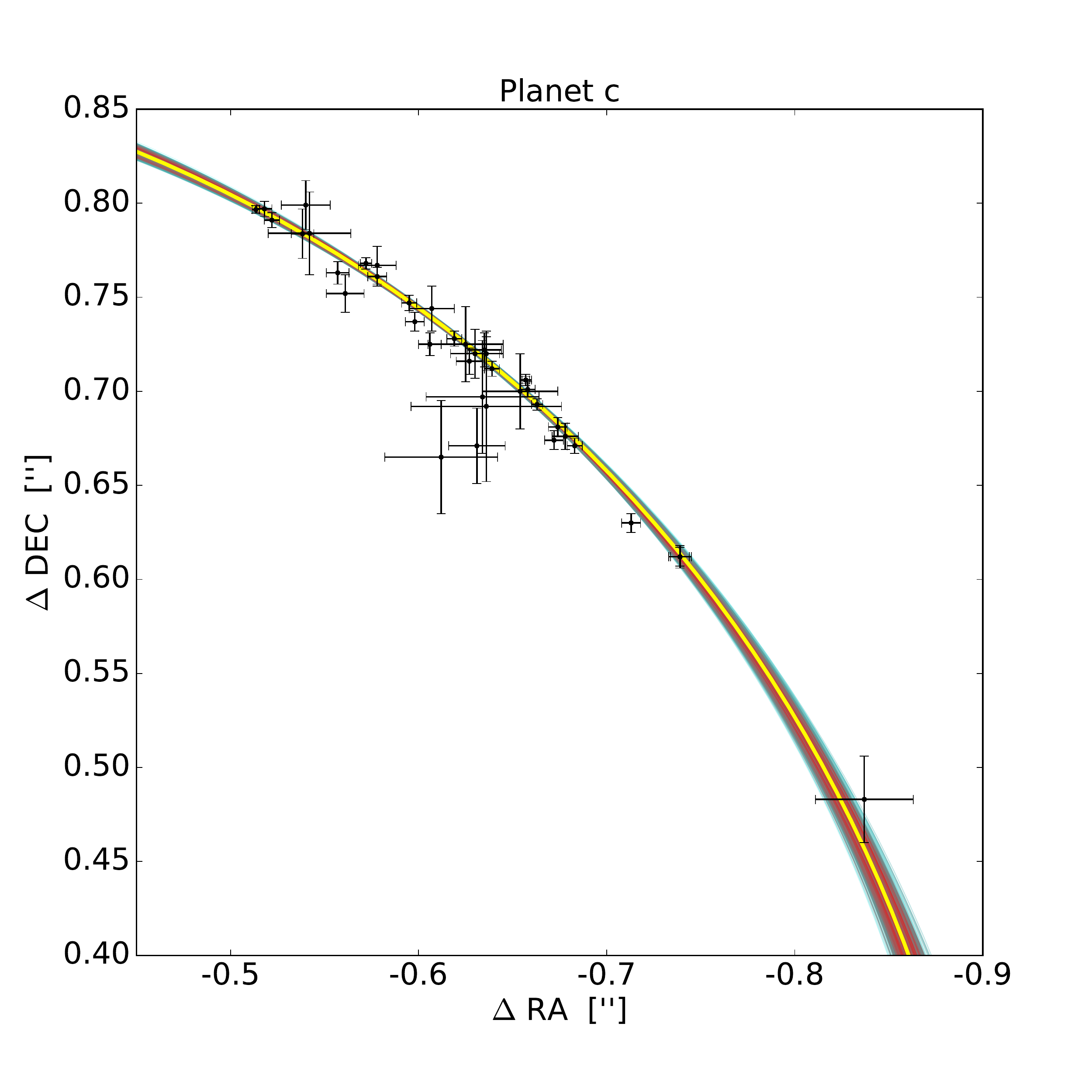}
   	\includegraphics[width=\zoom]{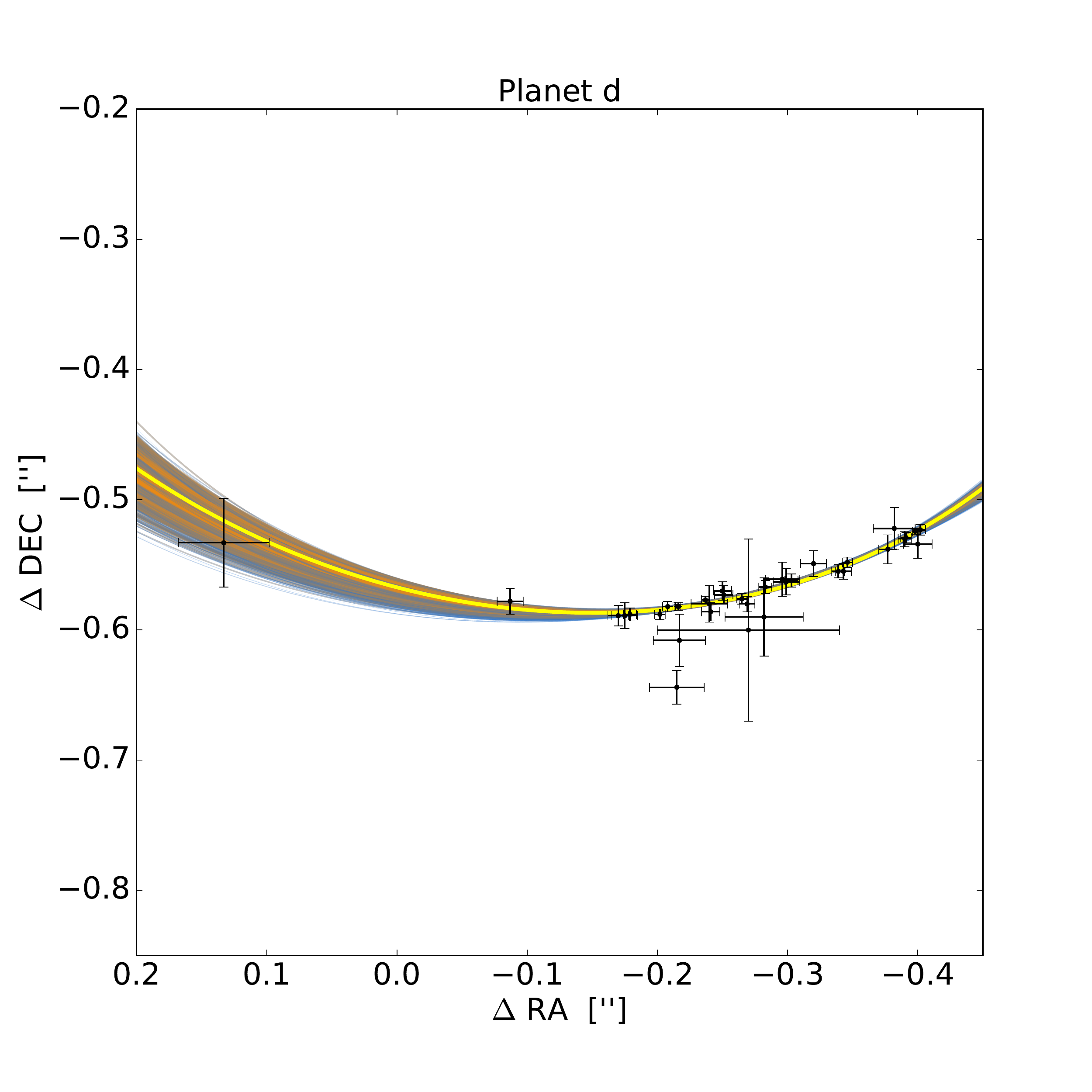}
   	\includegraphics[width=\zoom]{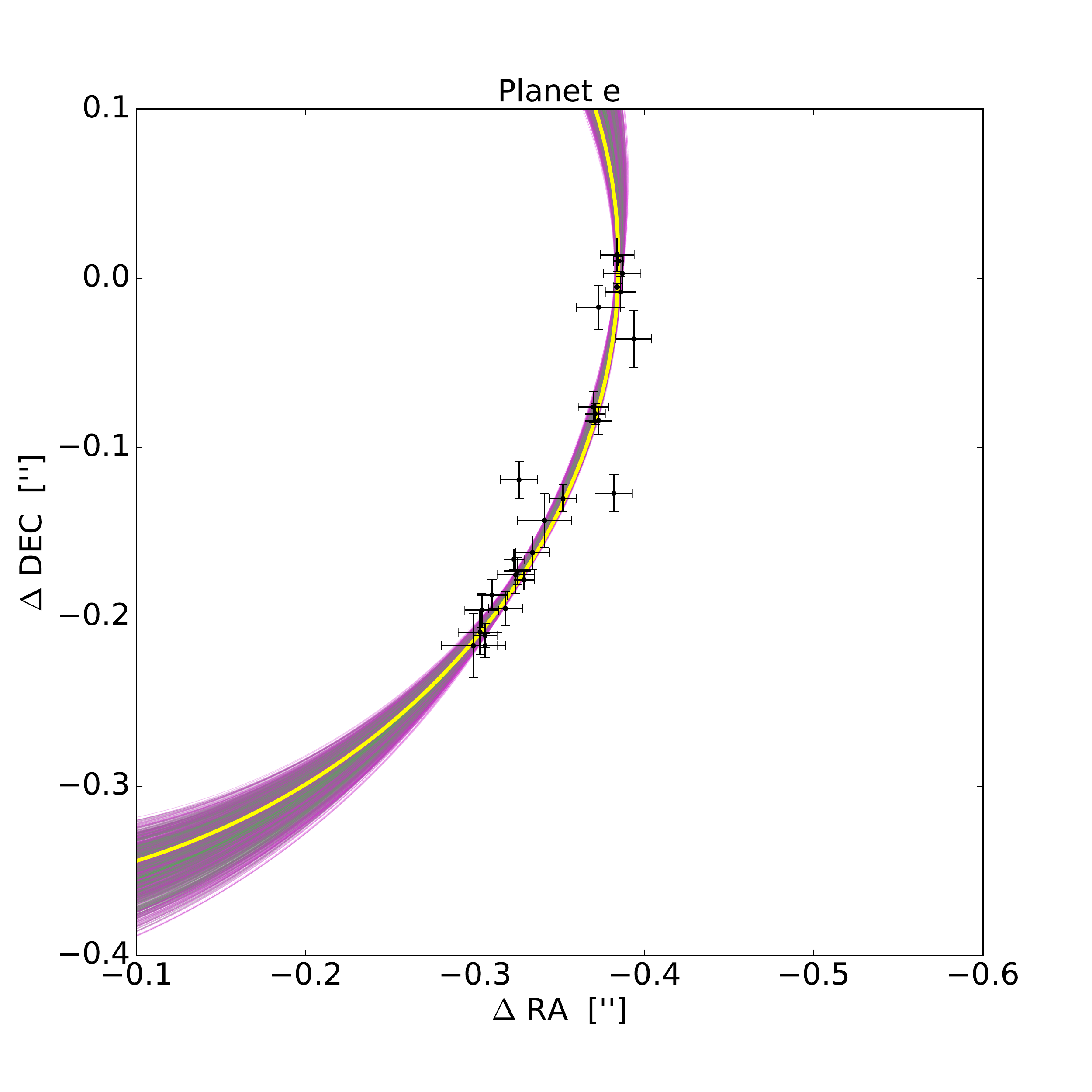}
     	\caption{Same as Fig.~\ref{figure:full_highly_probable_orbits}, zooming on the part of the orbit where astrometric measurements are available.}
       	\label{figure:zoom_highly_probable_orbits}       
\end{figure*}

\end{appendix}

\end{document}